\documentclass[10pt,letterpaper]{article}
\usepackage[applemac]{inputenc}
\usepackage{cite}
\usepackage{amssymb}
\usepackage{amsmath}
\usepackage{pstricks, pst-node, pst-tree}
\usepackage{fullpage}
\usepackage{relate}
\usepackage{changebar}
\usepackage{graphicx}
\usepackage{algorithmic}
\usepackage[linesnumbered,ruled,oldcommands,nokwfunc]{algorithm2e}

\newtheorem{lemma}{Lemma}
\newtheorem{theorem}{Theorem}
\newtheorem{corollary}{Corollary}

\newtheorem{prop}{Proposition}

\newcommand{\qed}{\hfill\ensuremath{\Box}\medskip\\\noindent}
\newenvironment{proof}{\noindent\emph{Proof. }}{\qed}
\newenvironment{invariant}{\noindent {\bf Invariant}}{}

\newcommand{\ceil}[1]{\left\lceil{#1}\right\rceil}

\newcommand{\showOld}[2]{#2}
\newcommand{\ignore}[1]{}

\newcommand{\norm}[1]{\ensuremath{| #1 |}}

\newcommand{\match}{\ensuremath{\textsc{match}}}
\newcommand{\mopc}{\ensuremath{\textsc{mop}}}
\newcommand{\ncac}{\ensuremath{\textsc{nca}}}
\newcommand{\parentc}{\ensuremath{\textsc{parent}}}
\newcommand{\flc}{\ensuremath{\textsc{fl}}}
\newcommand{\deepc}{\ensuremath{\textsc{deep}}}
\newcommand{\ancestorc}{\ensuremath{\textsc{ancestor}}}
\newcommand{\mc}[1]{\ensuremath{\mathcal{#1}}}

\newcommand{\rn}{\ensuremath{\textsc{right}}}
\newcommand{\leftn}{\ensuremath{\textsc{left}}}
\newcommand{\size}{\ensuremath{\textsc{size}}}
\newcommand{\sizes}[1]{\ensuremath{\textsc{size}(#1)}}
\newcommand{\restrict}[2]{\ensuremath{\mathop{#1|}_{#2}}}
\newcommand{\leftof}{\ensuremath{\textsc{leftof}}}
\newcommand{\rightof}{\ensuremath{\textsc{rightof}}}
\newcommand{\Pred}{\ensuremath{\mathsf{Pred}}}
\newcommand{\SuccL}{\ensuremath{\mathsf{Succ}_L}}
\newcommand{\PredL}{\ensuremath{\mathsf{Pred}_L}}
\newcommand{\SuccZ}{\ensuremath{\mathsf{Succ}_Z}}
\newcommand{\PredZ}{\ensuremath{\mathsf{Pred}_Z}}
\newcommand{\Succ}{\ensuremath{\mathsf{Succ}}}

\newcommand{\Eq}{\ensuremath{\textsc{eq}}}

\newcommand{\lightdepth}{\ensuremath{\mathrm{ldepth}}}
\newcommand{\heavy}{\ensuremath{\mathrm{heavy}}}

\newcommand{\depth}{\ensuremath{\mathrm{depth}}}

\newcommand{\roots}{\ensuremath{\mathrm{root}}}
\newcommand{\parent}{\ensuremath{\mathrm{parent}}}
\newcommand{\child}{\ensuremath{\mathrm{child}}}
\newcommand{\lab}{\ensuremath{\mathrm{label}}}
\newcommand{\nca}{\ensuremath{\mathrm{nca}}}
\newcommand{\mop}{\ensuremath{\mathrm{mop}}}
\newcommand{\fl}{\ensuremath{\mathrm{fl}}}

\newcommand{\pre}{\ensuremath{\mathrm{pre}}}
\newcommand{\post}{\ensuremath{\mathrm{post}}}
\newcommand{\first}{\ensuremath{\mathrm{first}}}
\newcommand{\emb}{\ensuremath{\mathrm{emb}}}

\newcommand{\Nca}{\ensuremath{\textsc{Nca}}}

\newcommand{\Fl}{\ensuremath{\textsc{Fl}}}
\newcommand{\Mop}{\ensuremath{\textsc{MopRight}}}
\newcommand{\MopLeft}{\ensuremath{\textsc{MopLeft}}}
\newcommand{\Mopsim}{\ensuremath{\textsc{MopSim}}}
\newcommand{\Match}{\ensuremath{\textsc{Match}}}

\newcommand{\Deep}{\ensuremath{\textsc{Deep}}}
\newcommand{\Parent}{\ensuremath{\textsc{Parent}}}
\newcommand{\Emb}{\ensuremath{\textsc{Emb}}}

\newcommand{\Cluster}{\ensuremath{\textsc{Cluster}}}

\title{The Tree Inclusion Problem: In Linear Space and Faster\thanks{An extended abstract of this paper appeared in Proceedings of the 32nd International Colloquium on Automata, Languages and Programming, Lecture Notes in Computer Science, vol. 3580, pp. 66-77, Springer-Verlag, 2005.}}

\author{Philip Bille\thanks{Technical University of Denmark,  Department of Informatics and Mathematical Modelling. This work is part of the DSSCV project supported by the IST Programme of the European Union (IST-2001-35443).} 
\and Inge Li G{\o}rtz\thanks{Corresponding author: Technical University of Denmark,  Department of Informatics and Mathematical Modelling,
Building 322, Office 124,
DK-2800 Kongens Lyngby,
Denmark. Phone: (+45) 45 25 36 73. Fax: (+45) 45 88 26 73. Email: {\tt ilg@imm.dtu.dk}.}}

\date{}

\begin{document}
\maketitle

\begin{abstract}
Given two rooted, ordered, and labeled trees $P$ and $T$ the tree
inclusion problem is to determine if $P$ can be obtained from $T$
by deleting nodes in $T$. This problem has recently been
recognized as an important query primitive in XML databases.
Kilpel\"{a}inen and Mannila [\emph{SIAM J. Comput. 1995}] presented the
first polynomial time algorithm using quadratic time and space.
Since then several improved results have been obtained for special
cases when $P$ and $T$ have a small number of leaves or small
depth. However, in the worst case these algorithms still use
quadratic time and space. 
Let $n_S$, $l_S$, and $d_S$ denote the
number of nodes, the number of leaves, and the 
depth of a tree
$S \in \{P, T\}$. In this paper we show that the tree inclusion
problem can be solved in space $O(n_T)$ and time:
\begin{equation*}
O\left( \min\left\{
\begin{array}{l}
    l_Pn_T \\
      l_Pl_T\log \log n_T + n_T \\
      \frac{n_Pn_T}{\log n_T} + n_{T}\log n_{T}
\end{array}\right\}
\right)
\end{equation*}

This improves or matches the best
known time complexities while using only linear space instead of
quadratic. This is particularly important in practical applications,
such as XML databases, where the space is likely to be a bottleneck.
\end{abstract}

\section{Introduction}
Let $T$ be a rooted tree. We say that $T$ is \emph{labeled} if
each node is assigned a character from an alphabet
$\Sigma$ and we say that $T$ is \emph{ordered} if a left-to-right
order among siblings in $T$ is given. All trees in this paper are rooted, ordered, and labeled.
A tree $P$ is \emph{included} in $T$, denoted $P\sqsubseteq T$, if $P$ can be
obtained from $T$ by deleting nodes of $T$. Deleting a node $v$ in
$T$ means making the children of $v$ children of the parent of $v$
and then removing $v$. The children are inserted in the place of
$v$ in the left-to-right order among the siblings of $v$. The \emph{tree inclusion problem}
is to determine if $P$ can be included in $T$ and if so report all subtrees of $T$ that include $P$.
\begin{figure}[t]
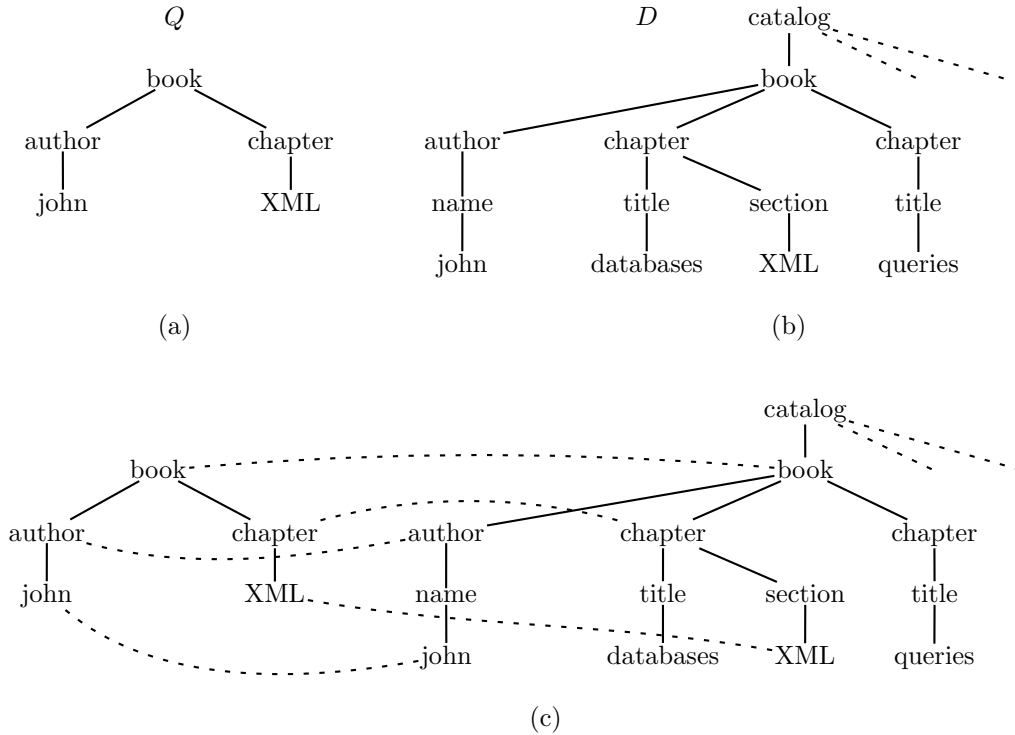

\begin{center}
  \begin{psmatrix}[colsep=0.6cm,rowsep=0.4cm,labelsep=1pt, nodesep=1pt]
  &&$Q$&&&&&$D$&[name=cat] catalog \\
  &&[name=book] book &&&&&&[name=book2] book & [name=book3] & [name=book4]\\
  & [name=author] author &&[name=chapter] chapter&&
  [name=author2] author && [name=chapter2] chapter && [name=chapter3] chapter \\
  & [name=john] john && [name=xml] XML &&
  [name=name] name && [name=title] title  & [name=section] section & [name=title2] title \\
  &&&&&  [name=john2] john && [name=DB] databases & [name=xml2] XML & [name=queries] queries \\
  && (a) &&&&&& (b) \\
  \ncline{author}{book} \ncline{chapter}{book}
  \ncline{john}{author} \ncline{xml}{chapter}
  \ncline{cat}{book2}\psset{linestyle=dashed,dash=2pt 4pt}\ncline{cat}{book3} \ncline{cat}{book4}
  \psset{linestyle=solid}
  \ncline{book2}{author2}\ncline{book2}{chapter2}\ncline{book2}{chapter3}
  \ncline{author2}{name}\ncline{chapter2}{title}\ncline{chapter2}{section}\ncline{chapter3}{title2}
  \ncline{name}{john2}\ncline{title}{DB}\ncline{section}{xml2}\ncline{title2}{queries}
  \end{psmatrix}

  \begin{psmatrix}[colsep=0.6cm,rowsep=0.4cm,labelsep=1pt, nodesep=1pt]
  &&&&&&&&[name=cat] catalog \\
  &&[name=book] book &&&&&&[name=book2] book & [name=book3] & [name=book4]\\
  & [name=author] author &&[name=chapter] chapter&&
  [name=author2] author && [name=chapter2] chapter && [name=chapter3] chapter \\
  & [name=john] john && [name=xml] XML &&
  [name=name] name && [name=title] title  & [name=section] section & [name=title2] title \\
  &&&&&  [name=john2] john && [name=DB] databases & [name=xml2] XML & [name=queries] queries \\
  &&&&&&(c)&&&& \\
  \ncline{author}{book} \ncline{chapter}{book}
  \ncline{john}{author} \ncline{xml}{chapter}
  \ncline{cat}{book2}\psset{linestyle=dashed,dash=2pt 4pt} \ncline{cat}{book3} \ncline{cat}{book4}
  \psset{linestyle=solid}
  \ncline{book2}{author2}\ncline{book2}{chapter2}\ncline{book2}{chapter3}
  \ncline{author2}{name}\ncline{chapter2}{title}\ncline{chapter2}{section}\ncline{chapter3}{title2}
  \ncline{name}{john2}\ncline{title}{DB}\ncline{section}{xml2}\ncline{title2}{queries}

  \psset{linestyle=dashed}
  \nccurve[angleA=5,angleB=175]{book}{book2}
  \nccurve[angleA=345,angleB=190]{author}{author2}
  \nccurve[angleA=320,angleB=190]{john}{john2}
  \nccurve[angleA=15,angleB=165]{chapter}{chapter2}
  \nccurve[angleA=350,angleB=170]{xml}{xml2}
  \end{psmatrix}
   \caption{(a) The tree $Q$ corresponding to the query. (b) A fragment of the tree $D$. Can the tree $Q$ be included in the tree $D$? It can and an embedding is given in (c).}
  \label{inclusionexample}
  \end{center}
\end{figure}

Recently, the problem has been recognized as an important query
primitive for XML data and has received considerable attention,
see e.g., \cite{SM2002, YLH2003,YLH2004, ZADR03, SN2000, TRS2002}.
The key idea is that an XML document can be viewed as a tree and queries on the document correspond to a tree
inclusion problem. As an example consider Figure~\ref{inclusionexample}. Suppose that we want to maintain
a catalog of books for a bookstore. A fragment of the tree,
denoted $D$, corresponding to the catalog is shown in (b). In addition to supporting
full-text queries, such as find all documents containing the word
"John", we can also utilize the tree structure of the catalog to
ask more specific queries, such as "find all books written by John
with a chapter that has something to do with XML". We can model
this query by constructing the tree, denoted $Q$, shown in (a) and
solve the tree inclusion problem: is $Q \sqsubseteq D$? The answer
is yes and a possible way to include $Q$ in $D$ is indicated by
the dashed lines in (c). If we delete all the nodes in $D$ not
touched by dashed lines the trees $Q$ and $D$ become isomorphic.
Such a mapping of the nodes from $Q$ to $D$ given by the dashed
lines is called an \emph{embedding} (formally defined in
Section~\ref{sec:recursion}). We note that the ordering
of the XML document, and hence the left-to-right order of siblings,
is important in many cases. For instance, in the above example, the
relative order of contents of the chapters is most likely important.
Also, in biological databases, order is of critical importance.
Consequently, standard XML query languages, such as XPath~\cite{CD99}  and
XQuery~\cite{BCFF01}, require the output of queries to be ordered.

The tree inclusion problem was initially introduced by Knuth
\cite[exercise 2.3.2-22]{Knuth1969} who gave a sufficient
condition for testing inclusion. Motivated by applications in
structured databases \cite{KM93, MR90} Kilpel\"{a}inen and Mannila
\cite{KM1995} presented the first polynomial time algorithm using
$O(n_Pn_T)$ time and space, where $n_P$ and $n_T$ is the number of
nodes in $P$ and $T$, respectively. During the last decade
several improvements of the original algorithm of \cite{KM1995}
have been suggested \cite{Kilpelainen1992,AS2001, Richter1997a,
Chen1998}. The previously best known bound is due to Chen
\cite{Chen1998} who presented an algorithm using $O(l_Pn_T)$ time
and $O(l_P \cdot \min\{d_T, l_T\})$ space. Here, $l_S$ and $d_S$ denote
the number of leaves and the 
depth of a tree $S$,
respectively. This algorithm is based on an algorithm of
Kilpel\"{a}inen \cite{Kilpelainen1992}. Note that the time and
space is still $\Theta(n_Pn_T)$ for worst-case input trees.

In this paper we present three algorithms which combined improve all of the previously known time and space bounds. To avoid trivial cases we always assume that $1 \leq n_P \leq n_T$. We show the following theorem:
\begin{theorem}\label{thm:main}
For trees $P$ and $T$ the tree inclusion problem can be solved in $O(n_T)$ space with the following running time:
\begin{equation*}
O\left( \min\left\{
\begin{array}{l}
    l_Pn_T \\
      l_Pl_T\log \log n_T + n_T \\
      \frac{n_Pn_T}{\log n_T} + n_{T}\log n_{T}
\end{array}\right\}
\right)
\end{equation*}
\end{theorem}
Hence, when $P$ has few leaves we obtain a fast algorithm and even faster if both $P$ and $T$ have few leaves.  
When both trees have many leaves and $n_{P} = \Omega (\log^{2} n_{T})$, we instead improve the previous quadratic time bound by a logarithmic factor. Most importantly, the space used is linear. In the context of XML databases this will likely make it possible to query larger trees and speed up the query time since more of the computation can be kept in main memory. 

The extended abstract of this paper~\cite{BG05} contained an
error. The algorithms in the paper~\cite{BG05} did not use linear space. The
problem was due to a recursive traversal of $P$ which stored too many
sets of nodes leading to a worst-case space complexity of $\Omega(d_P
l_T)$. In this paper we fix this problem by recursively visiting the
nodes such that the child with the largest number of descendant leaves
is visited first, and by showing that the size of the resulting stored
node sets exponentially decrease. With these ideas we show that all of
our algorithms use $O(n_T)$ space. Additionally, our improved analysis
of the sizes of the stored node sets also leads to an improvement in
the running time of the algorithm in the second case above. 
In the previous paper the  running time was $O(n_pl_T\log\log n_T+ n_T)$. 

\subsection{Techniques}
Most of the previous algorithms, including the best one
\cite{Chen1998}, are essentially based on a simple dynamic
programming approach from the original algorithm of \cite{KM1995}. The
main idea behind this algorithm is the following: Let $v$ be a node in $P$ with children $v_1, \ldots, v_i$ and let $w$ be a node in $T$. Consider the subtrees rooted at $v$ and $w$, denoted by $P(v)$ and $T(w)$. To decide if $P(v)$ can be included in $T(w)$ we try to find a sequence $w_{1}, \ldots, w_{i}$ of left-to-right ordered descendants of $w$ such that $P(v_k) \sqsubseteq T(w_{k})$ for all $k$, $1 \leq k \leq i$. The sequence is computed greedily from left-to-right in $T(w)$ effectively finding the \emph{leftmost inclusion} of $P(v)$ in $T(w)$. Applying this approach in a bottom-up fashion we can determine, if $P(v) \sqsubseteq T(w)$, for all pairs of nodes $v$ in $P$ and $w$ in $T$.


In this paper we take a different approach. The main
idea is to construct a data structure on $T$ supporting a small
number of procedures, called the \emph{set procedures}, on subsets
of nodes of $T$. We show that any such data structure implies an
algorithm for the tree inclusion problem. We consider various
implementations of this data structure which all use linear space.
The first simple implementation gives an algorithm with
$O(l_Pn_T)$ running time. As it turns out, the running time
depends on a well-studied problem known as the \emph{tree color
problem}. We show a direct connection between a data structure
for the tree color problem and the tree inclusion problem.
Plugging in a data structure of Dietz \cite{Die89} we obtain an
algorithm with $O(l_Pl_T\log \log n_T + n_T)$ running time.

Based on the simple algorithms above we show how to improve the
worst-case running time of the set procedures by a logarithmic
factor. The general idea used to achieve this is to divide $T$
into small trees called \emph{clusters} of logarithmic size which overlap with other clusters in at most $2$ nodes. Each cluster is represented by a constant number of nodes in a \emph{macro tree}. The nodes in the
macro tree are then connected according to the overlap of the cluster they represent. We show how to efficiently preprocess the clusters and the macro tree such that the set procedures use constant time for each cluster. Hence, the worst-case quadratic running time is improved by a logarithmic factor.

Our algorithms recursively traverse $P$
top-down. For each node $v \in V(P)$ we compute a set of nodes
representing all of the subtrees in $T$ that include $P(v)$. To avoid
storing too many of these node sets the traversal of $P$ visits the
child with the largest number of descendant leaves first. For the
first two algorithms this immediately implies a space complexity of
$O(l_T \log l_P)$, however, by carefully analyzing the sizes of stored
node sets we are able to show that they decrease exponentially leading
to the linear space bound. In the last algorithm the node sets are
compactly encoded in $O(n_T/\log n_T)$ space and therefore our
recursive traversal alone implies a space bound of $O(n_T/\log n_T
\cdot \log l_P) = O(n_T)$.

Throughout the paper we assume a unit-cost RAM model of computation with word size $\Theta(\log n_T)$ and a standard instruction set including bitwise boolean operations, shifts, addition, and multiplication. All space complexities refer to the number of words used by the algorithm.

\subsection{Related Work}
For some applications considering \emph{unordered} trees is more
natural. However, in \cite{MT1992,KM1995} this problem was proved
to be NP-complete. The tree inclusion problem is closely related
to the \emph{tree pattern matching problem} \cite{CO1982,
Kosaraju1989,DGM1990, CHI1999}. The goal is here to find an
injective mapping $f$ from the nodes of $P$ to the nodes of $T$
such that for every node $v$ in $P$ the $i$th child of $v$ is
mapped to the $i$th child of $f(v)$. The tree pattern matching
problem can be solved in $(n_P+n_T) \log^{O(1)} (n_P+n_T)$ time. Another similar problem is the \emph{subtree isomorphism} problem \cite{Chung1987, ST1999}, which is to determine if $T$ has
a subgraph isomorphic to $P$. The subtree isomorphism
problem can be solved efficiently for ordered and unordered trees.
The best algorithms for this problem use $O(\frac{n_P^{1.5}n_T}{\log
n_P} + n_T)$ time for unordered trees and $O(\frac{n_Pn_T}{\log n_P} + n_T)$ time for ordered
trees \cite{Chung1987, ST1999}. Both use $O(n_Pn_T)$ space. The
tree inclusion problem can be considered a special case of the
\emph{tree edit distance problem} \cite{Tai1979, ZS1989,
Klein1998, DMRW07}. Here one wants to find the minimum sequence of insert,
delete, and relabel operations needed to transform $P$ into $T$.
Currently the best algorithm for this problem uses $O(n_T n_P^2 (1 + \log \frac{n_T}{n_P}))$ time~\cite{DMRW07}. For more details and references see the survey \cite{Bille2005}.

\subsection{Outline}
In Section~\ref{def} we give notation and definitions used
throughout the paper. In Section~\ref{sec:recursion} a common
framework for our tree inclusion algorithms is given.
Section~\ref{simple} presents two simple algorithms and then, based on these results, we show how to get a faster algorithm in Section~\ref{micromacro}.

\section{Notation and Definitions}\label{def}
In this section we define the notation and definitions we will use
throughout the paper. For a graph $G$ we denote the set of nodes
and edges by $V(G)$ and $E(G)$, respectively. Let $T$ be a rooted
tree. The root of $T$ is denoted by $\roots(T)$. The \emph{size}
of $T$, denoted by $n_T$, is $|V(T)|$. The \emph{depth} of a node
$v\in V(T)$, $\depth(v)$, is the number of edges on the path from
$v$ to $\roots(T)$ and the depth of $T$, denoted $d_T$, is the
maximum depth of any node in $T$. The parent of $v$ is denoted $\parent(v)$ and the set of children of $v$ is denoted $\child(v)$. We define $\parent(\roots(T))=\bot$, where $\bot \not\in V(T)$ is a special \emph{null node}. 
A node with no children is a leaf and otherwise an internal node. The set of leaves of $T$ is
denoted $L(T)$ and we define $l_T = |L(T)|$. We say that $T$ is \emph{labeled} if
each node $v$ is a assigned a character, denoted $\lab(v)$, from an alphabet
$\Sigma$ and we say that $T$ is \emph{ordered} if a left-to-right
order among siblings in $T$ is given. Note that we \emph{do not} require that the size of the alphabet is bounded by a constant.
All trees in this paper are rooted, ordered, and labeled.

\paragraph{Ancestors and Descendants}
Let $T(v)$ denote the subtree of $T$ rooted at a node $v \in
V(T)$. If $w\in V(T(v))$ then $v$ is an ancestor of $w$, denoted
$v \preceq w$, and if $w\in V(T(v))\backslash \{v\}$ then $v$ is a
proper ancestor of $w$, denoted $v \prec w$. If $v$ is a (proper)
ancestor of $w$ then $w$ is a (proper) descendant of $v$. A node
$z$ is a common ancestor of $v$ and $w$ if it is an ancestor of
both $v$ and $w$. The nearest common ancestor of $v$ and $w$,
$\nca(v,w)$, is the common ancestor of $v$ and $w$ of greatest
depth. The \emph{first ancestor of $w$ labeled $\alpha$}, denoted
$\fl(w,\alpha)$, is the node $v$ such that $v \preceq w$, $\lab(v)
= \alpha$, and no node on the path between $v$ and $w$ is labeled
$\alpha$. If no such node exists then $\fl(w,\alpha) = \bot$.

\paragraph{Traversals and Orderings}
Let $T$ be a tree with root $v$ and let $v_1, \ldots ,v_k$ be the
children of $v$ from left-to-right. The \emph{preorder traversal}
of $T$ is obtained by visiting $v$ and then recursively visiting
$T(v_i)$, $1 \leq i \leq k$, in order. Similarly, the
\emph{postorder traversal} is obtained by first visiting $T(v_i)$,
$1 \leq i \leq k$, in order and then $v$. The \emph{preorder
number} and \emph{postorder number} of a node $w \in T(v)$,
denoted by $\pre(w)$ and $\post(w)$, are the number of nodes
preceding $w$ in the preorder and postorder traversal of $T$,
respectively. The nodes to the \emph{left} of $w$ in $T$ is the
set of nodes $u \in V(T)$ such that $\pre(u) < \pre(w)$ and
$\post(u) < \post(w)$. If $u$ is to the left of $w$, denoted by $u
\lhd w$, then $w$ is to the \emph{right} of $u$. If $u \lhd w$ or $u
\preceq w$ or $w \prec u$ we write $u \unlhd w$. The null node
$\bot$ is not in the ordering, i.e., $\bot \ntriangleleft v$ for
all nodes $v$.

\begin{figure}[t]
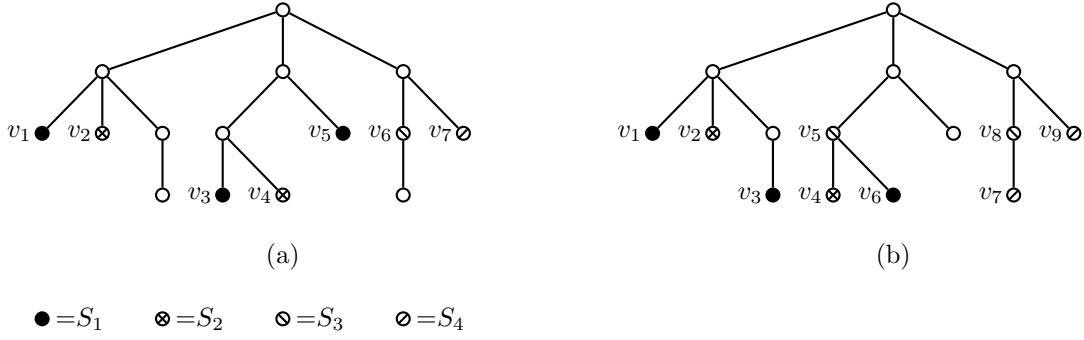

\begin{center}
\begin{psmatrix}[colsep=0.8cm,rowsep=0.4cm,labelsep=1pt]
  &&&& \cnode{.1}{root}\rput(0,.3){}
  \\
  & \cnode{.1}{l}\rput(0,.3){} &&&  \cnode{.1}{c}\rput(-.3,.3){}
  &&
  \cnode{.1}{r}\rput(0,.3){} \\
  \cnode*[fillcolor=black]{.1}{ll}\rput(-.3,0){$v_1$}\rput(0,-.3){} &
  \psset{fillstyle=crosshatch}\cnode{.1}{lc}\rput(-.3,0){$v_2$}\rput(0,-.3){} &
  \cnode{.1}{lr}\rput(-.2,-.3){} &  \cnode{.1}{cl}\rput(0,-.3){}
  & &
  \cnode*[fillcolor=black]{.1}{cr}\rput(0,-.3){}\rput(-.3,0){$v_5$} & \psset{fillstyle=vlines}
  \cnode{.1}{rl}\rput(-.2,-.3){}\rput(-.3,0){$v_6$}
  &
  \psset{fillstyle=hlines}\cnode{.1}{rr}\rput(0,-.3){}\rput(-.3,0){$v_7$}&&&
     \\
  && \cnode{.1}{lrc}\rput(0,-.3){} & \cnode*[fillcolor=black]{.1}{cll}\rput(0,-.3){}\rput(-.3,0){$v_3$}
  & \psset{fillstyle=crosshatch}\cnode{.1}{clr}\rput(0,-.3){}\rput(-.3,0){$v_4$}&&
  \cnode{.1}{rlc}\rput(0,-.3){}\\
  &&&&\rput(0,0){(a)} \\
  \ncline{root}{l}\ncline{root}{c}\ncline{root}{r}
  \ncline{l}{ll}\ncline{l}{lc}\ncline{l}{lr}
  \ncline{c}{cr}\ncline{c}{cl}
  \ncline{r}{rl}\ncline{r}{rr}
  \ncline{lr}{lcr}\ncline{cl}{cll}\ncline{cl}{clr}
  \ncline{rl}{rlc}\ncline{lr}{lrc}
  \cnode*[fillcolor=black]{.1}{1}\rput(.5,0){=$S_1$} &&
  \psset{fillstyle=crosshatch}\cnode{.1}{2}\rput(.5,0){=$S_2$} &&
  \psset{fillstyle=vlines}\cnode{.1}{3}\rput(.5,0){=$S_3$} &&
  \psset{fillstyle=hlines}\cnode{.1}{4}\rput(.5,0){=$S_4$} &&
  \end{psmatrix}
  \begin{psmatrix}[colsep=0.8cm,rowsep=0.4cm,labelsep=1pt]
  &&&& \cnode{.1}{root}\rput(0,.3){}
  \\
  & \cnode{.1}{l}\rput(0,.3){} &&&  \cnode{.1}{c}\rput(-.3,.3){}
  &&
  \cnode{.1}{r}\rput(0,.3){} \\
  \cnode*[fillcolor=black]{.1}{ll}\rput(-.3,0){$v_1$}\rput(0,-.3){} &
  \psset{fillstyle=crosshatch}\cnode{.1}{lc}\rput(-.3,0){$v_2$}\rput(0,-.3){} &
  \cnode{.1}{lr}\rput(-.2,-.3){} &  \psset{fillstyle=vlines}\cnode{.1}{cl}\rput(0,-.3){}\rput(-.3,0){$v_5$}
  &&
  \cnode{.1}{cr}\rput(0,-.3){} & \psset{fillstyle=vlines}
  \cnode{.1}{rl}\rput(-.2,-.3){}\rput(-.3,0){$v_8$}
  &
  \psset{fillstyle=hlines}\cnode{.1}{rr}\rput(0,-.3){}\rput(-.3,0){$v_9$}
     \\
  && \cnode*[fillcolor=black]{.1}{lrc}\rput(0,-.3){}\rput(-.3,0){$v_3$}
  & \psset{fillstyle=crosshatch}\cnode{.1}{cll}\rput(0,-.3){}\rput(-.3,0){$v_4$}
  & \cnode*[fillcolor=black]{.1}{clr}\rput(0,-.3){}\rput(-.3,0){$v_6$}&&
  \psset{fillstyle=hlines}\cnode{.1}{rlc}\rput(0,-.3){}\rput(-.3,0){$v_7$}\\
  &&&&\rput(0,0){(b)} \\
  \ncline{root}{l}\ncline{root}{c}\ncline{root}{r}
  \ncline{l}{ll}\ncline{l}{lc}\ncline{l}{lr}
  \ncline{c}{cr}\ncline{c}{cl}
  \ncline{r}{rl}\ncline{r}{rr}
  \ncline{lr}{lcr}\ncline{cl}{cll}\ncline{cl}{clr}
  \ncline{rl}{rlc}\ncline{lr}{lrc}
  \end{psmatrix}
     \caption{In (a) we have
      $\{(v_1,v_2,v_3,v_6,v_7),(v_1,v_2,v_5,v_6,v_7),
      (v_1,v_4,v_5,v_6,v_7),(v_3,v_4,v_5,v_6,v_7)\}=\Phi(S_1,S_2,S_1,S_3,S_4)$ and
      thus
      $\mop(S_1,S_2,S_1,S_3,S_4)=\{(v_3,v_7)\}$. In  (b) we have
      $\Phi(S_1,S_2,S_1,S_3,S_4)=\{(v_1,v_2,v_3,v_5,v_7),(v_1,v_2,v_6,v_8,v_9),
      (v_1,v_2,v_3,v_8,v_9),(v_1,v_2,v_3,v_5,v_9),(v_1,v_4,v_6,v_8,v_9),
      (v_3,v_4,v_6,v_8,v_9) \}$ and thus
      $\mop(S_1,S_2,S_1,S_3,S_4)=\{(v_1,v_7),(v_3,v_9)\}$.}
  \label{fig:mopexample}
\end{center}
\end{figure}

\paragraph{Minimum Ordered Pairs}
A set of nodes $X \subseteq V(T)$ is \emph{deep} if no node in $X$
is a proper ancestor of another node in $X$. For $k$ deep sets of
nodes $X_1, \ldots, X_k$ let $\Phi(X_1,\ldots,X_k) \subseteq (X_1
\times \cdots \times X_k)$, be the set of tuples such that $(x_1,
\ldots, x_k) \in \Phi(X_1,\ldots,X_k)$ iff $x_1 \lhd \cdots \lhd
x_k$. If $(x_1, \ldots, x_k) \in \Phi(X_1,\ldots,X_k) $ and there
is no $(x_1', \ldots, x_k') \in \Phi(X_1,\ldots,X_k) $, where
either $x_1 \lhd x_1' \lhd x_k' \unlhd x_k$ or $x_1 \unlhd x_1'
\lhd x_k' \lhd x_k$ then the pair $(x_1, x_k)$ is a \emph{minimum
ordered pair}. Intuitively, $(x_1, x_k)$ is a closest
pair of nodes from $X_1$ and $X_k$ in the left-to-right order for
which we can find $x_2, \ldots, x_{k-1}$ such that $x_1 \lhd \cdots
\lhd x_k$. The set of minimum ordered pairs for $X_1, \ldots,
X_k$ is denoted by $\mop(X_1, \ldots, X_k)$.
Figure~\ref{fig:mopexample} illustrates these concepts on a small
example.
For any set of pairs $Y$, let $\restrict{Y}{1}$ and
$\restrict{Y}{2}$ denote the \emph{projection} of $Y$ to the first
and second coordinate, that is, if $(y_1,y_2) \in Y$ then $y_1 \in
\restrict{Y}{1}$ and $y_2 \in \restrict{Y}{2}$. We say that $Y$ is
\emph{deep} if $\restrict{Y}{1}$ and $\restrict{Y}{2}$ are deep.
The following lemma shows that given deep sets $X_1, \ldots, X_k$
we can compute $\mop(X_1,\ldots,X_k)$ iteratively by first
computing $\mop(X_1,X_2)$ and then
$\mop(\restrict{\mop(X_1,X_2)}{2},X_3)$ and so on.

\begin{lemma}\label{lem:nnsmaller2}
  For any deep sets of nodes $X_1, \ldots, X_k$, $k>2$, we have,
  $(x_1,x_k) \in \mop(X_1, \ldots, X_k)$ iff there exists a node
  $x_{k-1}$  such that $(x_1,x_{k-1}) \in \mop(X_1, \ldots,
  X_{k-1})$ and $(x_{k-1},x_k) \in \mop(\restrict{\mop(X_1,
  \ldots, X_{k-1})}{2}, X_k)$.
\end{lemma}

\begin{proof}
  We start by showing that if $(x_1,x_k) \in \mop(X_1, \ldots,
  X_k)$ then there exists  a node $x_{k-1}$  such that $(x_1,x_{k-1}) \in
  \mop(X_1, \ldots, X_{k-1})$ and $(x_{k-1},x_k) \in
  \mop(\restrict{\mop(X_1, \ldots, X_{k-1})}{2}, X_k)$.

  First note that $(z_1,\ldots,z_k) \in
  \Phi(X_1,\ldots,X_k)$ implies $(z_1,\ldots,z_{k-1}) \in
  \Phi(X_1,\ldots,X_{k-1})$. Since $(x_1,x_k) \in
  \mop(X_1,\ldots,X_k)$ there must be a minimum node $x_{k-1}$ such that
  the tuple $(x_1,\ldots,x_{k-1})$ is in $\Phi(X_1,\ldots,X_{k-1})$.
  We have $(x_1, x_{k-1}) \in \mop(X_1,\ldots,X_{k-1})$. We need to
  show that $(x_{k-1},x_k) \in \mop(\restrict{\mop(X_1, \ldots, X_{k-1})}{2}, X_k)$.
  Since $(x_1,x_k) \in \mop(X_1, \ldots,
  X_k)$ there exists no node $z \in X_k$ such that  $x_{k-1} \lhd z
  \lhd x_k$. If such a $z$ existed we would have $(x_1,\ldots, x_{k-1},z)\in \Phi(X_1,\ldots,X_k)$, contradicting that $(x_1,x_k)\in  \mop(X_1, \ldots, X_k)$.   
   Assume there exists a node $z \in \restrict{\mop(X_1, \ldots,
  X_{k-1})}{2}$ such that  $x_{k-1} \lhd z
  \lhd x_k$. Since $(x_1,x_{k-1})\in \mop(X_1,\ldots,X_{k-1})$ this
  implies that there is a node $z' \rhd x_1$ such that $(z',z) \in
  \mop(X_1,\ldots,X_{k-1})$. But this implies that the tuple
  $(z',\ldots,z,x_k)$ is in $\Phi(X_1,\ldots,X_k)$ contradicting that
  $(x_1,x_k) \in \mop(X_1,\ldots,X_k)$.

  We will now show that if there exists a node $x_{k-1}$  such that $(x_1,x_{k-1}) \in
  \mop(X_1, \ldots, X_{k-1})$ and $(x_{k-1},x_k) \in
  \mop(\restrict{\mop(X_1, \ldots, X_{k-1})}{2}, X_k)$
  then the pair $(x_1,x_k) \in \mop(X_1, \ldots, X_k)$.
  Clearly, there exists a tuple $(x_1,\ldots,x_{k-1},x_k)\in \Phi(X_1,\ldots,X_k)$.
  Assume that there exists a tuple $(z_1,\ldots,z_k)\in
  \Phi(X_1,\ldots,X_k)$ such that $x_1 \lhd z_1 \lhd z_k \unlhd
  x_k$. Among the tuples satisfying these constraints let $(y_1, \ldots, y_{k-1}, y_k)$ be the one with maximum $y_1$, minimum $y_{k-1}$, and maximum $y_k$. It follows that $(y_1,y_{k-1})\in \mop(X_1,\ldots,X_{k-1})$.
  Since $(x_1,x_{k-1}) \in \mop(X_1,\ldots,X_{k-1})$ we must have $x_{k-1}\lhd y_{k-1}$. But this contradicts 
   $(x_{k-1},x_k) \in \mop(\restrict{\mop(X_1, \ldots, X_{k-1})}{2}, X_k)$, since node $y_{k-1}\in \restrict{\mop(X_1,\ldots,X_{k-1})}{2}$.
  
  Assume that there exists a tuple $(z_1,\ldots,z_k)\in
  \Phi(X_1,\ldots,X_k)$ such that $x_1 \unlhd z_1 \lhd z_k \lhd
  x_k$. Since $(x_1,x_{k-1})\in \mop(X_1,\ldots,X_{k-1})$ we have
  $x_{k-1} \unlhd z_{k-1}$ and thus $x_{k-1} \lhd z_k \lhd x_{k}$ contradicting
  $(x_{k-1},x_k) \in \mop(\restrict{\mop(X_1, \ldots, X_{k-1})}{2}, X_k)$.
\end{proof}
The following lemma is the reverse of the previous lemma and shows that given deep sets $X_1, \ldots, X_k$
we also can compute $\mop(X_1,\ldots,X_k)$ iteratively from right to left. The proof is similar to the proof of Lemma~\ref{lem:nnsmaller2}.
\begin{lemma}\label{lem:nnsmallerleft}
  For any deep sets of nodes $X_1, \ldots, X_k$, $k > 2$, we have,
  $(x_1,x_k) \in \mop(X_1, \ldots, X_k)$ iff there exists a node
  $x_{2}$  such that $(x_2,x_{k}) \in \mop(X_2, \ldots,
  X_{k})$ and $(x_1,x_2) \in \mop(X_1, \restrict{\mop(X_2,
  \ldots, X_{k})}{1})$.
\end{lemma}

\paragraph{Heavy Leaf Path Decomposition}
We construct a \emph{heavy leaf path decomposition} of $P$ as follows. We classify each node as \emph{heavy} or \emph{light}. The root is light. For each internal node $v$ we pick a child $v_j$ of $v$ with maximum $l_{P(v_j)}$ and classify it as heavy. The remaining children of $v$ are light. An edge to a light node is a \emph{light edge}, and an edge to a heavy node is a \emph{heavy edge}. The heavy child of a node $v$ is denoted heavy($v$). Let $\lightdepth(v)$ denote the number of light edges on the path from $v$ to root($P$).

Note that a heavy leaf path decomposition is the same as the
classical \emph{heavy path decomposition} \cite{HT1984}
except that the heavy child is defined as the child with largest
number of the descendant leaves and not the child with the largest
number of descendants. This distinction is essential for achieving the
linear space bound of our algorithms. Note that heavy path decompositions have
previously been used in algorithms for the related tree edit distance
problem~\cite{Klein1998}.

\begin{lemma}
For any tree $P$ and node $v \in V(P)$, $$l_{P(v)}\leq \frac{l_P}{2^{\lightdepth(v)}}\;.$$
\end{lemma}
\begin{proof}
By induction on $\lightdepth(v)$. For $\lightdepth(v)=0$ it is trivially true. Let $\lightdepth(v)=\ell$. Assume wlog. that $v$ is light. Let $w$ be the unique light ancestor of $v$ with $\lightdepth(w)=\ell-1$. By the induction hypothesis $l_{P(w)}\leq l_P/2^{\ell-1}$. Now $v$ has a sibling $\heavy(\parent(v))$ and thus at most half of the leaves in $P(\parent(v))$ can be in the subtree rooted at $v$. Therefore, $l_{P(v)}\leq l_{P(w)}/2 \leq l_P/2^{\ell}$.
\end{proof}

\begin{corollary}\label{cor:ldepth}
For any tree $P$ and node $v \in V(P)$, $\lightdepth(v) \leq \log l_P$.
\end{corollary}

\paragraph{Notation}
When we want to specify which tree we mean in the above relations we add a subscript. For instance, $v \prec_{T} w$ indicates that $v$ is an ancestor of $w$ \emph{in $T$}.

\section{Computing Deep Embeddings}\label{sec:recursion}
In this section we present a general framework for answering tree
inclusion queries. As in \cite{KM1995} we solve the equivalent
\emph{tree embedding problem}. Let $P$ and $T$ be rooted labeled
trees. An \emph{embedding} of $P$ in $T$ is an injective function
$f : V(P) \rightarrow V(T)$ such that for all nodes $v,u \in
V(P)$,
\begin{itemize}
  \item[(i)] $\lab(v) = \lab(f(v))$. (label preservation condition)
  \item[(ii)] $v \prec u$  iff $f(v) \prec f(u)$. (ancestor condition)
  \item[(iii)] $v \lhd u$ iff $f(v) \lhd f(u)$. (order condition)
\end{itemize}
An example of an embedding is given in
Figure~\ref{inclusionexample}(c).
\begin{lemma}[Kilpel\"{a}inen and Mannila~\cite{KM1995}]
For any trees $P$ and $T$, $P \sqsubseteq T$ iff there exists an
embedding of $P$ in $T$.
\end{lemma}
We say that the embedding $f$ is \emph{deep} if there is no
embedding $g$ such that $f(\roots(P)) \prec g(\roots(P))$. The
\emph{deep occurrences} of $P$ in $T$, denoted $\emb(P, T)$ is the
set of nodes,
\begin{equation*}
\emb(P,T) = \{f(\roots(P)) \mid \text{$f$ is a deep embedding of $P$ in $T$}\}.
\end{equation*}
By definition the set of ancestors of nodes in $\emb(P,T)$ is exactly the set of nodes $\{u \mid P \sqsubseteq T(u)\}$. Hence, to solve the tree inclusion problem it is sufficient to compute $\emb(P,T)$ and then, using additional $O(n_T)$ time, report all ancestors of this set. We note that Kilpel\"{a}inen and Mannila~\cite{KM1995} used the
similar concept of \emph{left embeddings} in their algorithms. A left
embedding of $P$ in $T$ is an embedding such that the root of $P$ is
mapped to the node in $T$ with the smallest postorder number, i.e.,
the deepest node among the nodes furthest to the left. Our definition
of $\emb(P,T)$ only requires that the root is mapped to a deepest
node.

\begin{figure}
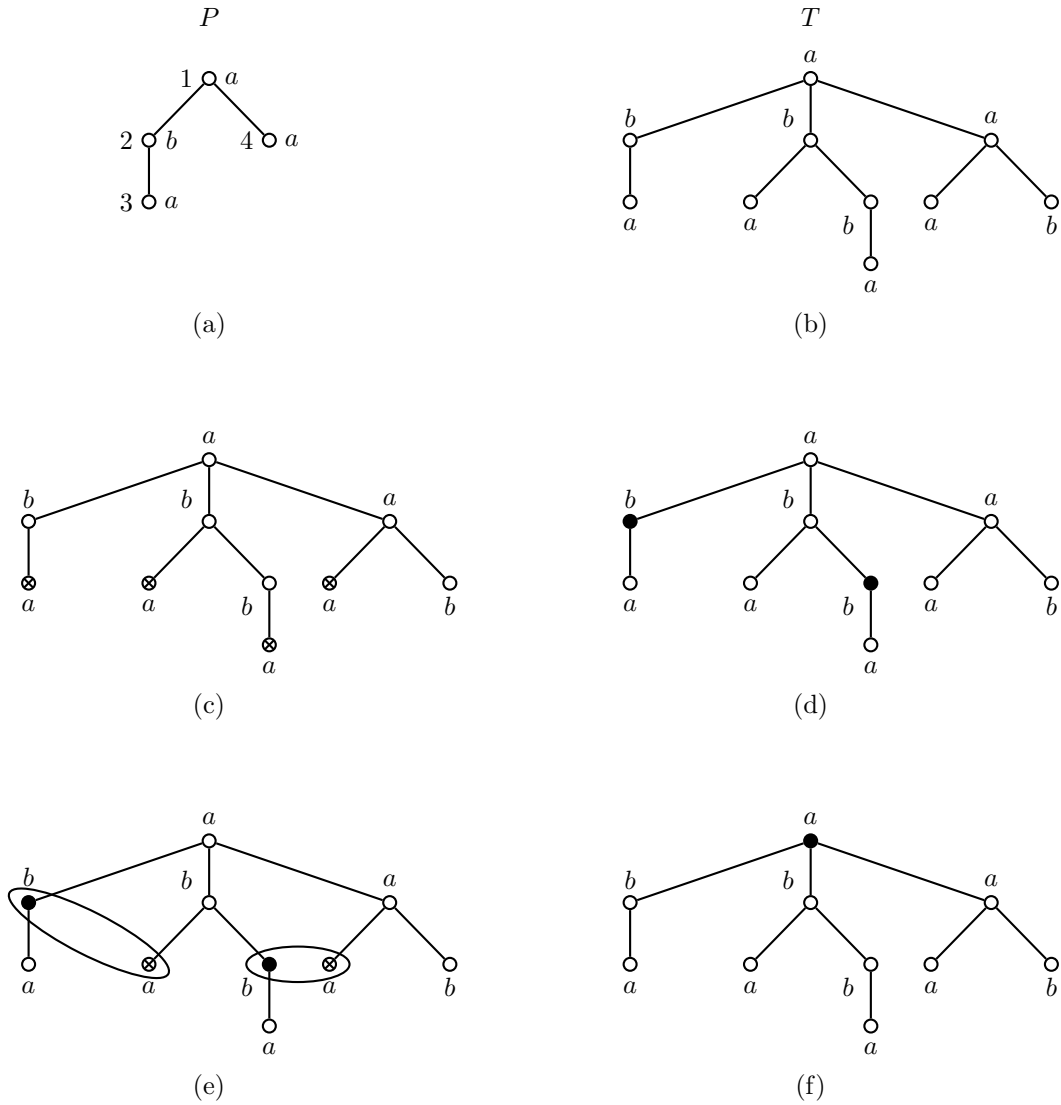

\begin{center}
  \begin{psmatrix}[colsep=0.8cm,rowsep=0.4cm,labelsep=1pt]
  &&&\rput(0,0){$P$} &&&&&&&&&& \rput(0,0){$T$} \\
  &&& \cnode{.1}{root1}\rput(.3,0){$a$}\rput(-.3,0){$1$} &&&&&&&&&&
  \cnode{.1}{root2}\rput(0,.3){$a$}\\
  && \cnode{.1}{l1}\rput(.3,0){$b$}\rput(-.3,0){$2$} &&  \cnode{.1}{r1}\rput(.3,0){$a$}\rput(-.3,0){$4$} &&&&&&
  \cnode{.1}{l2}\rput(0,.3){$b$} &&&  \cnode{.1}{c2}\rput(-.3,.3){$b$} &&&
  \cnode{.1}{r2}\rput(0,.3){$a$} \\
  && \cnode{.1}{ll1}\rput(.3,0){$a$}\rput(-.3,0){$3$} && &&&&&&
  \cnode{.1}{ll2}\rput(0,-.3){$a$} && \cnode{.1}{lc2}\rput(0,-.3){$a$} &&
  \cnode{.1}{rc2}\rput(-.3,-.3){$b$} & \cnode{.1}{lr2}\rput(0,-.3){$a$} &&
  \cnode{.1}{rr2}\rput(0,-.3){$b$} \\
  &&&&&&&&&&&&&&\cnode{.1}{e2}\rput(0,-.3){$a$}\\
  &&&\rput(0,0){(a)} &&&&&&&&&& \rput(0,0){(b)} \\\\
  \ncline{root1}{l1}\ncline{root1}{r1}
  \ncline{l1}{ll1}
  \ncline{root2}{l2}\ncline{root2}{c2}\ncline{root2}{r2}
  \ncline{l2}{ll2}\ncline{c2}{rc2}\ncline{c2}{lc2}
  \ncline{r2}{lr2}\ncline{r2}{rr2}
  \ncline{rc2}{e2}
  \end{psmatrix}
  \begin{psmatrix}[colsep=0.8cm,rowsep=0.4cm,labelsep=1pt]
  &&& \cnode{.1}{root1}\rput(0,.3){$a$} &&&&&&&&&&
  \cnode{.1}{root2}\rput(0,.3){$a$}\\
  \cnode{.1}{l1}\rput(0,.3){$b$} &&&  \cnode{.1}{c1}\rput(-.3,.3){$b$} &&&
  \cnode{.1}{r1}\rput(0,.3){$a$} &&&&
  \cnode*[fillcolor=black]{.1}{l2}\rput(0,.3){$b$} &&&  \cnode{.1}{c2}\rput(-.3,.3){$b$} &&&
  \cnode{.1}{r2}\rput(0,.3){$a$} \\
  \psset{fillstyle=crosshatch}\cnode{.1}{ll1}\rput(0,-.3){$a$} && \psset{fillstyle=crosshatch}\cnode{.1}{lc1}\rput(0,-.3){$a$} &&
  \cnode{.1}{rc1}\rput(-.3,-.3){$b$} & \psset{fillstyle=crosshatch}\cnode{.1}{lr1}\rput(0,-.3){$a$} &&
  \cnode{.1}{rr1}\rput(0,-.3){$b$} &&&
  \cnode{.1}{ll2}\rput(0,-.3){$a$} && \cnode{.1}{lc2}\rput(0,-.3){$a$} &&
  \cnode*[fillcolor=black]{.1}{rc2}\rput(-.3,-.3){$b$} & \cnode{.1}{lr2}\rput(0,-.3){$a$} &&
  \cnode{.1}{rr2}\rput(0,-.3){$b$} \\
  &&&& \psset{fillstyle=crosshatch}\cnode{.1}{e1}\rput(0,-.3){$a$} &&&&&&&&&&
  \cnode{.1}{e2}\rput(0,-.3){$a$}\\
  &&&\rput(0,0){(c)} &&&&&&&&&& \rput(0,0){(d)} \\\\
  \ncline{root1}{l1}\ncline{root1}{c1}\ncline{root1}{r1}
  \ncline{l1}{ll1}\ncline{c1}{rc1}\ncline{c1}{lc1}
  \ncline{r1}{lr1}\ncline{r1}{rr1}
  \ncline{rc1}{e1}
  \ncline{root2}{l2}\ncline{root2}{c2}\ncline{root2}{r2}
  \ncline{l2}{ll2}\ncline{c2}{rc2}\ncline{c2}{lc2}
  \ncline{r2}{lr2}\ncline{r2}{rr2}
  \ncline{rc2}{e2}
  \end{psmatrix}
  \begin{psmatrix}[colsep=0.8cm,rowsep=0.4cm,labelsep=1pt]
  &&& \cnode{.1}{root1}\rput(0,.3){$a$} &&&&&&&&&&
  \cnode*[fillcolor=black]{.1}{root2}\rput(0,.3){$a$}\\
   \rput{-27}(0,0){\psellipse(.9,0)(1.2,.3)}\cnode*[fillcolor=black]{.1}{l1}\rput(0,.35){$b$} &&&  \cnode{.1}{c1}\rput(-.3,.3){$b$} &&&
  \cnode{.1}{r1}\rput(0,.3){$a$} &&&&
  \cnode{.1}{l2}\rput(0,.3){$b$} &&&  \cnode{.1}{c2}\rput(-.3,.3){$b$} &&&
  \cnode{.1}{r2}\rput(0,.3){$a$} \\
  \cnode{.1}{ll1}\rput(0,-.3){$a$} &&\psset{fillstyle=crosshatch}\cnode{.1}{lc1}\rput(0,-.3){$a$} &&
\psellipse(.38,0)(.7,.25) \cnode*[fillcolor=black]{.1}{rc1}\rput(-.3,-.3){$b$} & \psset{fillstyle=crosshatch}\cnode{.1}{lr1}\rput(0,-.3){$a$} &&
  \cnode{.1}{rr1}\rput(0,-.3){$b$} &&&
  \cnode{.1}{ll2}\rput(0,-.3){$a$} && \cnode{.1}{lc2}\rput(0,-.3){$a$} &&
  \cnode{.1}{rc2}\rput(-.3,-.3){$b$} & \cnode{.1}{lr2}\rput(0,-.3){$a$} &&
  \cnode{.1}{rr2}\rput(0,-.3){$b$} \\
  &&&& \cnode{.1}{e1}\rput(0,-.3){$a$} &&&&&&&&&&
  \cnode{.1}{e2}\rput(0,-.3){$a$}\\
  &&&\rput(0,0){(e)} &&&&&&&&&& \rput(0,0){(f)} \\
  \ncline{root1}{l1}\ncline{root1}{c1}\ncline{root1}{r1}
  \ncline{l1}{ll1}\ncline{c1}{rc1}\ncline{c1}{lc1}
  \ncline{r1}{lr1}\ncline{r1}{rr1}
  \ncline{rc1}{e1}
  \ncline{root2}{l2}\ncline{root2}{c2}\ncline{root2}{r2}
  \ncline{l2}{ll2}\ncline{c2}{rc2}\ncline{c2}{lc2}
  \ncline{r2}{lr2}\ncline{r2}{rr2}
  \ncline{rc2}{e2}
  \end{psmatrix}
   \caption{Computing the deep occurrences of $P$ into $T$ depicted in (a) and (b)
   respectively. The nodes in $P$ are numbered $1$--$4$ for easy reference.
   (c) Case 1 of $\Emb$: The crossed nodes are the nodes in the set $\Emb(3)$.
   Since $3$ and $4$ are leaves and $\lab(3) = \lab(4)$ we have $\Emb(3) = \Emb(4)$. (d)
   Case 2 of $\Emb$: The black nodes are the nodes in the set $\Emb(2)$.
   Note that the middle child of the root of $T$ is not in the set since it is not a
   deep occurrence. (e) and (f) illustrates the computation of $\Emb(1)$ and case 3 of $\Emb$: (e) The two minimal ordered pairs of the sets from
   (d) and (c). In the procedure $R_1$ is the set from (d) and $R_2$ is the set from (c). The set $U_1=\{(v,v)\mid v \in R_1\}$ and the set $U_2=\Mop(U_1,R_2)$ which corresponds to the pairs shown in (e). The black nodes in the pairs are the nodes from $R_1$ and the crossed nodes are the nodes from $R_2$. Since $k=2$ we set $U_1=U_2$. (f) The nearest common ancestors of both pairs shown in (e) is
   the root node of $T$ which is the only (deep) occurrence of $P$.}
  \label{fig:embexample}
\end{center}
\end{figure}

In the following we show how to compute deep embeddings. The key
idea is to build a data structure for $T$ allowing a fast
implementation of the following procedures. For all $X \subseteq
V(T)$, $Y \subseteq V(T)\times V(T)$, and $\alpha \in \Sigma$
define:
\begin{relate}
    \item[$\Parent(X)$:] Return the set
    $\{\parent(x) \mid x \in X\}$.
    \item[$\Nca(Y)$:]
    Return the set $\{\nca(y_1,y_2) \mid (y_1,y_2) \in Y\}$.
    \item[$\Deep(X)$:] Return the set $\{x \in X\mid \text{there is no }
    z \in X \text{ such that } x \prec z\}$.
    \item[$\Mop(Y,X)$:]
    Return the set of pairs $R$ such that for any pair $(y_1,y_2)
    \in Y$, $(y_1,x) \in R$ iff $(y_2, x) \in
    \mop(\restrict{Y}{2}, X)$.
   \item[$\MopLeft(X,Y)$:]
    Return the set of pairs $R$ such that for any pair $(y_1,y_2)
    \in Y$, $(x,y_2) \in R$ iff $(x,y_1) \in
    \mop(X,\restrict{Y}{1})$.
    \item[$\Fl(X, \alpha)$:]
    Return the set $\Deep(\{\fl(x, \alpha) \mid x \in X\})$.
\end{relate}
Collectively we call these procedures the \emph{set procedures}.
The procedures $\Parent$ and $\Nca$ are self-explanatory.
$\Deep(X)$ returns the set of all nodes in $X$ that have no
descendants in $X$. Hence,  the returned set is always deep.
$\Mop$ and $\MopLeft$ are used to iteratively compute minimum ordered pairs. $\Fl(X, \alpha)$ returns the 
deep set of first  ancestors with label $\alpha$ of all nodes in $X$. If we
want to specify that a procedure applies to a certain tree $T$ we
add the subscript $T$. With the set procedures we can compute deep
embeddings. The following procedure $\Emb(v)$, $v \in V(P)$,
recursively computes the set of deep occurrences of $P(v)$ in $T$.
Figure~\ref{fig:embexample} illustrates how $\Emb$ works on a
small example.
\\\\
%
%
%
%
%
%
%

%

\begin{procedure}[H]
\SetProcNameSty{textnormal}
Let $v_1, \ldots, v_k$ be the sequence of
children of $v$ ordered from left to right. There are three cases: 

\uCase($\quad /\!/ \; v$ is a leaf){{\bf 1.} $k=0$}
{Compute 
$R := \Fl(L(T),\lab(v))$. }
\uCase{{\bf 2.} $k=1$}{
Recursively compute 
$R_1 := \Emb(v_1)$. 

Compute $R := \Fl(\Deep(\Parent(R_1)),\lab(v))$.}
\uCase{{\bf 3.} $k > 1$}
{ Let $v_j$ be the heavy child of $v$. 

Recursively compute $R_j := \Emb(v_j)$ and set $U_j := \{(r,r) \mid r \in R_j\}.$ 

\For{$i := j+1$ to $k$} {
Recursively compute 
$R_i := \Emb(v_i)$ and set $U_i := \Mop(U_{i-1}, R_i)$.}

Set  $U_j:=U_k$. 
   
\For{$i := j-1$ downto $1$} {
Recursively compute 
 $R_i := \Emb(v_i)$ and set $U_i := \MopLeft(R_i,U_{i+1})$. }
Compute 
$R := \Fl(\Deep(\Nca(U_1)), \lab(v))$.}

\uIf{$R = \emptyset$}
{stop and report that there is no deep embedding of $P(v)$ in $T$.}
\uElse{Return $R$.}
\caption{\Emb($v$)
}
\end{procedure}

\showOld{
\begin{relate}
\item[$\Emb(v)$:] Let $v_1, \ldots, v_k$ be the sequence of
children of $v$ ordered from left to right. There are three cases:
    \begin{enumerate}
    \item $k=0$ ($v$ is a leaf). Compute $R := \Fl(L(T),\lab(v))$.
    \item $k=1$. Recursively compute $R_1 := \Emb(v_1)$. 
    
    Compute $R := \Fl(\Deep(\Parent(R_1)),\lab(v))$.
    \item $k > 1$.  Let $v_j$ be the heavy child of $v$. 
    
   Recursively compute  $R_j := \Emb(v_j)$ and set $U_j := \{(r,r) \mid r \in R_j\}$.
    
    For $i := j+1$ to $k$, recursively compute $R_i := \Emb(v_i)$ and $U_i := \Mop(U_{i-1}, R_i)$.
    
    Set $U_j:=U_k$.
    
    For $i := j-1$ to $1$, recursively compute $R_i := \Emb(v_i)$ and $U_i := \MopLeft(R_i,U_{i+1})$.
    
    Finally, compute $R := \Fl(\Deep(\Nca(U_1)), \lab(v))$.
%
%
%
     \end{enumerate}
If $R = \emptyset$ stop and report that there is no deep embedding
of $P(v)$ in $T$. Otherwise return $R$.
\end{relate}
}{}

\begin{figure}
\begin{center}
\includegraphics[scale=.5]{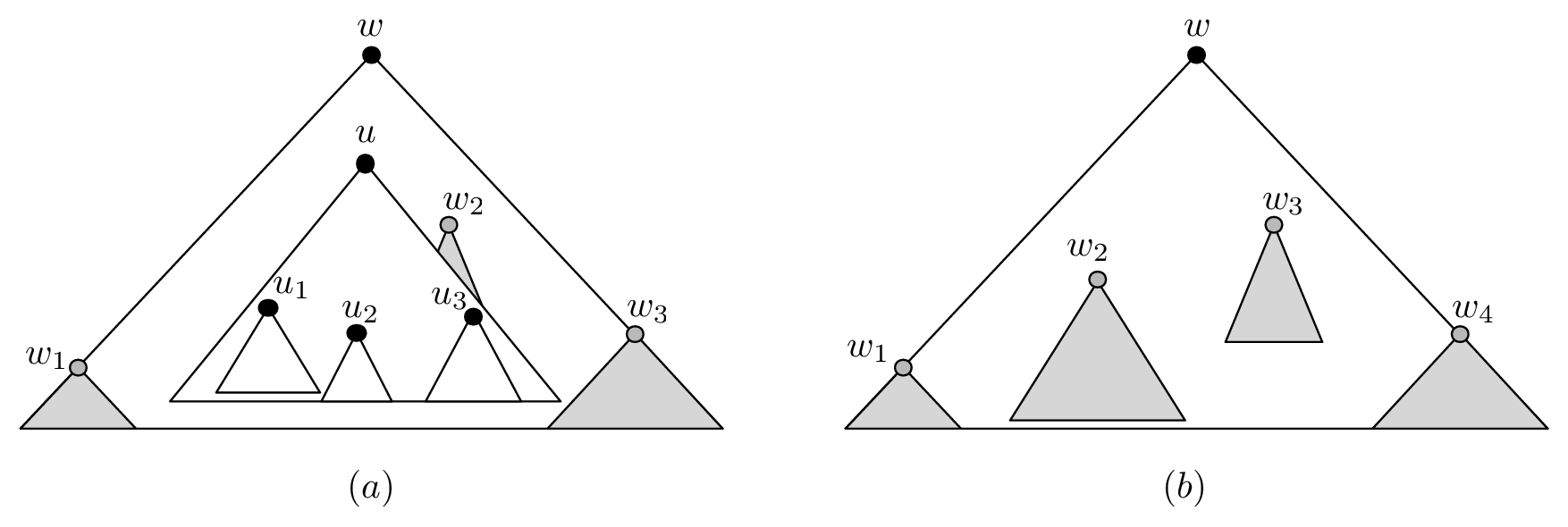}
\end{center}
\caption{(a) For all $i$, $w_i$ and $u_i$ are roots of occurrences of $P(v_i)$ in $T$, and $w$ and $u$ is the nearest common ancestor of $(w_1,w_3)$ and $(u_1,u_3)$, respectively. Since $w_1 \lhd u_1$ and $u_3 \lhd w_3$ we cannot have $u \prec w$.
(b) For all $i$, $w_i$ is an embedding of $P(v_i)$ in $T$, $(w_1,w_4)$ is a minimum ordered pair, and $w$ is the nearest common ancestor of all the $w_i$'s. The number of leaves in $T(w)$ is at least $\sum_{i=1}^4 l_{T(w_i)} \geq \sum_{i=1}^4 l_{P(v_i)}$.}\label{fig:MopSubtree}
\end{figure}

To prove the correctness of the
$\Emb$ procedure we need the following two propositions. The first proposition characterizes
for node $v \in V(P)$ the set $\emb(v,T)$ using $\mop$, $\nca$, and
$\fl$. The second proposition shows that the set $U_1$ computed in
case 3 of the $\Emb$ procedure is the set $\mop(\Emb(v_1), \ldots,  \Emb(v_k))$.
\begin{prop}\label{prop:embchildren}
Let $v$ be a node in $P$ and let $v_1, \ldots ,v_k$ be the sequence of children of $v$ ordered from left to right, where $k\geq 2$. 
For any node $w\in \emb(P(v),T)$, there exists a pair of nodes 
$(w_1,w_k)\in \mop(\emb(P(v_1),T),\ldots,\emb(P(v_k),T))$ such that $w=\fl(\nca(w_1,w_k),\lab(v))$.
\end{prop}
\begin{proof}
Since $w$ is the root of an occurrence of $P(v)$ in $T$ there must exist a set of disjoint occurrences of $P(v_1),\ldots,P(v_k)$ in $T(w)$ with roots $w_1 \lhd \ldots \lhd w_k$, such that $w$ is an ancestor of $w_1,\ldots, w_k$. Since the $w_i$'s are ordered $w$ must be an ancestor of $\nca(w_1,w_k)$. Since $w$ is the root of a \emph{deep} occurrence of  $P(v)$ in $T$ it follows that $w=\fl(\nca(w_1,w_k),\lab(v))$.

It remains to show that we can assume $(w_1,w_k)\in \mop(\emb(P(v_1),T),\ldots,\emb(P(v_k),T))$. It follows from the previous discussion that $(w_1,\ldots,w_k)\in \Phi(\emb(P(v_1),T),\ldots,\emb(P(v_k),T))$.
Assume for the sake of contradiction that $(w_1,w_k)$ is \emph{not} a minimum ordered pair. Then there exists a set of disjoint occurrences of $P(v_1),\ldots,P(v_k)$ in $T(w)$ with roots $u_1 \lhd \ldots \lhd u_k$, such that either $w_1 \lhd u_1 \lhd u_k \unlhd w_k$ or $w_1 \unlhd u_1 \lhd u_k \lhd w_k$, 
and $(u_1,u_k)\in\mop(\emb(P(v_1),T),\ldots,\emb(P(v_k),T))$. Therefore $u=\fl(\nca(u_1,u_k),\lab(v))$ is an embedding of $P(v)$ in $T$. Now either $w \prec u$ contradicting the assumption that $w$ is a deep embedding or $w=u$ in which case $(u_1,u_k)$ satisfies the properties of the lemma (see also Figure~\ref{fig:MopSubtree}(a)).
\end{proof}

%
\begin{prop}\label{prop:Mop}
For $j+1\leq l \leq k$, 
\begin{equation}\label{eq:uk}
U_l =
\mop(\Emb(v_j), \ldots, \Emb(v_l)),
\end{equation}
For $1\leq l \leq j-1$, 
\begin{equation}\label{eq:uj}
U_l =
\mop(\Emb(v_l), \ldots, \Emb(v_k)). 
\end{equation}
\end{prop}
\begin{proof}
We first show Equation~\eqref{eq:uk} by induction on $l$.
For $l=j+1$ it follows from the definition of
$\Mop$ that $U_l$ is the set of minimum ordered pairs of $\Emb(v_j)$ and $\Emb(v_{j+1})$, i.e., $U_l = \mop(\Emb(v_j), \Emb(v_{l}))$. Hence,
assume that $l > j+1$. By the induction hypothesis we have $$U_l = \Mop(U_{l-1}, \Emb(v_l)) =
\Mop(\mop(\Emb(v_j), \ldots, \Emb(v_{l-1})), R_l)\;.$$ 
By
definition of $\Mop$, $U_l$ is the set of pairs such that for
any pair  $(r_j, r_{l-1}) \in \mop(\Emb(v_j), \ldots,
\Emb(v_{l-1}))$, $(r_j, r_l) \in U_l$ iff $(r_{l-1}, r_l) \in
\mop(\restrict{\mop(\Emb(v_j), \ldots, \Emb(v_{l-1}))}{2},
R_l)$. By Lemma~\ref{lem:nnsmaller2} it follows that $(r_j, r_l)
\in U_l$ iff $(r_j, r_l) \in \mop(\Emb(v_j), \ldots,
\Emb(v_l))$.

We can now similarly show Equation~\eqref{eq:uj} by induction on $j'=j-l$.  By Equation~\eqref{eq:uk} we have $U_j= \mop(\Emb(v_j), \ldots, \Emb(v_k))$ when we begin computing $U_{j-1}$. For $j'=1$ ($l=j-1$) it follows from the definition of
$\MopLeft$ that $U_{j-1} = \mop(\Emb(v_{j-1}), \Emb(v_j))$. Hence,
assume that $j' > 1$. Using Lemma~\ref{lem:nnsmallerleft} the Equation follows similarly to the proof of Equation~\eqref{eq:uk}.
\end{proof}
By Proposition~\ref{prop:Mop}, $U_1=\mop(\Emb(v_1),\ldots,\Emb(v_k))$. We can now show the correctness of procedure $\Emb$.

\begin{lemma}\label{lem:correctEmb}
For trees $P$ and $T$ and node $v\in V(P)$, $\Emb(v)$ computes the set of
deep occurrences of $P(v)$ in $T$.
\end{lemma}
\begin{proof}
By induction on the size of the subtree $P(v)$. If $v$ is a leaf,
$\emb(v,T)$ is the deep set of nodes in $T$ with label $\lab(v)$. It immediately follows that
$\emb(v,T) = \Fl(L(T),
\lab(v))$ and thus case 1 follows. 

Suppose that $v$ is an
internal node with $k\geq1$ children $v_1, \ldots, v_k$. We show
that $\emb(P(v),T) = \Emb(v)$. Consider cases 2 and 3 of the
algorithm.

For $k=1$ we have that $w\in \Emb(v)$ implies that $\lab(w) =
\lab(v)$ and there is a node $w_1 \in \Emb(v_1)$ such that
$\fl(\parent(w_1), \lab(v)) = w$, that is, no node on the path
between $w_1$ and $w$ is labeled $\lab(v)$.  By induction
$\Emb(v_1) = \emb(P(v_1),T)$ and therefore $w$ is the root of an
embedding of $P(v)$ in $T$. Since $\Emb(v)$ is the deep set of
all such nodes it follows that $w \in \emb(P(v),T)$. Conversely,
if $w \in \emb(P(v),T)$ then $\lab(w) = \lab(v)$, there is a node
$w_1 \in \emb(P(v_1),T)$ such that $w \prec w_1$, and no node on
the path between $w$ and $w_1$ is labeled $\lab(v)$, that is,
$\fl(w_1, \lab(v)) = w$. Hence, $w \in \Emb(v)$.

Next consider the case $k>1$.
By Proposition~\ref{prop:Mop} and the induction hypothesis  $$U_1=\mop(\emb(P(v_1),T),\ldots,\emb(P(v_k),T))\;.$$ 
We first show that $w \in
\emb(P(v),T)$ implies that $w\in \Emb(v)$. By Proposition~\ref{prop:embchildren} there exists a pair of nodes $(w_1,w_k)\in \mop(\emb(P(v_1),T),\ldots,\emb(P(v_k),T))$ such that $w=\fl(\nca(w_1,w_k),\lab(v))$. We have $(w_1,w_k)\in U_1$ and it follows directly from the implementation that $w\in \Emb(v)$. To see that we do not loose $w$ by taking \Deep\ of $\Nca(U_1)$ assume that $w'=\nca(w_1,w_k)$ is removed from the set in this step. This means there is a node $u$ in $\Nca(U_1)$ which is a descendant of  $w'$ and which is still in the set.  Since $w$ is the root of a \emph{deep} occurrence we must have $w=\fl(w',\lab(v))=\fl(u,\lab(v))$.

Let $w \in \Emb(v)$. Then $w$ is the first ancestor with label $\lab(v)$ of a nearest common ancestor of a pair in $U_1$. That is, $\lab(w) = \lab(v)$ and there exists nodes $(w_1, w_k) \in
\mop(\emb(P(v_1),T), \ldots, \emb(P(v_k),T))$ such that
$w=\fl(\nca(w_1, w_k), \lab(v))$. Clearly, $w$ is the root of an
embedding of $P(v)$ in $T$. 
Assume for contradiction that $w$ is
not a deep embedding, that is, $w \prec u$ for some node $u \in
\emb(P(v),T)$. We have just shown that this implies $u\in \Emb(v)$. Since $\Emb(v)$ is a deep set this contradicts $w \in \Emb(v)$.
\end{proof}
The set $L(T)$ is deep and in all three cases of $\Emb(V)$ the returned set is also deep. By induction it follows that the input to $\Parent$, $\Fl$, $\Nca$, and $\Mop$ is always deep. We will use this fact to our advantage in the following algorithms.

\section{A Simple Tree Inclusion Algorithm}\label{simple}
In this section we a present a simple implementation of the set
procedures which leads to an efficient tree inclusion algorithm.
Subsequently, we modify one of the procedures to obtain a family
of tree inclusion algorithms where the complexities depend on the
solution to a well-studied problem known as the \emph{tree color
problem}.

\subsection{Preprocessing}\label{sec:simplepreprocessing}
To compute deep embeddings we require a data structure
for $T$ which allows us, for any $v,w \in V(T)$, to compute
$\nca_T(v,w)$ and determine if $v \prec w$ or $v \lhd w$. In
linear time we can compute $\pre(v)$ and $\post(v)$ for all nodes
$v \in V(T)$, and with these it is straightforward to test the two
conditions. Furthermore,
\begin{lemma}[Harel and Tarjan~\cite{HT1984}]
For any tree $T$ there is a data structure using $O(n_T)$ space
and preprocessing time which supports nearest common ancestor
queries in $O(1)$ time.
\end{lemma}
Hence, our data structure uses linear preprocessing time and space (see also~\cite{BFC2000,AGKR2004} for more recent nearest common ancestor data structures). 

\subsection{Implementation of the Set Procedures}\label{implementationsimple}
To answer tree inclusion queries we give an efficient
implementation of the set procedures. The idea is to represent
sets of nodes and sets of pairs of nodes in a left-to-right order
using linked lists. For this purpose we introduce some helpful
notation. Let $X = [x_1, \ldots, x_k]$ be a linked list of nodes.
The \emph{length} of $X$, denoted $|X|$, is the number of elements
in $X$ and the list with no elements is written $[]$. The $i$th
node of $X$, denoted $X[i]$, is $x_i$. Given any node $y$ the list
obtained by \emph{appending} $y$ to $X$, is the list $X \circ y =
[x_1, \ldots, x_k, y]$. If for all $i$, $1 \leq i \leq |X|-1$,
$X[i] \lhd X[i+1]$ then $X$ is \emph{ordered} and if $X[i] \unlhd
X[i+1]$ then $X$ is \emph{semiordered}. Recall that $X[i] \unlhd X[i+1]$ means that we can have $X[i] \lhd X[i+1]$ or either of the nodes can be an ancestor of the other ($X[i] \lhd X[i+1]$ or $X[i] \preceq X[i+1]$ or $X[i] \succeq X[i+1]$). A list $Y = [(x_1, z_k),
\ldots, (x_k, z_k)]$ is a \emph{node pair list}. By analogy, we
define length, append, etc. for $Y$. For a pair $Y[i] = (x_i,
z_i)$ define $Y[i]_1 = x_i$ and $Y[i]_2 = z_i$. If the lists
$[Y[1]_1, \ldots, Y[k]_1]$ and $[Y[1]_2, \ldots, Y[k]_2]$ are both
ordered or semiordered then $Y$ is \emph{ordered} or
\emph{semiordered}, respectively.

The set procedures are implemented using node lists. All lists
used in the procedures are either ordered or semiordered. As noted
in Section~\ref{sec:recursion} we may assume that the inputs to all
of the procedures, except $\Deep$, represent deep sets, that is,
the corresponding node list or node pair list is ordered. We
assume that the input list given to $\Deep$ is semiordered and the
output, of course, is ordered. Hence, the output of all the other
set procedures must be semiordered. In the following let $X$ be a
node list, $Y$ a node pair list, and $\alpha$ a character in
$\Sigma$. The detailed implementation of the set procedures is
given below. We show the correctness in
Section~\ref{sec:simplecorrectness} and discuss the complexity in
Section~\ref{sec:simplecomplexity}.

\showOld{
\begin{relate}[simple]
\item[$\Parent(X)$:] Return the list $[\parent(X[1]), \ldots,
\parent(X[\norm{X}])]$.

\item[$\Nca(Y)$:] Return the list $[\nca(Y[1]), \ldots, \nca(Y[\norm{Y}])]$.

\item[$\Deep(X)$:] Initially, set $x := X[1]$ and $R := []$.

For $i:=2$ to $|X|$ do:
\begin{itemize}
\item[] Compare $x$ and $X[i]$. There are three cases:
    \begin{enumerate}
    \item 
    $x \lhd X[i]$. Set $R:= R \circ x$ and $x:= X[i]$.

    \item 
    $x \prec X[i]$. Set $x := X[i]$.  

    \item $X[i] \preceq x$. Do nothing.
    \end{enumerate}
\end{itemize}
Return $R \circ x$.
\end{relate}
}{}

\begin{procedure}[H]
\SetProcNameSty{textnormal}
Return the list $[\parent(X[1]), \ldots,
\parent(X[\norm{X}])]$.
\caption{\Parent($X$)}
\end{procedure}

\begin{procedure}[h]
\SetProcNameSty{textnormal}
Return the list $[\nca(Y[1]), \ldots, \nca(Y[\norm{Y}])]$.
\caption{\Nca($Y$)}
\end{procedure}

\begin{procedure}[h]
\SetProcNameSty{textnormal}
Initially, set $x := X[1]$ and $R := []$.

\For{$i:=2$ to $|X|$}
{
Compare $x$ and $X[i]$. There are three cases:

\uCase{{\bf 1.} $x \lhd X[i]$} {Set $R:= R \circ x$ and $x:= X[i]$.}
\uCase{{\bf 2.} $x \prec X[i]$} {Set $x := X[i]$.}  
\uCase{{\bf 3.} $X[i] \preceq x$} {Do nothing.}}

Return $R \circ x$.

\caption{\Deep($X$)}
\end{procedure}

The implementation of procedure \Deep\ takes advantage of the fact
that the input list is semiordered. In case 1 the node $X[i]$ is to the
right of our "potential output node" $x$. Since any node that is a
descendant of $x$ must also be to the left of $X[i]$ it cannot 
appear later in the list $X$ than $X[i]$. We can thus safely add
$x$ to $R$ at this point. In case 2 the node $x$ is an ancestor of
$X[i]$ and thus $x$ cannot be in $\Deep(X)$. In case 3 the node
$X[i]$ is an ancestor of $x$ and can therefore not be in $\Deep(X)$.

\ignore{
\begin{relate}[simple]
\item[$\Mop(Y,X)$:]
    Initially, set $R:=[]$.

    Find the smallest $j$ such that $Y[1]_2 \lhd X[j]$ and set
    $y:=Y[1]_1$, $x:= X[j]$. 
    If no such $j$ exists stop and return $R$.

    For $i :=2$ to $\norm{Y}$ do:
    \begin{itemize}
    \item[] 
    Until $Y[i]_2 \lhd X[j]$ or $j>\norm{X}$ set $j:=j+1$.

    If $j> \norm{X}$
    stop and return $R := R \circ (y,x)$.
    Otherwise, compare $X[j]$ and $x$.
    There are two cases:
    \begin{enumerate}
    \item If $x \lhd X[j]$ set
    $R:=R\circ (y,x)$, $y:=Y[i]_1$, and $x:=X[j]$.
    \item If $x=X[j]$
    set $y:=Y[i]_1$.
    \end{enumerate}
    \end{itemize}
    Return $R := R \circ (y,x)$.
\end{relate}
}

\begin{procedure}[h]
  \SetProcNameSty{textnormal}
    Initially, set $R:=[]$.

    Find the smallest $j$ such that $Y[1]_2 \lhd X[j]$ and set
    $y:=Y[1]_1$, $x:= X[j]$. 
    If no such $j$ exists stop and return $R$.

    \For{$i :=2$ to $\norm{Y}$}{
    {\bf until} $Y[i]_2 \lhd X[j]$ or $j>\norm{X}$ {\bf do}

    $\quad$ set $j:=j+1$.


    \uIf{$j> \norm{X}$}
    {stop and return $R := R \circ (y,x)$.}
    \uElse{ Compare $X[j]$ and $x$.
    There are two cases:
    
    \uCase{{\bf 1.} $x \lhd X[j]$} {set
    $R:=R\circ (y,x)$, $y:=Y[i]_1$, and $x:=X[j]$.}
    \uCase{{\bf 2.} If $x=X[j]$}{
    set $y:=Y[i]_1$.}
    }
    	}
    Return $R := R \circ (y,x)$.
\caption{\Mop($Y$,$X$)
}
\end{procedure}
In procedure \Mop\ we have a "potential pair" $(y,x)$ where
$y=Y[i']_1$ for some $i'$ and $Y[i']_2 \lhd x$. In case 1 we have $x \lhd X[j]$ and also
$Y[i']_2 \lhd Y[i]_2$ since the input lists are ordered and $i'<i$ (see
Figure~\ref{fig:mopimplexample}(a)). Therefore, $(y,x)$ is
inserted into $R$. In case 2 we have $x=X[j]$, i.e., $Y[i]_2\lhd
x$, and as before $Y[i']_2 \lhd Y[i]_2$ (see
Figure~\ref{fig:mopimplexample}(b)). Therefore $(y,x)$ cannot be
in $\Mop(Y,X)$, and we set $(Y[i]_1,x)$ to be the new potential
pair.
\begin{figure}[t]
\begin{center}
\begin{psmatrix}[colsep=0.8cm,rowsep=0.4cm,labelsep=2pt]
  &&&& \cnode{.1}{root}\rput(0,.3){}
  \\
  & \cnode{.1}{l}\rput(0,.3){} &&&  \cnode{.1}{c}\rput(-.3,.3){}
  &&
  \cnode{.1}{r}\rput(0,.3){} \\
  \cnode{.1}{ll}&
  \cnode*{.1}{lc}\rput(0,-.3){$Y[i']_2$} &
  \cnode{.1}{lr}\rput(-.2,-.3){} &  \cnode*{.1}{cl}\rput(.6,0){$Y[i]_2$}
  & &
  \cnode{.1}{cr}& 
  \cnode{.1}{rl}
  &
  \psset{fillstyle=crosshatch}\cnode{.1}{rr}\rput(0,-.3){$X[j]$}&&&
     \\
  && \cnode{.1}{lrc}& \psset{fillstyle=crosshatch}\cnode{.1}{cll}\rput(0,-.3){$x$}
  & \cnode{.1}{clr}&&
  \cnode{.1}{rlc}\rput(0,-.3){}\\[.2cm]
  &&&&\rput(0,0){(a)}
  \ncline{root}{l}\ncline{root}{c}\ncline{root}{r}
  \ncline{l}{ll}\ncline{l}{lc}\ncline{l}{lr}
  \ncline{c}{cr}\ncline{c}{cl}
  \ncline{r}{rl}\ncline{r}{rr}
  \ncline{lr}{lcr}\ncline{cl}{cll}\ncline{cl}{clr}
  \ncline{rl}{rlc}\ncline{lr}{lrc}
  \end{psmatrix}
  \begin{psmatrix}[colsep=0.8cm,rowsep=0.4cm,labelsep=1pt]
  &&&& \cnode{.1}{root}
  \\
  & \cnode{.1}{l} &&&  \cnode{.1}{c}
  &&
  \cnode{.1}{r} \\
  \cnode{.1}{ll} &
  \cnode*{.1}{lc}\rput(0,-.3){$Y[i']_2$} &
  \cnode{.1}{lr} &  \cnode{.1}{cl}
  &&
  \cnode{.1}{cr} &
  \cnode{.1}{rl}
  &
  \cnode{.1}{rr}
     \\
  && \cnode{.1}{lrc}
  & \cnode{.1}{cll}
  & \cnode*{.1}{clr}\rput(0,-.3){$Y[i]_2$}&&
  \psset{fillstyle=crosshatch}\cnode{.1}{rlc}\rput(.3,-.3){$x=X[j]$}\\[.2cm]
  &&&&\rput(0,0){(b)}
  \ncline{root}{l}\ncline{root}{c}\ncline{root}{r}
  \ncline{l}{ll}\ncline{l}{lc}\ncline{l}{lr}
  \ncline{c}{cr}\ncline{c}{cl}
  \ncline{r}{rl}\ncline{r}{rr}
  \ncline{lr}{lcr}\ncline{cl}{cll}\ncline{cl}{clr}
  \ncline{rl}{rlc}\ncline{lr}{lrc}
  \end{psmatrix}
     \caption{Case 1 and 2 from the implementation of \Mop.
     (a) We have $x \lhd X[j]$ and therefore $Y[i]_2 \ntriangleleft x$. So $(y,x)$ is inserted in $R$.
     (b) We have $Y[i']_2 \lhd Y[i]_2 \lhd x = X[j]$.
      }
  \label{fig:mopimplexample}
\end{center}
\end{figure}

We can implement $\MopLeft(X,Y)$ similarly to \Mop\ replacing smallest by largest, $\lhd$ by $\rhd$, and traversing the lists backwards: 

\showOld{
\begin{relate}[simple]
\item[$\MopLeft(X,Y)$:]
    Initially, set $R:=[]$.

    Find the largest $j$ such that $Y[\norm{Y}]_1 \rhd X[j]$ and set
    $y:=Y[\norm{Y}]_2$ and $x:= X[j]$. If no such $j$ exists stop and return $R$.

    For $i :=\norm{Y}-1$ to $1$ do:
    \begin{itemize}
    \item[]  Until $Y[i]_1 \rhd X[j]$ or $j<1$ set $j:=j-1$.

    If $j<1$
    stop and return $R := (x,y) \circ R$.
    Otherwise, compare $X[j]$ and $x$.
    There are two cases:
    \begin{enumerate}
    \item If $x \rhd X[j]$ set
    $R:=(x,y) \circ R$, $y:=Y[i]_2$, and $x:=X[j]$.
    \item If $x=X[j]$
    set $y:=Y[i]_2$.
    \end{enumerate}
    \end{itemize}
    Return $R := (x,y)\circ R $.
\end{relate}
\begin{relate}[simple]
\item[$\Fl(X,\alpha)$:]
Initially, set $L:=X$, $Z:=L$.

    Repeat until $Z=[]$:
    \begin{itemize}
    \item[] For $i:=1$ to $\norm{Z}$ do: 
    	\begin{enumerate}
	\item If $\lab(Z[i])=\alpha$: Delete $Z[i]$ from $Z$ (but keep it in $L$).
	\item If $\lab(Z[i])\neq \alpha$ and $\parent(Z[i])\neq \bot$: Replace $Z[i]$ with $\parent(Z[i])$ in both $Z$ and $L$.
	\item If $\lab(Z[i])\neq \alpha$ and $\parent(Z[i])= \bot$: Delete $Z[i]$ from both $Z$ and $L$.
	\end{enumerate}
    \item[] Set $Z:=\Deep^*(Z,L)$.
    \end{itemize}
    Return $L$.
\end{relate}}{}

\begin{procedure}[h]
  \SetProcNameSty{textnormal}
    Initially, set $R:=[]$.

    Find the largest $j$ such that $Y[\norm{Y}]_1 \rhd X[j]$ and set
    $y:=Y[\norm{Y}]_2$ and $x:= X[j]$. If no such $j$ exists stop and return $R$.

    \For{$i :=\norm{Y}-1$ to $1$}{
     {\bf until} $Y[i]_1 \rhd X[j]$ or $j<1$ {\bf do} 

     $\quad$ set $j:=j-1$.

    \uIf{$j<1$}
    {stop and return $R := (x,y) \circ R$.}
    \Else{compare $X[j]$ and $x$.
    There are two cases:
    
    \uCase{{\bf 1.} $x \rhd X[j]$} {set
    $R:=(x,y) \circ R$, $y:=Y[i]_2$, and $x:=X[j]$.}
    
    \uCase{{\bf 2.} $x=X[j]$} {
    set $y:=Y[i]_2$.}
    		}
        }
    Return $R := (x,y) \circ R$.

\caption{\MopLeft($X$,$Y$)}
\end{procedure}

\begin{procedure}[H]
  \SetProcNameSty{textnormal}
  Initially, set $L:=X$, $Z:=L$.

    \While{$Z\neq[]$}{
        \For{$i:=1$ to $\norm{Z}$}{
       \uCase{{\bf 1.} $\lab(Z[i])=\alpha$}{Delete $Z[i]$ from $Z$ (but keep it in $L$).}
	\uCase{{\bf 2.} $\lab(Z[i])\neq \alpha$ and $\parent(Z[i])\neq \bot$} {Replace $Z[i]$ with $\parent(Z[i])$ in both $Z$ and $L$.}
	\uCase{{\bf 3.} $\lab(Z[i])\neq \alpha$ and $\parent(Z[i])= \bot$} {Delete $Z[i]$ from both $Z$ and $L$.}
%
		}
    Set $(Z,L):=\Deep^*(Z,L)$.
	}	
    Return $L$.
\caption{\Fl($X$,$\alpha$)}
\end{procedure}

The procedure \Fl\ computes the set $\Deep(\{\fl(x,\alpha)| x\in X \})$ bottom up. The list $Z$ contains ancestors of the elements of $X$ for which we have not yet found an ancestor with label $\alpha$. In each step it considers each node $z$ in the list $Z$. If it has the right label then $x \in \Fl(X,\alpha)$ and we remove it from $Z$ but keep it in $L$. Otherwise we replace it with its parent (unless it is the root). Thus $L$ contains both the elements in $Z$ and the part of $\Fl(X,\alpha)$ found until now. 

To keep the running time down we wish to maintain the invariant that $L$ is deep at the beginning of each iteration of the outer loop. 
To do this procedure $\Fl$ calls an auxiliary procedure $\Deep^*(Z,L)$ which takes two
ordered lists $Z$ and $L$, where $Z\subseteq L$, and returns two
ordered lists representing the set $\Deep(L) \cap Z$ and $\Deep(L)$, i.e., $\Deep^*(Z,L)=([z\in Z|\nexists
x \in L:z\prec x], \Deep(L))$. If we use the procedure \Deep\ to calculate $\Deep^*$ it takes time $O(\norm{Z}+\norm{L})=O(\norm{L})$. Instead we will show how to calculate it in time $O(\norm{Z})$ using a linked list representation for $Z$ and $L$. We will need this in the proof of Lemma~\ref{lem:fl}, which shows that the total running time of $\emph{all}$ calls to \Fl\ from \Emb\ takes time $O(n_T)$.
Below we describe in more detail how to implement \Fl\ together with the auxiliary procedures.

We use a doubly linked list to represent $L$ and extra pointers in
this list to represent $Z$. Each element in the list has pointers
\SuccL\ and \PredL\ pointing to its predecessor and successor in $L$. Similarly, each element in $Z$ has pointers \SuccZ\ and \PredZ\ pointing to its predecessor and successor in $Z$ (right after the initialization these are equal to \SuccL\ and \PredL). In the for loop we use the \SuccZ\ pointers to find the next element in $Z$. To delete $Z[i]$ from $Z$ in case 1 we set $\SuccZ(\PredZ(Z[i]))=\SuccZ(Z[i])$ and $\PredZ(\SuccZ(Z[i]))=\PredZ(Z[i])$. The $L$ pointers stay the same. In case 2 we simply replace $Z[i]$ with its parent in the linked list. The \Succ\ and \Pred\ pointers stay the same. To delete  $Z[i]$ from both $Z$ and $L$ in case 3 we set $\Succ_j(\Pred_j(Z[i]))=\Succ_j(Z[i])$ and $\Pred_j(\Succ_j(Z[i]))=\Pred_j(Z[i])$ for $j \in \{Z,L\}$. 
Finally, to compute $\Deep^*(Z,L)$ walk through $Z$
following the \SuccZ\ pointers. At each node $z$ compare $\PredL(z)$
and $\SuccL(z)$ with $z$. If one of them is a descendant of $z$
remove $z$ from the doubly linked list $Z$ and $L$ as in case 3. Note that instead of calling $\Deep^*(Z,L)$ this comparison can also be done directly in step 2, which is the only place where we insert nodes that might be an ancestor of another node in $L$. 
We will show in the next section that it is enough to compare $z$ to its neighbors in the list $L$.

\subsection{Correctness of the Set Procedures}\label{sec:simplecorrectness}
Clearly, \Parent\ and \Nca\ are correct. The following lemmas show
that $\Deep$, $\Fl$, and $\Mop$ are also correctly implemented.
For notational convenience we write $x\in X$, for a list $X$, if
$x = X[i]$ for some $i$, $1 \leq i \leq \norm{X}$.
\begin{lemma}\label{lem:deep}
    Procedure $\Deep(X)$ is correct.
\end{lemma}

\begin{proof} Let $x$ be the variable in the procedure. We will first
  prove the following invariant on $x$: 
  \\\\
  \begin{invariant}
    At the beginning of each iteration of the for loop in line 2 we have $x \not\prec X[j]$ for any $1\leq j \leq i-1$.\\
  \end{invariant}
  
  We prove the invariant by induction on $i$. The invariant obviously
  holds for the base case $i=2$.
  \\
  For the induction step let $i\geq 3$. Let iteration $k$ denote the
  iteration of the for loop when $i=k$. By the induction hypothesis we
  have $x \not\prec X[j]$ for any $1\leq j \leq i-2$ at the beginning
  of iteration $i-1$.

  Let $x'$ denote the value of
  the variable $x$ at the beginning of iteration
  $i-1$.  
  Consider the value of variable $x$ at the beginning of the
  iteration $i$. 
  There are two cases:
  \begin{enumerate}
  \item If $x=x'$ then by the induction hypothesis $x = x' \not\prec X[j]$
    for any $1\leq j \leq i-2$. Since $x$ was not changed in
    iteration $i-1$ we have $X[i-1]\preceq x$ (case 3 of the procedure) and thus $x \not\prec X[j]$ for any $1\leq j \leq i-1$.
  \item If $x\neq x'$ then $x$ was set in either case 1 ($x'\lhd x$) or case 2 ($x' \prec x$)
    in iteration $i-1$. Therefore, $x=X[i-1]$ and by the induction
    hypothesis $x' \not\prec X[j]$ for any $1\leq j \leq i-2$. There are
    two subcases:
    \begin{enumerate}
    \item If $x'\prec x$ it follows immediately from the induction
      hypothesis that $x \not\prec X[j]$ for any $1\leq j \leq i-1$, since
      all descendants of $x$ also are descendants of $x'$. 
    \item If $x'\lhd x$ we note that $x$ is the first node to the left of $x'$ occuring
      after $x'$ in $X$ (otherwise $x$ would have been reset in case 1
      of the procedure in an earlier iteration, contradicting that
      $x'$ is the value of variable $x$ at the beginning of iteration
      $i-1$). 
Since $X$ is semiordered no node $X[j]$ with smaller index in $X$ than $x'$ can be to the right of $x'$. Thus no node $X[j]$, $1\leq j < i-2$, can be to the right of $x'$.
      Since all descendants of $x$ must be to the right of $x'$ we have $x
      \not\prec X[j]$ for any $1\leq j \leq i-1$.
    \end{enumerate}
  \end{enumerate}
  
  We are now ready to prove that $y\in \Deep(X)$ iff there exists no $z\in X$ such that $y\prec z$.
  We first argue that if $y\in \Deep(X)$ then $\nexists z\in X$ such
  that $y\prec z$. Let $y$ be an element in $\Deep(X)$. Only elements that have been assigned to $x$ during the procedure are in the output. Consider the iteration where $x = y$ is appended to $R$. This only happens in case 1 of the procedure and thus $y=x \lhd X[i]$. Since $X$ is semiordered this implies that $x \lhd X[j]$ for $i \leq j \leq \norm{X}$, and therefore $y=x \not \prec X[j]$ for $i \leq j \leq \norm{X}$. 
  By the above invariant it follows that $y=x \not \prec X[j]$ for $1 \leq j \leq i-1$. 
  Thus if $y\in \Deep(X)$ then $\nexists z\in X$ such that $y\prec z$. 
  
  Let $y\in X$ be an element such that $X\cap V(T(y))=\{y\}$. Let
  $j$ be the smallest index such that $X[j]=y$. When comparing $y$ and
  $x$ during the iteration where $i=j$ we are in case 1 or 2, since $j$ is the smallest index such that $X[j]=y$ (implying $x\neq y$) and $X\cap V(T(y))=\{y\}$ (implying $y\not \prec x$). In either case $x$ is set to $y$. Since there are no descendants of $y$ in $X$, the variable $x$ remains equal to $y$ until added to $R$. If
  $y$ occurs several times in $X$ we will have $x=y$ each time
  we meet a copy of $y$ (except the first) and it follows from
  the implementation that $y$ will occur exactly once in $R$.
\end{proof}

To show that the implementation of \Mop\ is correct we will use the following proposition.

\begin{prop}\label{prop:mopInvariant}
Before the first iteration of the for loop in line 3 of \Mop\ we have $y=Y[1]_1$, $x=X[j]$ and either $X[j-1] \lhd Y[1]_2 \lhd X[j]$ (if $j>1$) or $Y[1]_2 \lhd X[j]$ if  ($j=1$).

At the end of each iteration of the for loop then, unless $Y[i]_2 \not\!\!\lhd X[|X|]$, we have  $y=Y[i]_1$, $x=X[j]$ and either $X[j-1] \lhd Y[i]_2 \lhd X[j]$ (if $j>1$) or $Y[i]_2 \lhd X[j]$ if  ($j=1$). 
\end{prop}
\begin{proof} The first statement ($y=Y[1]_1$, $x=X[j]$ and either
  $X[j-1] \lhd Y[1]_2 \lhd X[j]$ (if $j>1$) or $Y[1]_2 \lhd X[j]$ if
  ($j=1$)) follows immediately from the implementation of the
  procedure line 2  and the fact that the input lists are ordered.

We prove the second statement by induction on $i$. Base case $i=2$.  By the first statement we have $y=Y[1]_1$, $x=X[j]$ and either $X[j-1] \lhd Y[1]_2 \lhd X[j]$ (if $j>1$) or $Y[1]_2 \lhd X[j]$ if  ($j=1$) before this iteration. Let $j'$ be the value of $j$ before this iteration.  It follows immediately from the implementation that $y=Y[2]_1$ since $y$ is set to this in both case 1 and 2. If $Y[2]_2 \lhd X[j']$ then $j=j'$. Since $Y$ is ordered it follows that  $X[j-1] \lhd Y[1]_2\lhd Y[2]_2 \lhd X[j]$ (if $j>1$) or $Y[2]_2 \lhd X[j]$ if  ($j=1$). If $Y[2]_2 \not\!\!\lhd X[j']$ then $j$ is increased until $Y[2]_2 \lhd X[j]$ implying $X[j-1] \lhd Y[2]_2 \lhd X[j]$ unless $j>\norm{X}$, since $X$ is ordered.

Induction step $i >2$. It follows immediately from the implementation that $y=Y[i]_1$ since $y$ is set to this in both case 1 and 2. By the induction hypothesis we have $y=Y[i]_1$, $x=X[j]$ and $Y[i]_2 \lhd X[j]$ right before this iteration. 
Let $j'$ be the value of $j$ before this iteration. If $Y[i]_2 \lhd X[j']$ then $j=j'$. Since $Y$ is ordered it follows that  $X[j-1] \lhd Y[i-1]_2\lhd Y[i]_2 \lhd X[j]$ (if $j>1$) or $Y[i]_2 \lhd X[j]$ if  ($j=1$). If $Y[i]_2 \not\!\!\lhd X[j']$ then $j$ is increased until $Y[i]_2 \lhd X[j]$ implying $X[j-1] \lhd Y[i]_2 \lhd X[j]$ unless $j>\norm{X}$.
\end{proof}

\begin{lemma}\label{lem:nnm}
    Procedure $\Mop(Y,X)$ is correct.
\end{lemma}
\begin{proof}
    We want to show that for any $1\leq i' \leq \norm{Y}$, $1 \leq j' \leq
    \norm{X}$: 
    $$(Y[i']_1,X[j'])\in R \quad \Leftrightarrow \quad (Y[i']_2,X[j'])
    \in \mop(Y|_2,X)\;.$$
    
    \noindent Since $\restrict{Y}{2}$ and $X$ are ordered lists we
    have
    $(Y[i']_2,X[j']) \in \mop(Y|_2,X)$ if and only if
    \begin{itemize}
    \item[(1)]      $\arg\min_j Y[i']_2 \lhd X[j] = j'$
    \item[(2)] $\arg\max_i Y[i]_2 \lhd X[j'] = i'$
    \end{itemize}

    We will first show that (1) and (2) implies  $(Y[i']_2,X[j']) \in
    R$. We start by showing that when $i$ is about to be incremented
    to $i'+1$ then $y=Y[i']_1$ and $x = X[j']$. There are two cases to
    consider.
    \begin{itemize}
    \item $i'=1$. After line 2 is executed, $y$ is set to $Y[1]_1$,
      $j$ is set to $j'$ and $x$ is set to $X[j']$.
    \item $i'>1$. Consider the step in the iteration when $i=i'$. At
      the beginning of this iteration, $y=Y[i'-1]$ and $j$ is the
      minimal index such that $Y[i'-1] \lhd X[j]$. By (1) this implies
      that $j\leq j'$, and that after line 4 and 5 are executed, $j$ is
      set to $j'$. At the end of the
      iteration $y=Y[i']$ ($y$ is assigned to $Y[i']$ in both cases)
      and $x=X[j']$. If $j=j'$ then $x$ set to $X[j']$ in case 1,
      otherwise we had  $j=j'$ (case 2) and then $x$ was set to
      $X[j']$ already). 
    \end{itemize}
    We have established that when $i$ is about to be incremented to
    $i'+1$ then $y=Y[i']_1$ and $x = X[j']$. To show that
    $(Y[i']_2,X[j']) \in R$ we consider the following two cases.
    \begin{itemize}
    \item $i' < |Y|$. Consider the $(i'+1)^{\textit{th}}$
      iteration. By condition (2) $X[j'] \unlhd Y[i'+1]_2$ and therefore $j$ is
      increased in line 5. So now $j>j'$. If $j>|X|$ then
      $(y,x)=(Y[i']_2,X[j'])$ is added to $R$ in line 7. Otherwise,
      since $X$ is ordered, $x =X[j'] \lhd X[j]$. We are therefore in
      case 1 and $(y,x) =(Y[i']_2,X[j'])$ is added to $R$.
    \item $i'= |Y|$. Then $(y,x)=(Y[i']_2,X[j'])$ is added to $R$ in
      line 15.
    \end{itemize}

     We will now show that $(Y[i']_1,X[j'])\in R$ implies (1) and (2).
     Since $(Y[i']_1,X[j'])\in R$ we had $(y,x)=(Y[i']_1,X[j'])$ at some point during the execution. The pair $(y,x)$ can be added to $R$ only in the for loop before changing the values of $y$ and $x$ or at the execution of the last line of the procedure. Therefore $(y,x)=(Y[i']_1,X[j'])$ at the beginning of  some execution of the for loop, or after the last iteration ($i=|Y|$). It follows by Proposition~\ref{prop:mopInvariant} that
     $X[j-1] \lhd Y[i]_2 \lhd X[j]$ if $j>1$ or $Y[i]_2 \lhd X[j]$ if $j=1$. It remains to show that $X[j']\lhd Y[i'+1]_2$ for $i'<|Y|$. It follows from the implementation that $(y,x)$ only is added to $R$ inside the for loop if $j$ is increased. Thus $j$ was increased in the next iteration ($i=i'+1$) implying $X[j']\lhd Y[i'+1]_2$.

\end{proof}
\begin{lemma}
    Procedure $\MopLeft(X,Y)$ is correct.
\end{lemma}
\begin{proof}
Similar to the proof of  Lemma~\ref{lem:nnm}.
\end{proof}
To show that \Fl\ is correct we need the following proposition.
\begin{prop}\label{prop:deeplist}
Let $X$ be an ordered list and let $x$ be an ancestor of $X[i]$
for some $i \in \{1,\ldots,k\}$. If $x$ is an ancestor of some
node in $X$ other than $X[i]$ then $x$ is an ancestor of $X[i-1]$
or $X[i+1]$.
\end{prop}
\begin{proof}
Recall that $u\lhd v$ iff $\pre(u)<\pre(v)$ and $\post(u)< \post(v)$. Since $x\prec X[i]$ we have $\pre(x)<\pre(X[i])$ and $\post(X[i])<\post(x)$. Assume there exists a descendant $X[j]$ of $x$ such that $j\not \in \{i-1,i,i+1\}$. If $j<i-1$ we have 
$$\pre(x)\leq \pre(X[j]) <\pre(X[i-1]),$$
where the first inequality follows from $x\prec X[j]$ and the second from $X$ being ordered. And $$\post(X[i-1])<\post(X[i])\leq \post(x),$$ where the first inequality follows from $X$ being ordered and the second from  $x\prec X[i]$. Thus $x \prec X[i-1]$. 

Similarly, for $j> i+1$, we have $\pre(x)\leq \pre(X[i]) <\pre(X[i+1])$ and $\post(X[i+1])<\post(X[j])\leq \post(x)$ implying that $x\prec X[i+1]$.
%
\end{proof}
Proposition~\ref{prop:deeplist} shows that the doubly linked list
implementation of $\Deep^*$ is correct. Since all changes to the list are either deletions or insertions of a parent in the place of its child, the list $L$ (and thus also $Z$) is ordered at the beginning of each iteration of the outer loop.
\begin{lemma}\label{lem:flcorrect}
    Procedure $\Fl(X,\alpha)$ is correct.
\end{lemma}
\begin{proof}
    Let $F=\{\fl(x,\alpha)\mid x \in X\}$. We first show that $\Fl(X,\alpha) \subseteq F$. Consider a node $x \in \Fl(X,\alpha)$.
    Since $x$ is in $L$ after the final iteration, $x$ was deleted from $Z$ during some iteration. Thus $\lab(x)=\alpha$. 
    For any $y\in X$ we follow the path from $y$ to the root and stop the first time we meet a node with label $\alpha$ or even earlier since we keep the list deep. Thus $x\in F$. 
              
    The set $\Fl(X,\alpha)$ is a deep set, and therefore $\Deep(F)\subseteq
\Fl(X,\alpha) \subseteq F  \Rightarrow \Deep(F)=\Fl(X,\alpha)$.
Hence, 
it remains to show that $\Deep(F)\subseteq \Fl(X,\alpha)$.
    Let $x$ be a
    node in $\Deep(F)$, let $z \in X$ be a node such that
    $x=\fl(z,\alpha)$, and let $z=x_1,x_2,\ldots,x_k=x$
    be the nodes on the
    path from $z$ to $x$. We will argue that after each iteration of the algorithm we have $x_i \in L$
    for some $i$. Since $\lab(x_i)\neq \alpha$ for $i<k$ this is the
    same as $x_i\in Z$ for $i<k$. Before the first iteration we have
    $x_1 \in X =Z$. As long as $i<k$ we replace $x_i$ with
    $x_{i+1}$ in case 2 of the for loop, since $\lab(x_i)\neq
    \alpha$. When $i=k$ we remove $x_k$ from $Z$ but keep it in
    $L$. It remains to show that we do not delete $x_i$ in the
    computation of $\Deep^*(Z,L)$ in any iteration. If $x_i$ is
    removed then there is a node $y\in L$ that is a descendant of
    $x_i$ and thus also a descendant of $x$. We argued above that
    $L\setminus Z\subseteq F$ and thus $y \in Z$ since $x \in
    \Deep(F)$. But since $x \in \Deep(F)$ no node on the path from $y$
    to $x$ can have label $\alpha$ and therefore $x_i$ will eventually be reinserted in $Z$.  
\end{proof}

\subsection{Complexity of the Set Procedures}\label{sec:simplecomplexity}
For the running time of the node list implementation observe that,
given the data structure described in
Section~\ref{sec:simplepreprocessing}, all set procedures, except
$\Fl$, perform a single pass over the input using constant time at
each step. Hence we have,
\begin{lemma}\label{lem:auxprocedures}
For any tree $T$ there is a data structure using $O(n_T)$ space
and preprocessing which supports each of the procedures $\Parent$,
$\Deep$, $\Mop$, $\MopLeft$, and $\Nca$ in linear time (in the size of their
input).
\end{lemma}
The running time of a single call to \Fl\ might take time
$O(n_T)$. Instead we will divide the calls to \Fl\ into groups and
analyze the total time used on such a group of calls. The
intuition behind the division is that for a path in $P$ the calls
made to \Fl\ by \Emb\ are done bottom up on disjoint lists of
nodes in $T$.

\begin{lemma}\label{lem:fl}
For disjoint ordered node lists $X_1, \ldots, X_k$ and labels
$\alpha_1, \ldots, \alpha_k$, such that any node in $X_{i+1}$ is
an ancestor of some node in $\Fl_T(X_i, \alpha_i)$, $1 \leq
i < k$, all of $\Fl_T(X_1, \alpha_1), \ldots , \Fl_T(X_k,
\alpha_k)$ can be computed in $O(n_T)$ time.
\end{lemma}
\begin{proof}
Let $Z$ and $L$ be as in the implementation of the
procedure. Since $\Deep^*$ takes time $O(\norm{Z})$ and each of the steps in the for loop takes constant time, we
only need to show that the total length of the lists $Z$---summed
over all the calls---is $O(n_T)$ to analyze the total time usage. 
We will show that any node in $T$ can be in $Z$ at the beginning of the while loop at most twice during
all calls to \Fl. The size of $Z$ cannot increase in the iterations of
the for loop (line 3--10), and thus the size of $Z$ when $\Deep^*$ is
called (line 11) is at most the size of $Z$ at the beginning of this
iteration of the while loop.

Consider a single call to \Fl. 
Except for the first iteration, a node can be in $Z$ only if one
of its children were in $Z$ in the last iteration. Note that $Z$ is ordered 
at the beginning of each for loop. 
Thus if a node is in $Z$ at the beginning of the while loop none of its children 
are in $Z$ and thus in one
call to \Fl\ a node can be in $Z$ only once.

Look at a node $z$ the first time it appears in $Z$ at the beginning of an execution of the while loop. Assume that
this is in the call $\Fl(X_i,\alpha_i)$. 
\begin{itemize}
\item
If $z \in X_i$ then $z$
cannot be in $Z$ in any later calls, since no node in $X_j$ where
$j>i$ can be a descendant of a node in $X_i$.
\item
 If $\lab(z)\neq
\alpha_i$ then $z$ is removed from $Z$ in case 2 or case 3 of
the procedure and
cannot be in $Z$ in any of the later calls.
To see this consider the time when $z$ is removed from $Z$ (case 2 or case 3). 
Since the set $L$ is deep at the beginning of the while  loop 
and $Z\subseteq L$, no descendant of $z$ will appear in
$Z$ later in this call to \Fl, and no node in the output from $\Fl(X_i,\alpha_i)$ can be a
descendant of $z$. Since any node in $X_j$, $j>i$, is an ancestor
of some node in $\Fl(X_i,\alpha_i)$ neither $z$ or any
descendant of $z$ can be in any $X_j$, $j>i$. Thus $z$ cannot
appear in $Z$ in any later calls to \Fl. 
\item Now if $\lab(z)=\alpha_i$ then we
might have $z \in X_{i+1}$. In that case, $z$ will appear in $Z$
in the first iteration of the procedure call
$\Fl(X_{i+1},\alpha_i)$, but not in any later calls since the
lists are disjoint, and since no node in $X_j$ where $j>i+1$ can
be a descendant of a node in $X_{i+1}$. If $\lab(z)=\alpha_i$ and $z \not
\in X_{i+1}$ then clearly $z$ cannot appear in $Z$ in any later
call. 
\end{itemize}
Thus a node in $T$ is in $Z$ at the beginning of an execution of the while loop at most twice during all the
calls.
\end{proof}

\subsection{Complexity of the Tree Inclusion Algorithm}
Using the node list implementation of the set procedures we get:

\begin{lemma}\label{lem:simpletime}
For trees $P$ and $T$ the tree inclusion problem can be solved in
$O(l_Pn_T)$ time.
\end{lemma}
\begin{proof}
By Lemma~\ref{lem:auxprocedures} we can preprocess $T$ in $O(n_T)$
time and space. Let $g(n)$ denote the time used by $\Fl$ on a
list of length $n$. Consider the time used by $\Emb(\roots(P))$.
We bound the contribution for each node $v\in V(P)$. If $v$ is a leaf we are in case 1 of \Emb. The cost of computing $\Fl(L(T), \lab(v))$ is $O(g(l_T))$, and by Lemma~\ref{lem:fl} (with $k=1$) we get $O(g(l_T))=O(n_T)$. Hence, the total cost of all leaves is $O(l_P  n_T)$.
If $v$ has a single child $w$ we are in case 2 of \Emb, and by Lemma~\ref{lem:auxprocedures} the cost is $O(g(|\Emb(w)|))$. If
$v$ has more than one child the cost of $\Mop$, $\Nca$, and $\Deep$ is bounded by $\sum_{w \in \child(v)} O(|\Emb(w)|)$. Furthermore, since the length of the output of
$\Mop$ (and thus $\Nca$) is at most $z = \min_{w \in \child(v)}
|\Emb(w)|$ the cost of $\Fl$ is $O(g(z))$. Hence, the total cost
for internal nodes is,
\begin{equation}
\label{eq:internal}  \sum_{v \in V(P)\backslash L(P)}
O\bigg(g(\min_{w \in \child(v)} |\Emb(w)|) + \sum_{w \in
\child(v)} |\Emb(w)|\bigg) = \sum_{v \in V(P)}
O(g(|\Emb(v)|))\;.
\end{equation}
Next we bound (\ref{eq:internal}). For any $w \in \child(v)$ we have that $\Emb(w)$ and $\Emb(v)$ are disjoint ordered lists. Furthermore we have that any node in $\Emb(v)$ must be an ancestor of some node in $\Fl(\Emb(w), \lab(v))$. Hence, by Lemma~\ref{lem:fl}, for any leaf to root path $\delta = v_1, \ldots, v_k$ in $P$, we have that $\sum_{u \in \delta} g(|\Emb(u)|) = O(n_T)$. Let
$\Delta$ denote the set of all root to leaf paths in $P$. It
follows that,
\begin{equation*}
\sum_{v \in V(T)} g(|\Emb(v)|) \leq \sum_{p \in \Delta} \sum_{u
\in p} g(|\Emb(u)|) = O(l_Pn_T)\;.
\end{equation*}
Since this time is the same as the time spent at the leaves the time
bound follows.
\end{proof}
%
To analyze the space used by the algorithm we first bound the size of $\Emb(v)$ for each node $v \in V(P)$. We then   use this to bound the total the size of embeddings stored in the recursion stack in the computation of $\Emb(\roots(P))$, i.e., the total size of embeddings stored by recursive calls during the computation.
\begin{lemma}\label{lem:emb_size}
For any tree $P$ we have $\forall v \in V(P)$:
$$\norm{\Emb_T(v)} \leq \frac{l_T}{l_{P(v)}}\;.$$
\end{lemma}
\begin{proof}
By Lemma~\ref{lem:correctEmb} $\Emb(v)$ is the set of deep occurrences of $P(v)$ in $T$. By the definition of deep the occurrences are disjoint and no node in one occurrence can be an ancestor of a node in another occurrence. Each occurrence has at least $l_{P(v)}$ descendant leaves and each of these leaves is an ancestor of at least one distinct leaf in $T$ (see also Figure~\ref{fig:MopSubtree}(b)). Thus the number of occurrences is bounded by $l_T/l_{P(v)}$.
\end{proof}

\begin{figure}
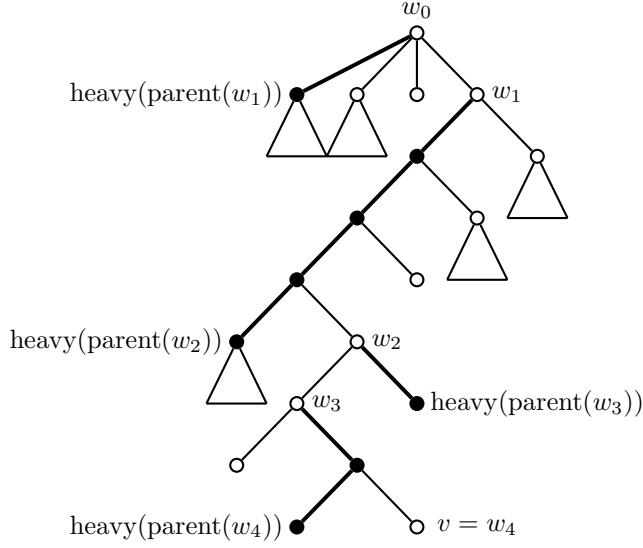

\begin{center}
 \begin{psmatrix}[colsep=0.4cm,rowsep=0.4cm,labelsep=1pt]
 & &&&&&& \cnode{.1}{w0}\rput(0,.3){$w_0$}\\
  &&&\cnode*{.1}{h1}\rput(-1.6,0){$\heavy(\parent(w_1))$}& &\cnode{.1}{a1}&&\cnode{.1}{a2}&&  \cnode{.1}{w1}\rput(.4,0){$w_1$} \\  
 & &\pnode{h11}&&\pnode{h12}&&\pnode{a12}&\cnode*{.1}{bh} &&&&\cnode{.1}{bl} \\
  &&&&&\cnode*{.1}{ch} &&&&\cnode{.1}{cl} & \pnode{b11}&&\pnode{b12} \\
  &&&\cnode*{.1}{dh} &&&&\cnode{.1}{dl}& \pnode{c11}&&\pnode{c12} \\
  &\cnode*{.1}{h2}\rput(-1.6,0){$\heavy(\parent(w_2))$} &&&&\cnode{.1}{w2}\rput(.4,0){$w_2$}  \\
   \pnode{h21}&&\pnode{h22}&\cnode{.1}{w3}\rput(.4,0){$w_3$} &&&&\cnode*{.1}{h3}\rput(1.6,0){$\heavy(\parent(w_3))$} \\
   &\cnode{.1}{el} &&&&\cnode*{.1}{eh} \\
 &&&\cnode*{.1}{h4}\rput(-1.6,0){$\heavy(\parent(w_4))$} &&&&\cnode{.1}{w4}\rput(.8,0){$v=w_4$}
  \ncline{w0}{a1}\ncline{w0}{a2}\ncline{w0}{w1}
  \ncline{w1}{bl}\ncline{bh}{cl}\ncline{ch}{dl}
  \ncline{dh}{w2}\ncline{w2}{w3}
  \ncline{w3}{el}\ncline{eh}{w4}
  \ncline{h1}{h11}\ncline{h1}{h12}\ncline{h12}{h11}  
  \ncline{a1}{a12}\ncline{a1}{h12}\ncline{h12}{a12}
  \ncline{bl}{b11}\ncline{bl}{b12}\ncline{b12}{b11} 
  \ncline{cl}{c11}\ncline{cl}{c12}\ncline{c12}{c11} 
  \ncline{h2}{h21}\ncline{h2}{h22}\ncline{h22}{h21}
\psset{linewidth=1.5pt}
 \ncline{w0}{h1}
 \ncline{w1}{bh} 
 \ncline{bh}{ch}\ncline{ch}{dh}
 \ncline{dh}{h2}\ncline{w2}{h3}
 \ncline{w3}{eh}\ncline{eh}{h4}
  \end{psmatrix}
\end{center}
\caption{Path from root to $v$. The heavy nodes are black and the light nodes are white. The heavy edges are the thick edges and the light edges are thin.}\label{fig:heavypath}
\end{figure}

\begin{lemma}\label{lem:lin_space}
The total size of saved embeddings on the recursion stack at any time during the computation of $\Emb(\roots (P))$ is at most $O(l_T)$.
\end{lemma}
\begin{proof}
Let node $v$ be the node for which we are currently computing \Emb. 
Let  $p$ be the path from the root to $v$ and let $w_0,\ldots,w_\ell$ be the light nodes on this path. Let $\ell=\lightdepth(v)$. 
There is one embedding on the stack for each light node on the path (see Figure~\ref{fig:heavypath}): For the heavy nodes on the path there can be no saved embeddings in the recursion as the algorithm always recurses on the heavy child first.
For each light node $w_i$ on the path $p$ except the root $w_0$ the stack will contain either $\Emb(\heavy(\parent(w_i)))$, or $U_j=\Mop(U_{j-1},R_j)$, where $v_j$ is $w_i$'s left sibling, or $U_j=\MopLeft(U_{j-1},R_j)$, where $v_j$ is $w_i$'s right sibling. The computation of $U_j$ is a series of \Mop\ (or \MopLeft) computations that started with the pair of node lists $(\Emb(\heavy(\parent(w_i))),\Emb(\heavy(\parent(w_i))))$ as the first argument to \Mop\ (or \MopLeft). As the output of \Mop\ (or \MopLeft) can be no larger than the input to the procedure we have $\norm{U_j}=O(\norm{\Emb(\heavy(\parent(w_i))))}$ and thus the total space used at any time during the recursion is 
$$
O\big (\sum_{i=1}^{\lightdepth(v)} \norm{\Emb(\heavy(\parent(w_i)))}\big )\;.$$
By Lemma~\ref{lem:emb_size} we have 
$$\norm{\Emb(\heavy(\parent(w_i)))} \leq \frac{l_T}{l_{P(\heavy(\parent(w_i)))}}\;,$$
and thus 
\begin{equation}\label{eq:totalspace}
\sum_{i=1}^{\lightdepth(v)} \norm{\Emb(\heavy(\parent(w_i)))}\leq l_T\sum_{i=1}^{\lightdepth(v)}\frac{1}{l_{P(\heavy(\parent(w_i)))}}\;.
\end{equation}
By the definition of heavy the node $\heavy(\parent(w_i))$ has more leaves in its subtree than $w_i$, i.e.,
\begin{equation}\label{eq:sibling}
l_{P(w_i)} \leq l_{P(\heavy(\parent(w_i)))}\;.
\end{equation}
Obviously, $\heavy(\parent(w_i))$ has no more leaves in its subtree than its parent, i.e.,
\begin{equation}\label{eq:heavy_parent}
l_{P(\heavy(\parent(w_i)))}\leq l_{P(\parent(w_i))}\;.
\end{equation}
Since $w_i$ is light it has at most half the number of leaves in its subtree as its parent, that is
\begin{equation}\label{eq:half_parent}
l_{P(w_i)}\leq l_{P(\parent(w_i))}/2\;.
\end{equation}
Combining this with the fact that $w_i$ is an ancestor of $w_{i+1}$ and $\heavy(\parent(w_{i+1}))$ we get,
\begin{alignat*}{2}
l_{P(\heavy(\parent(w_j))} &\leq  l_{P(\parent(w_j))} &\quad& \textrm{by }  \eqref{eq:heavy_parent}\\
&\leq  l_{P(w_{j-1})} &&  \textrm{since } w_{j-1} \textrm{ is an ancestor of } w_j \\
&\leq l_{P(\parent(w_{j-1}))}/2 && \textrm{by } \eqref{eq:half_parent} \\
&\leq l_{P(w_{j-2})}/2 && \textrm{since } w_{j-2} \textrm{ is an ancestor of } \parent(w_{j-1}) \\
&\leq l_{P(\heavy(\parent(w_{j-2})))}/2, && \textrm{by } \eqref{eq:sibling}
\end{alignat*}
for any $2<j\leq \lightdepth(v)$. Let $l_i=l_{P(\heavy(\parent(w_i)))}$ for all $i$. 
To bound the sum in~\eqref{eq:totalspace} we will use that $l_i \leq l_{i-2}/2$, $l_i < l_{i-1}$, and $l_{\lightdepth(v)} \geq 1$. We have 
$$\sum_{i=1}^{\lightdepth(v)}\frac{1}{l_i}\leq  2\sum_{i=2, i \textrm{ odd}}^{\lightdepth(v)}\frac{1}{l_i} \leq 2 \cdot 2=4,$$ since the $l_i$'s in the last sum are decreasing with a factor of 2. Combining this with Equation~\eqref{eq:totalspace} we get 
$$\sum_{i=1}^{\lightdepth(v)} \norm{\Emb(\heavy(\parent(w_i)))}\leq l_T\sum_{i=1}^{\lightdepth(v)}\frac{1}{l_{P(\heavy(\parent(w_i)))}} \leq 4 l_T \;.$$
\end{proof}

\begin{theorem}\label{thm:simple}
For trees $P$ and $T$ the tree inclusion problem can be solved in
$O(l_Pn_T)$ time and $O(n_T)$  space.
\end{theorem}
\begin{proof}
The time bound follows from Lemma~\ref{lem:simpletime}. Next consider the space used by
$\Emb(\roots(P))$.
  The preprocessing of
Section~\ref{sec:simplepreprocessing} uses only $O(n_T)$ space.
By Lemma~\ref{lem:lin_space} the space used
for the saved embeddings is $O(l_T) = O(n_T)$.  \end{proof}

\subsection{An Alternative Algorithm}\label{sec:alt}
In this section we present an alternative algorithm. Since the
time complexity of the algorithm in the previous section is
dominated by the time used by $\Fl$, we present an implementation
of this procedure which leads to a different complexity. Define a
\emph{firstlabel data structure} as a data structure supporting
queries of the form $\fl(v, \alpha)$, $v\in V(T)$, $\alpha \in
\Sigma$. Maintaining such a data structure is known as the
\emph{tree color problem}. This is a well-studied problem, see
e.g. \cite{Die89,MM1996, FM1996,AHR1998}. With such a data
structure available we can compute $\Fl$ as follows,
\begin{relate}
\item[$\Fl(X, \alpha)$:] Return the list $\Deep([\fl(X[1], \alpha),
\ldots, \fl(X[\norm{X}], \alpha)])$.
\end{relate}
\begin{theorem}\label{thm:simple2}
Let $P$ and $T$ be trees. Given a firstlabel data structure using
$s(n_T)$ space, $p(n_T)$ preprocessing time, and $q(n_T)$ time for
queries, the tree inclusion problem can be solved in $O(p(n_T) +
l_Pl_T\cdot q(n_T))$ time and $O(s(n_T) + n_T)$ space.
\end{theorem}
\begin{proof}
Constructing the firstlabel data structures uses $O(s(n_T))$ space and
$O(p(n_T))$ time. The total cost of the leaves is bounded by $O(l_p l_T\cdot q(n_T))$, since the cost of a single leaf is $O(l_T\cdot q(n_T))$. As in the proof of Theorem~\ref{thm:simple} we
have that the total time 
used by the internal nodes is bounded by
$\sum_{v \in V(P)} g(|\Emb(v)|)$, where $g(n)$ is the time used
by $\Fl$ on a list of length $n$, that is, $g(n)\leq n\cdot q(n_T)$.
By Lemma~\ref{lem:auxprocedures} and Lemma~\ref{lem:lin_space} for any leaf to root path $\delta = v_1, \ldots, v_k$ in $P$, we have that $\sum_{u \in \delta} |\Emb(u)| \leq O(l_T)$. Let
$\Delta$ denote the set of all root to leaf paths in $P$. It
follows that,
\begin{equation*}
\sum_{v \in V(P)} g(|\Emb(v)|) \leq \sum_{p \in \Delta} \sum_{u
\in p} g(|\Emb(u)|) \leq \sum_{p \in \Delta}O(l_T \cdot q(n_T))\leq O(l_Pl_T \cdot q(n_T)) .
\end{equation*}
Since this time is the same as the time spent at the leaves the time
bound follows.
\end{proof}
Several firstlabel data structures are available, for instance, if we want to maintain linear space we have,
\begin{lemma}[Dietz~\cite{Die89}]\label{lem:dietz}
For any tree $T$ there is a data structure using $O(n_T)$ space,
$O(n_T)$ expected preprocessing time which supports firstlabel
queries in $O(\log \log n_T)$ time.
\end{lemma}
The expectation in the preprocessing time is due to perfect hashing. Since our data structure does not need to support efficient updates we can remove the expectation by using the deterministic dictionary of Hagerup et al.~\cite{HMP2001}. This gives a worst-case preprocessing time of $O(n_T \log n_T)$. However, using a simple two-level approach this can be reduced to $O(n_T)$ (see e.g. \cite{Thorup2003}). Plugging in this data structure we obtain,
\begin{corollary}\label{cor:simple}
For trees $P$ and $T$ the tree inclusion problem can be solved in
$O(l_Pl_T\log\log n_T + n_T)$ time and $O(n_T)$ space.
\end{corollary}

\section{A Faster Tree Inclusion Algorithm}\label{micromacro}
In this section we present a new tree inclusion algorithm which
has a worst-case subquadratic running time. As discussed in the introduction, the general idea is to divide $T$ into clusters of logarithmic size which we can efficiently preprocess and then use this to speed up the computation with a logarithmic factor.

\subsection{Clustering}
In this section we describe how to divide $T$ into clusters and
how the macro tree is created. For simplicity in the presentation
we assume that $T$ is a binary tree. If this is not the case it is
straightforward to construct a binary tree $B$, where $n_{B} \leq
2n_T$, and a mapping $g : V(T) \rightarrow V(B)$ such that for any
pair of nodes $v,w \in V(T)$, $\lab(v) = \lab(g(v))$, $v \prec w$
iff $g(v) \prec g(w)$, and $v \lhd w$ iff $g(v) \lhd g(w)$. The
nodes in the set $U = V(B)\backslash \{g(v) \mid v \in V(T)\}$ are
assigned a special label $\beta \not\in \Sigma$. It follows that
for any tree $P$, $P \sqsubseteq T$ iff $P \sqsubseteq B$.

Let $C$ be a connected subgraph of $T$. A node in $V(C)$ adjacent
to a node in $V(T)\backslash V(C)$ is a \emph{boundary} node. The
boundary nodes of $C$ are denoted by $\delta C$. We have 
$\roots(T) \in \delta C$ if $\roots(T) \in
V(C)$. A \emph{cluster}
of $C$ is a connected subgraph of $C$ with at most two boundary
nodes. A set of clusters $CS$ is a \emph{cluster partition} of $T$
iff $V(T) = \cup_{C\in CS} V(C)$, $E(T) = \cup_{C\in CS} E(C)$,
and for any $C_1 ,C_2 \in CS$, $E(C_1) \cap E(C_2) = \emptyset$,
$|E(C_1)| \geq 1$.  If $|\delta C| = 1$ we call $C$ a \emph{leaf cluster} and
otherwise an \emph{internal cluster}.

We use the following recursive procedure $\Cluster_T(v,s)$,
adopted from \cite{AR2002c}, which creates a cluster partition
$CS$ of the tree $T(v)$ with the property that $|CS| = O(s)$ and
$|V(C)| \leq \ceil{n_T/s}$ for each $C \in CS$. A similar cluster partitioning
achieving the same result follows from \cite{AHT2000, AHLT1997,
Frederickson1997}.
\begin{relate}
\item[$\Cluster_T(v,s)$:] For each child $u$ of $v$ there are two
cases:
\begin{enumerate}
  \item $|V(T(u))| + 1 \leq \ceil{n_T/s}$. Let the nodes $\{v\} \cup V(T(u))$ be a leaf cluster with boundary node $v$.
  \item $|V(T(u))| \geq \ceil{n_T/s}$. Pick a node $w \in V(T(u))$ of
  maximum depth such that $|V(T(u))| + 2 - |V(T(w))| \leq \ceil{n_T/s}$.
  Let the nodes $V(T(u)) \backslash V(T(w)) \cup \{v,w\}$ be an internal
  cluster with boundary nodes $v$ and $w$. Recursively, compute $\Cluster_T(w, s)$.
\end{enumerate}
\end{relate}

\begin{lemma}\label{lem:clustering}
Given a tree $T$ with $n_T>1$ nodes, and a parameter $s$, where
$\ceil{n_T/s} \geq 2$, we can build a cluster partition $CS$ in
$O(n_T)$ time, such that $|CS| = O(s)$ and $|V(C)| \leq
\ceil{n_T/s}$ for any $C \in CS$.
\end{lemma}
\begin{proof}
The procedure $\Cluster_T(\roots(T), s)$ clearly creates a cluster
partition of $T$ and it is straightforward to implement in
$O(n_T)$ time. Consider the size of the clusters created. There
are two cases for $u$. In case $1$, $|V(T(u))| + 1 \leq
\ceil{n_T/s}$ and hence the cluster $C = \{v\} \cup V(T(u))$
has size $|V(C)| \leq \ceil{n_T/s}$. In case $2$, $|V(T(u))| + 2 -
|V(T(w))| \leq \ceil{n_T/s}$ and hence the cluster $C = V(T(u))
\backslash V(T(w)) \cup \{v,w\}$ has size $|V(C)| \leq
\ceil{n_T/s}$.

Next consider the size of the cluster partition.  Let $c=
\ceil{n_T/s}$. We say that a cluster $C$ is \emph{bad} if
$\norm{V(C)} \leq c/2$ and \emph{good} otherwise. We will show
that at least a constant fraction of the clusters in the cluster
partition are good. 
It is easy to verify that the cluster
partition created by procedure \Cluster\ has the following
properties:
\begin{itemize}
\item[(i)] Let $C$ be a bad internal cluster with boundary nodes
$v$ and $w$ ($v \prec
w$). Then $w$ has two children with at least $c/2$ descendants
each.
\item[(ii)] Let $C$ be a bad leaf cluster with boundary node
$v$. Then the boundary node $v$ is contained in  a good cluster.
\end{itemize}
By (ii) the number of bad leaf clusters is at most twice the number of good internal clusters and by (i) each bad internal cluster has two child clusters. 
Therefore, the number of bad internal clusters is bounded by the number of leaf clusters. Let $b_i$ and $g_i$ denote the number of bad and good internal clusters, respectively, and let $b_l$ and $g_l$ denote the number of bad and good leaf clusters, respectively. We have $$ b_i \leq b_l + g_l  \leq 2 g_i + g_l,$$
and therefore the number of bad clusters is bounded by
$$ b_l + b_i \leq 2 g_i + g_l + 2 g_i = 4g_i + g_l \;.$$ 
Thus the number of bad clusters is at most 4 times the number of good clusters, and therefore at most a constant fraction of the total
number of clusters. Since a good cluster is of size more than
$c/2$, there can be at most $2s$ good clusters and thus
$\norm{CS}=O(s)$.
\end{proof}

Let $C\in CS$ be an internal cluster with $v, w \in \delta C$. The
\emph{spine path} of $C$ is the path between $v,w$
excluding $v$ and $w$. A node on the spine path is a \emph{spine
node}. A node to the left and right of $v$ 
or of any node on the spine path is a \emph{left node} and \emph{right node}, respectively.
If $C$ is a leaf cluster with $v \in \delta C$ then any proper
descendant of $v$ is a \emph{leaf node}.

\begin{figure}[t]
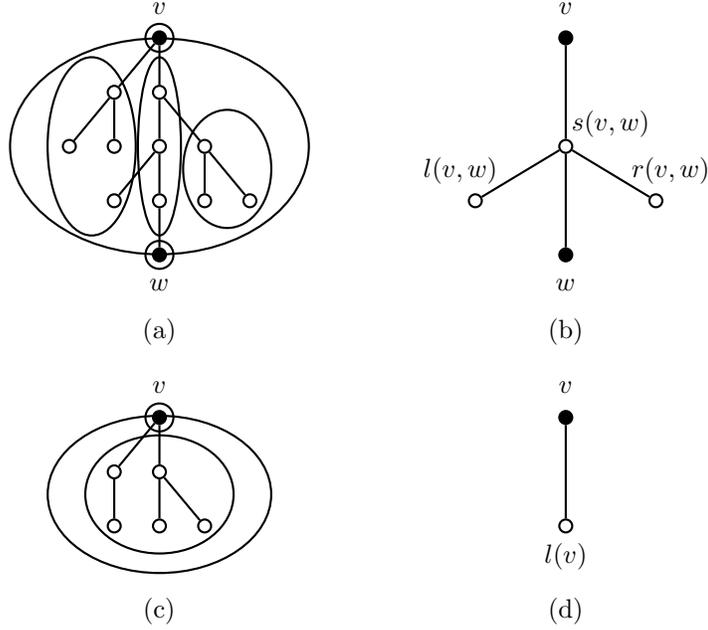

\begin{center}
  \begin{psmatrix}[colsep=0.6cm,rowsep=0.3cm,labelsep=1pt]
  &&& \psellipse(0,0)(.2,.2)\cnode*[fillcolor=black]{.1}{root1}\rput(0,.4){$v$} &&&&&&&&&
  \cnode*[fillcolor=black]{.1}{v}\rput(0,.4){$v$}\\
  && \cnode{.1}{a1} &  \cnode{.1}{a2} &&&&&& \\
  & \cnode{.1}{b1} & \psellipse(-.3,0)(.6,1.2)\cnode{.1}{b2} & \cnode{.1}{b3}\psellipse(0,0)(.3,1.2) \psellipse(0,0)(2,1.45) & \psellipse(.3,-.3)(.6,.8)\cnode{.1}{b4}&&&&&&&& \cnode{.1}{s}\rput(.6,.3){$s(v,w)$} \\
   & & \cnode{.1}{c1} & \cnode{.1}{c2} & \cnode{.1}{c3} & \cnode{.1}{c4} &&&&&
   \cnode{.1}{l}\rput(-.2,.4){$l(v,w)$} &&&& \cnode{.1}{r}\rput(.2,.4){$r(v,w)$}\\
  & && \psellipse(0,0)(.2,.2)\cnode*[fillcolor=black]{.1}{d2}\rput(0,-.4){$w$} &&&&&&&&& \cnode*[fillcolor=black]{.1}{w}\rput(0,-.4){$w$}\\
  &&& \rput(0,-.3){(a)} &&&&&&&&& \rput(0,-.3){(b)} \\\\
  \ncline{root1}{a1}\ncline{root1}{a2}
  \ncline{a1}{b1}\ncline{a1}{b2} \ncline{a2}{b3}\ncline{a2}{b4}
  \ncline{b3}{c1}\ncline{b3}{c2} \ncline{b4}{c3}\ncline{b4}{c4}
  \ncline{c2}{d2}
  \ncline{v}{s}\ncline{w}{s}\ncline{l}{s}\ncline{r}{s}
  &&& \psellipse(0,0)(.2,.2)\cnode*[fillcolor=black]{.1}{root1}\rput(0,.4){$v$} &&&&&&&&&
  \cnode*[fillcolor=black]{.1}{v}\rput(0,.4){$v$}\\
  && \cnode{.1}{a1} &  \psellipse(0,-.3)(1,.8)\psellipse(0,-.3)(1.5,1.06)\cnode{.1}{a2} &&&&&& \\
  &&\cnode{.1}{b2} & \cnode{.1}{b3} & \cnode{.1}{b4}&&&&&&&& \cnode{.1}{s}\rput(0,-.4){$l(v)$} \\
  &&& \rput(0,-.4){(c)} &&&&&&&&& \rput(0,-.4){(d)} \\
  \ncline{root1}{a1}\ncline{root1}{a2}
  \ncline{a1}{b2} \ncline{a2}{b3}\ncline{a2}{b4}
  \ncline{v}{s}
  \end{psmatrix}
   \caption{The clustering and the macro tree. (a) An internal cluster. The black nodes are the
   boundary nodes and the internal ellipses correspond to the boundary nodes,
   the right and left nodes, and spine path. (b) The macro tree corresponding to
   the cluster in (a). (c) A leaf cluster. The internal ellipses are the boundary node
   and the leaf nodes. (d) The macro tree corresponding to the cluster in (c).}
  \label{clusterexample}
\end{center}
\end{figure}

Let $CS$ be a cluster partition of $T$ as described in Lemma
\ref{lem:clustering}. We define an ordered \emph{macro tree}
$M$. Our definition of $M$ may be viewed as an ''ordered''
version of the macro tree defined in \cite{AR2002c}. The node set $V(M)$ consists of the boundary nodes in $CS$. Additionally, for each internal cluster $C \in CS$ with $v,w \in \delta C$, $v \prec w$, we have the nodes $s(v,w)$, $l(v,w)$ and $r(v,w)$ and edges $(v, s(v,w)), (s(v,w),l(v,w)), (s(v,w), w)$, and $(s(v,w),r(v,w))$. That is, the nodes $l(v,w)$,  $r(v,w)$ and $w$ are all children of $s(v,w)$. The nodes are ordered so that $l(v,w) \lhd w \lhd r(v,w)$. For each leaf cluster $C$, $v \in \delta C$, we have the node $l(v)$ and edge $(v,l(v))$.  Since $\roots(T)$ is a boundary node, $M$ is rooted at $\roots(T)$. Figure~\ref{clusterexample} illustrates these definitions.

With each node $v \in V(T)$ we associate a unique macro node denoted
$c(v)$. Let $u \in V(C)$, where $C \in CS$.
\begin{equation*}
c(u) =
\begin{cases}
        u & \text{if $u$ is boundary node}, \\
        l(v) & \text{if $u$ is a leaf node and $v \in \delta C$}, \\
        s(v,w) & \text{if $u$ is a spine node, $v,w \in \delta C$, and $v\prec w$}, \\
    l(v,w) & \text{if $u$ is a left node, $v,w \in \delta C$, and $v\prec w$}, \\
    r(v,w) & \text{if $u$ is a right node, $v,w \in \delta C$, and $v\prec w$}.
\end{cases}
\end{equation*}

Conversely, for any macro node $i \in V(M)$ define the
\emph{micro forest}, denoted $C(i)$, as the induced
subgraph of $T$ of the set of nodes $\{v \mid v \in V(T), i =
c(v)\}$. We also assign a \emph{set} of labels to $i$ given by
$\lab(i) = \{\lab(v) \mid v \in V(C(i))\}$. If $i$ is a spine node
or a boundary node the unique node in $V(C(i))$ of greatest depth
is denoted by $\first(i)$. Finally, for any set of nodes $\{i_1,
\ldots, i_k\} \subseteq V(M)$ we define $C(i_1, \ldots, i_k)$ as the induced subgraph of the set of nodes $V(C(i_1)) \cup \cdots \cup V(C(i_k))$.

\begin{figure}[t]
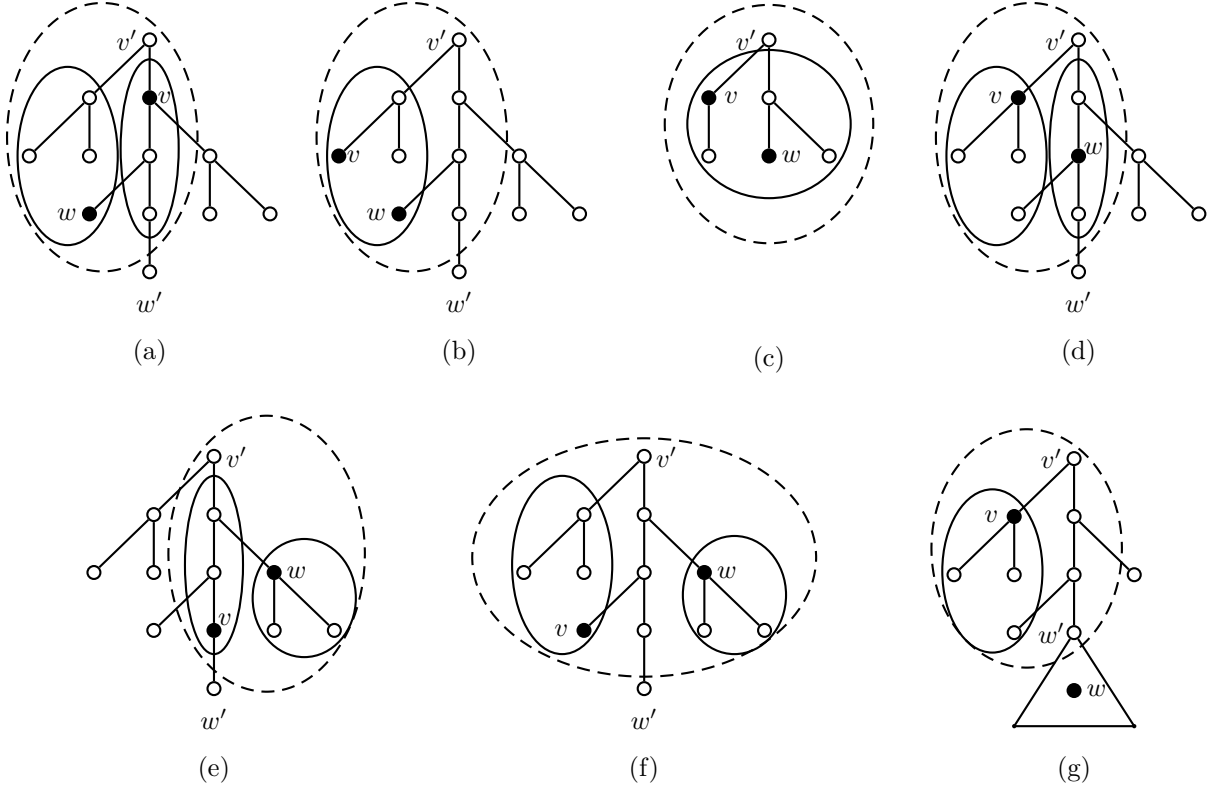

\begin{center}
  \begin{psmatrix}[colsep=0.8cm,rowsep=0.35cm,labelsep=1pt]
  &&& 
  \cnode{.1}{root1}\rput(-.3,0){$v'$}
  \\
  && \cnode{.1}{a1} &  \cnode*{.1}{a2}\rput(.2,0){$v$}
  \\
  & \cnode{.1}{b1} & \psellipse(-.29,0)(.68,1.2) 
  \cnode{.1}{b2} &
  \cnode{.1}{b3} \psellipse(0,.1)(.4,1.2) 
    \psellipse[linestyle=dashed](-.63,.25)(1.28,1.8)
  &
  \cnode{.1}{b4}
 \\
   & & \cnode*{.1}{c1}\rput(-.3,0){$w$} & \cnode{.1}{c2} & \cnode{.1}{c3} & \cnode{.1}{c4} \\
  & && 
  \cnode{.1}{d2}\rput(0,-.4){$w'$}
 \\
  &&&
  \rput(0,-.3){(a)}
  \\\\
  \ncline{root1}{a1}\ncline{root1}{a2}
  \ncline{a1}{b1}\ncline{a1}{b2} \ncline{a2}{b3}\ncline{a2}{b4}
  \ncline{b3}{c1}\ncline{b3}{c2} \ncline{b4}{c3}\ncline{b4}{c4}
  \ncline{c2}{d2}
  \end{psmatrix}
  \begin{psmatrix}[colsep=0.8cm,rowsep=0.35cm,labelsep=1pt]
  &&& 
  \cnode{.1}{root1}\rput(-.3,0){$v'$}
  \\
  && \cnode{.1}{a1} &  \cnode{.1}{a2}
  \\
  & \cnode*{.1}{b1} \rput(.2,0){$v$}& \psellipse(-.29,0)(.68,1.2) 
  \cnode{.1}{b2} &
  \cnode{.1}{b3}
    \psellipse[linestyle=dashed](-.63,.25)(1.28,1.8)
  &
  \cnode{.1}{b4}
 \\
   & & \cnode*{.1}{c1}\rput(-.3,0){$w$} & \cnode{.1}{c2} & \cnode{.1}{c3} & \cnode{.1}{c4} \\
  & && 
  \cnode{.1}{d2}\rput(0,-.4){$w'$}
 \\
  &&&
  \rput(0,-.3){(b)}
  \\\\
  \ncline{root1}{a1}\ncline{root1}{a2}
  \ncline{a1}{b1}\ncline{a1}{b2} \ncline{a2}{b3}\ncline{a2}{b4}
  \ncline{b3}{c1}\ncline{b3}{c2} \ncline{b4}{c3}\ncline{b4}{c4}
  \ncline{c2}{d2}
  \end{psmatrix}
  \begin{psmatrix}[colsep=0.8cm,rowsep=0.35cm,labelsep=1pt]
  &&&
  \cnode{.1}{root1}\rput(-.3,0){$v'$} 
  \\
   &&\cnode*{.1}{a1}\rput(.3,0){$v$} &  \psellipse(0,-.35)(1.1,1)\psellipse[linestyle=dashed](0,-.35)(1.4,1.6)\cnode{.1}{a2} \\ 
  &&\cnode{.1}{b2} & \cnode*{.1}{b3}\rput(.3,0){$w$} & \cnode{.1}{b4} &
  \\\\\\
  &&& \rput(0,-.4){(c)}    \\\\
  \ncline{root1}{a1}\ncline{root1}{a2}
  \ncline{a1}{b2} \ncline{a2}{b3}\ncline{a2}{b4}
  \end{psmatrix}
  \begin{psmatrix}[colsep=0.8cm,rowsep=0.35cm,labelsep=1pt]
  &&& 
  \cnode{.1}{root1}\rput(-.3,0){$v'$}
  \\
  && \cnode*{.1}{a1}\rput(-.3,0){$v$} &  \cnode{.1}{a2}
  \\
  & \cnode{.1}{b1} & \psellipse(-.29,0)(.68,1.2) 
  \cnode{.1}{b2} &
  \cnode*{.1}{b3}\rput(.2,.1){$w$} \psellipse(0,.1)(.4,1.2) 
    \psellipse[linestyle=dashed](-.63,.25)(1.28,1.8)
  &
  \cnode{.1}{b4}
 \\
   & & \cnode{.1}{c1} & \cnode{.1}{c2} & \cnode{.1}{c3} & \cnode{.1}{c4} \\
  & && 
  \cnode{.1}{d2}\rput(0,-.4){$w'$}
 \\
  &&&
  \rput(0,-.3){(d)}
  \\\\
  \ncline{root1}{a1}\ncline{root1}{a2}
  \ncline{a1}{b1}\ncline{a1}{b2} \ncline{a2}{b3}\ncline{a2}{b4}
  \ncline{b3}{c1}\ncline{b3}{c2} \ncline{b4}{c3}\ncline{b4}{c4}
  \ncline{c2}{d2}
  \end{psmatrix}
  \begin{psmatrix}[colsep=0.8cm,rowsep=0.35cm,labelsep=1pt]
  &&& 
  \cnode{.1}{root1}\rput(.3,0){$v'$}
  \\
  && \cnode{.1}{a1} &  \cnode{.1}{a2}
  \\
  & \cnode{.1}{b1} & 
  \cnode{.1}{b2} &
  \cnode{.1}{b3} \psellipse(0,.1)(.4,1.2) 
    \psellipse[linestyle=dashed](.7,.25)(1.32,1.85)
  &
  \psellipse(.4,-.34)(.7,.8) 
  \cnode*{.1}{b4}\rput(.3,0){$w$}
 \\
   & & \cnode{.1}{c1} & \cnode*{.1}{c2}\rput(.15,.15){$v$} & \cnode{.1}{c3} & \cnode{.1}{c4} && \\
  & && 
  \cnode{.1}{d2}\rput(0,-.4){$w'$}
 \\
  &&&
  \rput(0,-.3){(e)}
  \\
  \ncline{root1}{a1}\ncline{root1}{a2}
  \ncline{a1}{b1}\ncline{a1}{b2} \ncline{a2}{b3}\ncline{a2}{b4}
  \ncline{b3}{c1}\ncline{b3}{c2} \ncline{b4}{c3}\ncline{b4}{c4}
  \ncline{c2}{d2}
  \end{psmatrix}
  \begin{psmatrix}[colsep=0.8cm,rowsep=0.35cm,labelsep=1pt]
  &&& 
  \cnode{.1}{root1}\rput(.3,0){$v'$}
  \\
  && \cnode{.1}{a1} &  \cnode{.1}{a2}
  \\
  & \cnode{.1}{b1} & \psellipse(-.29,0.1)(.68,1.2) 
  \cnode{.1}{b2} &
  \cnode{.1}{b3} 
    \psellipse[linestyle=dashed](0,0.2)(2.3,1.6)
  &
  \psellipse(.4,-.3)(.7,.8) 
  \cnode*{.1}{b4}\rput(.3,0){$w$}
 \\
   & & \cnode*{.1}{c1}\rput(-.3,0){$v$}& \cnode{.1}{c2} & \cnode{.1}{c3} & \cnode{.1}{c4} &&\\
  & && 
  \cnode{.1}{d2}\rput(0,-0.4){$w'$}
 \\
  &&&
  \rput(0,-.3){(f)}
  \\
  \ncline{root1}{a1}\ncline{root1}{a2}
  \ncline{a1}{b1}\ncline{a1}{b2} \ncline{a2}{b3}\ncline{a2}{b4}
  \ncline{b3}{c1}\ncline{b3}{c2} \ncline{b4}{c3}\ncline{b4}{c4}
  \ncline{c2}{d2}
  \end{psmatrix}
   \begin{psmatrix}[colsep=0.8cm,rowsep=0.35cm,labelsep=1pt]
  &&& 
  \cnode{.1}{root1}\rput(-.3,0){$v'$}
  \\
  && \cnode*{.1}{a1}\rput(-.3,0){$v$} &  \cnode{.1}{a2}
  \\
  & \cnode{.1}{b1} & \psellipse(-.29,0.05)(.68,1.1) 
  \cnode{.1}{b2} &
  \cnode{.1}{b3} 
    \psellipse[linestyle=dashed](-.63,.35)(1.28,1.6)
  &
  \cnode{.1}{b4}
 \\
   & & \cnode{.1}{c1} & \cnode{.1}{c2}\rput(-.3,0){$w'$} 
   \\
  & & & 
  \cnode*{.1}{d3}\rput(.3,0){$w$}
 \\[-.3cm]
  && \cnode{0}{d2} && \cnode{0}{d4}
  \\[-.5cm]
  &&&
  \rput(0,-.3){(g)}
  \\
  \ncline{root1}{a1}\ncline{root1}{a2}
  \ncline{a1}{b1}\ncline{a1}{b2} \ncline{a2}{b3}\ncline{a2}{b4}
  \ncline{b3}{c1}\ncline{b3}{c2} 
  \ncline{c2}{d2}\ncline{c2}{d4}\ncline{d2}{d4}
  \end{psmatrix}
   \caption{Examples from the propositions. In all cases $v'$ and $w'$ are top and bottom
   boundary nodes of the cluster, respectively. (a)
   Proposition~\ref{lem:ancestorlemma}(ii).
   Here $c(v)=s(v',w')$ and $c(w)=l(v',w')$ (solid ellipses). The dashed ellipse corresponds to
   $C(c(w),s(v',w'),v')$.
   (b) Proposition~\ref{lem:orderlemma}(i) and~\ref{lem:ncalemma}(ii). Here $c(v)=c(w)=l(v',w')$ (solid ellipse). 
   The dashed ellipse corresponds to
   $C(c(w),s(v',w'),v')$. 
   (c) Proposition~\ref{lem:orderlemma}(ii) and~\ref{lem:ncalemma}(i). Here $c(v)=c(w)=l(v')$ (solid ellipse). 
   The dashed ellipse corresponds to   $C(c(v),v')$. 
   (d)
    Proposition~\ref{lem:orderlemma}(iii). Here $c(v)=l(v',w')$ and $c(w)=s(v',w')$ (solid ellipses). The dashed ellipse corresponds to
   $C(c(v),c(w),v')$.
   (e) Proposition~\ref{lem:orderlemma}(iv). Here $c(v)=s(v',w')$ and $c(w)=r(v',w')$ (solid ellipses). The dashed ellipse corresponds to
   $C(c(v),c(w),v')$. (f) Proposition~\ref{lem:ncalemma}(iv). Here $c(v)=l(v',w')$ and $c(w)=r(v',w')$ (solid ellipses). The dashed ellipse corresponds to $C(c(v),c(w),s(v',w'),v')$.  (g) Proposition~\ref{lem:ncalemma}(v). Here $c(v)=l(v',w')$ (solid ellipse) and $w' \preceq_M c(w)$. The dashed ellipse corresponds to $C(c(v),s(v',w'),v',w'))$.}
  \label{fig:propositions}
\end{center}
\end{figure}

The following propositions state useful properties of ancestors,
nearest common ancestor, and the left-to-right ordering in the
micro forests and in $T$. The propositions follow directly
from the definition of the clustering. See also
Figure~\ref{fig:propositions}.
\begin{prop}[Ancestor relations]\label{lem:ancestorlemma}
For any pair of nodes $v, w \in V(T)$, the following hold
\begin{itemize}
  \item[(i)] If $c(v) = c(w)$ then $v \prec_T w$ iff $v \prec_{C(c(v))} w$.
  \item[(ii)] If $c(v) \neq c(w)$, and for some boundary nodes $v',w'$ we have $c(v) =s(v',w')$, and $c(w)
  \in \{l(v', w'), r(v',w')\}$, then $v \prec_T w$ iff $v \prec_{C(c(w),s(v',w'), v')} w$.
  \item[(iii)] In all other cases, $v \prec_T w$ iff $c(v) \prec_{M} c(w)$.
\end{itemize}
\end{prop}
Case (i) says that if $v$ and $w$ belong to the same macro node
then $v$ is an ancestor of $w$ iff $v$ is an ancestor of $w$ in
the micro forest for that macro node. Case (ii) says that if $v$
is a spine node 
and
$w$ is a left or right node in the same cluster then
$v$ is an ancestor of $w$ iff $v$ is an ancestor of $w$ in the
micro tree induced by that cluster (Figure~\ref{fig:propositions}(a)). Case (iii) says that in all
other cases $v$ is an ancestor of $w$ iff the macro node $v$
belongs to is an ancestor of the macro node $w$ belongs to in the
macro tree.

\begin{prop}[Left-of relations]\label{lem:orderlemma}
For any pair of nodes $v, w \in V(T)$, the following hold
\begin{itemize}
  \item[(i)] If $c(v) = c(w) \in \{r(v',w'),l(v',w')\}$ for some boundary nodes $v',w'$, then $v \lhd w$
     iff $v \lhd_{C(c(v),v',s(v',w'))}
    w$.
  \item[(ii)] If $c(v) = c(w)=l(v')$ for some boundary node $v'$, then $v \lhd w$ iff $v \lhd_{C(c(v),v')} w$.
  \item[(iii)] If $c(v) = l(v', w')$ and $c(w)=s(v',w')$  for some boundary nodes $v',w'$, then
  $v \lhd w$ iff $v \lhd_{C(c(v), c(w), v')} w$.
  \item[(iv)] If  $c(v)=s(v',w')$  and $c(w) = r(v', w')$ for some boundary nodes $v',w'$, then
  $v \lhd w$ iff $v \lhd_{C(c(v), c(w), v')} w$.
  \item[(v)] In all other cases, $v \lhd w$ iff $c(v) \lhd_{M} c(w)$.
\end{itemize}
\end{prop}
Case (i) says that if $v$ and $w$  are both either left or right nodes in the same cluster then $v$ is to the left of $w$ iff
$v$ is to the left of $w$ in the micro tree induced by their macro
node together with the spine and top boundary node of the cluster (Figure~\ref{fig:propositions}(b)).
Case (ii) says that if $v$ and $w$ are both leaf nodes in the same cluster then $v$ is to the left of $w$ iff $v$ is to the
left of $w$ in the micro tree induced by that leaf cluster (Figure~\ref{fig:propositions}(c)). Case (iii) says that if $v$ is a left node and $w$ is a spine node in the
same cluster then $v$ is to the left of $w$ iff $v$ is to the left
of $w$ in the micro tree induced by their two macro nodes and the
top boundary node of the cluster (Figure~\ref{fig:propositions}(d)). Case (iv) says that if $v$ is a spine node and $w$  is a  right node in the same cluster then $v$ is to the left of $w$ iff $v$ is to the left
of $w$ in the micro tree induced by their two macro nodes  and the top boundary node of the cluster (Figure~\ref{fig:propositions}(e)). In all other cases $v$ is to the
left of $w$ if the macro node $v$ belongs to is to the left of the
macro node of $w$ in the macro tree (Case (v)).

\begin{prop}[Nca relations]\label{lem:ncalemma}
    For any pair of nodes $v, w \in V(T)$, the following hold
    \begin{itemize}
    \item[(i)] If $c(v) = c(w)=l(v')$ for some boundary node $v'$,
        then $\nca_T(v,w)=\nca_{C(c(v),v')}(v,w)$.
    \item[(ii)] If $c(v) = c(w)\in\{l(v',w'),r(v',w')\}$ for some boundary nodes $v',w'$,
        then  \\ $\nca_T(v,w)=\nca_{C(c(v), s(v',w'), v')}(v,w)$.
    \item[(iii)] If $c(v) = c(w)=s(v',w')$ for some boundary nodes $v',w'$,
        then $\nca_T(v,w)=\nca_{C(c(v))}(v,w)$.
    \item[(iv)] If $c(v) \neq c(w)$ and $c(v),c(w)\in\{l(v',w'),r(v',w'),s(v',w')\}$ for some boundary nodes $v',w'$,
        then \\ $\nca_T(v,w)=\nca_{C(c(v),c(w),s(v',w'),v')}(v,w)$.
    \item[(v)] If $c(v) \neq c(w)$,
        $c(v)\in\{l(v',w'),r(v',w'),s(v',w')\}$, and $w' \preceq_{M}
        c(w)$ for some boundary nodes $v',w'$, then  $\nca_T(v,w)=\nca_{C(c(v),s(v',w'),v',w')}(v,w')$.
    \item[(vi)] In all other cases, $\nca_T(v,w)=\nca_{M}(c(v),c(w))$.
    \end{itemize}
\end{prop}
Case (i) says that if $v$ and $w$ are leaf nodes in the same cluster
then the nearest common ancestor of $v$ and $w$ is the nearest
common ancestor of $v$ and $w$ in the micro tree induced by that
leaf cluster (Figure~\ref{fig:propositions}(c)). Case (ii) says
that if $v$ and $w$ are both either left nodes
or right nodes then the nearest common ancestor of $v$ and $w$ is
the nearest common ancestor in the micro tree induced by their
macro node together with the spine and top boundary node of the
cluster (Figure~\ref{fig:propositions}(b)). Case (iii) says that if $v$ and $w$ are both
spine nodes in the same cluster then the nearest common ancestor of $v$ and $w$ is the
nearest common ancestor of $v$ and $w$ in the micro tree induced
by their macro node. Case (iv) says that if $v$ and $w$ are in different macro nodes but are right, left, or spine nodes in the same cluster then the
nearest common ancestor of $v$ and $w$ is the nearest common
ancestor of $v$ and $w$ in the micro tree induced by that cluster
(we can omit the bottom boundary node) (Figure~\ref{fig:propositions}(f)). Case (v) says that if $v$
is a left, right, or spine node, and the bottom boundary
node $w'$ of $v$'s cluster is an ancestor in the macro tree of the
macro node containing $w$, then the nearest common ancestor of $v$
and $w$ is the nearest common ancestor of $v$ and $w'$ in the
micro tree induced by the macro node of $v$, the spine node, and the top and bottom boundary nodes of $v$'s cluster (Figure~\ref{fig:propositions}(g)). 
In all other cases the
nearest common ancestor of $v$ and $w$ is the nearest common
ancestor of their macro nodes in the macro tree (Case (vi)).

\subsection{Preprocessing}\label{sec:preprocessing}
In this section we describe how to preprocess $T$. First build a
cluster partition $CS$ of the tree $T$ with clusters of size $s$,
to be fixed later, and the corresponding macro tree $M$ in
$O(n_T)$ time. The macro tree is preprocessed as in Section~\ref{sec:simplepreprocessing}. However, since nodes in $M$ contain a set of labels, we now store a dictionary for $\lab(v)$ for each node $v \in V(M)$. Using the deterministic dictionary of Hagerup et al.~\cite{HMP2001} all these dictionaries can be constructed in $O(n_{T}\log n_{T})$ time and $O(n_{T})$ space.
Furthermore, we extend the definition of $\fl$ such that $\fl_{M}(v, \alpha)$ is the nearest ancestor $w$ of $v$ such that $\alpha \in \lab(w)$.

Next we show how to preprocess the micro forests. For any cluster $C\in CS$, deep sets $X, Y, Z\subseteq V(C)$, $i \in \mathbb{N}$, and $\alpha \in \Sigma$ define the following procedures.

\begin{relate}[clusterprocedures]

\item[$\size(X)$:] Return the number of nodes in $X$.

\item[$\leftn(i,X)$:] Return the leftmost $i$ nodes in $X$.

\item[$\rn(i,X)$:] Return the rightmost $i$ nodes in $X$.

\item[$\leftof(X,Y)$:] Return all nodes of $X$ to the left of the leftmost node in $Y$.

\item[$\rightof(X,Y)$:] Return all nodes of $X$ to the right of the rightmost node in $Y$.

\item[$\match(X,Y,Z)$,] where $X=\{m_1 \lhd \cdots \lhd m_k\}$,
$Y=\{v_1 \lhd \cdots \lhd v_k\}$, and $Z \subseteq Y$.
Return $R:=\{m_{j} \mid v_j \in Z \}$.

\item[$\mopc(X,Y)$] Return
the pair $(R_1,R_2)$, where
$R_1=\restrict{\mop(X,Y)}{1}$ and $R_2=\restrict{\mop(X,Y)}{2}$.

\end{relate}
 If we want to specify that a procedure applies to a certain cluster $C$ we add the subscript $C$. 
In addition to these procedures we also define the set procedures on clusters, that is, $\parentc$, $\ncac$, $\deepc$, 
and $\flc$, as in Section~\ref{sec:recursion}. Collectively, we will call these the \emph{cluster procedures}. We represent the input and output sets in the 
procedures as bit strings indexed by preorder numbers. Specifically, a
subset $X$ in a cluster $C$ is given by a bit string $b_{1} \ldots
b_{s}$, such that $b_{i} = 1$ iff the $i$th node in a preorder
traversal of $C$ is in $X$. If $C$ contains fewer than $s$ nodes we
set the remaining bits to $0$.

The procedure $\size(X)$ is the number of ones in the bit string.
The procedure $\leftn(i,X)$  corresponds to setting all bits in $X$ larger than the $i$th set bit to zero. Similarly, $\rn(i,X)$ corresponds to setting all bits smaller than the $i$th largest set bit to zero. Similarly, the procedures $\leftof(X,Y)$,  
$\rightof(X,Y)$, 
$\mopc(X,Y)$,  and  $\match(X,Y,Z)$ only depend on the preorder of the nodes and thus only on the bit string and not any other information about the cluster. 


Next we show how to implement the cluster procedures efficiently. We precompute the value of all procedures, except $\flc$, for all possible inputs and clusters. By definition, these procedures do not depend on any specific labeling of the nodes in the cluster. Hence, it suffices to precompute the value for all rooted, ordered trees with at most $s$ nodes. The total number of these is less than $2^{2s}$ (consider e.g. an encoding using balanced parenthesis). Furthermore, the number of possible input sets is at most $2^{s}$. Since at most $3$ sets are given as input to a cluster procedure, it follows that we can tabulate all solutions using less than $2^{3s}\cdot 2^{2s} = 2^{5s}$ bits of memory. Hence, choosing $s \leq 1/10\log n$ we use $O(2^{\frac{1}{2}\log n}) = O(\sqrt{n})$ bits. Using standard bit wise operations each solution is easily implemented in $O(s)$ time giving a total time of $O(\sqrt{n}\log n)$. 

Since the procedure $\flc$ depends on the alphabet, which may be of size $n_{T}$, we cannot efficiently apply the same trick as above. Instead define for any cluster $C \in CS$, $X \subseteq V(C)$, and $\alpha \in \Sigma$:
\begin{relate}[simple]
\item[$\ancestorc(X)$:] Return the set $\{x \mid \text{$x$ is an ancestor of a node in $X$}\}$.
\item[$\Eq_C(\alpha)$:] Return the set $\{x \mid x \in V(C), \lab(x) = \alpha\}$.
\end{relate}
Clearly, $\ancestorc$ can be implemented as above. For $\Eq_{C}$ note that the
 total number of distinct labels in $C$ is at most $s$. Hence, $\Eq_{C}$ can be stored in a dictionary with at most $s$ entries each of which is a bit string of length $s$. Thus, (using again the result of \cite{HMP2001}) the total time to build all such dictionaries is $O(n_{T}\log n_{T})$.

By the definition of $\flc$ we have that,
\begin{equation*}
\flc_C(X,\alpha) = \deepc_C(\ancestorc_C(X) \cap \Eq_C(\alpha)).
\end{equation*}
Since intersection can be implemented using a binary \emph{and}-operation, $\flc_{C}(X, \alpha)$ can be computed in constant time. Later, we will also need to compute union of sets represented as bit strings and we note that this can be done using a binary \emph{or}-operation.

To implement the set procedures in the following section we often need to ``restrict'' the cluster procedures to work on a subtree of a cluster. Specifically, for any set of macro nodes $\{i_{1}, \ldots, i_{k}\}$ in the \emph{same} cluster $C$ (hence, $k \leq 5$), we will replace the subscript $C$ with $C(i_{1}, \ldots, i_{k})$. For instance, $\parentc_{C(s(v,w), l(v,w))}(X) = \{\parent(x) \mid x\in X \cap V(C(s(v,w), l(v,w))\} \cap V(C(s(v,w), l(v,w))$. To implement all restricted versions of the cluster procedures, we compute for each cluster $C \in CS$ a bit string representing the set of nodes in each micro forest. Clearly, this can be done in $O(n_{T})$ time. Since there are at most $5$ micro forests in each cluster it follows that we can compute any restricted version using an additional constant number of and-operations.

Note that the total preprocessing time and space is dominated by the construction of deterministic dictionaries which use $O(n_{T}\log n_{T})$ time and $O(n_T)$ space.


\subsection{Implementation of the Set Procedures}
Using the preprocessing from the previous section we show how to
implement the set procedures in sublinear time. First we define a
compact representation of node sets. Let $T$ be a tree with macro tree
$M$. For simplicity, we identify nodes in $M$ with a number almost
equal to their preorder number, which we denote their \emph{macro tree
  number}: All nodes nodes except spine and left nodes are identified
with their preorder number. Spine nodes are identified with their
preorder number + 1 if they have a left node as a child and with their
preorder number otherwise, and left nodes are identified with their
preorder number - 1. Hence, we swap the order of left and spine nodes
in the macro tree numbering. We will explain the reason for using macro tree numbers below.  Note that the macro tree numbers are the same as the preorder numbers would be if we had let $l(v,w)$ and $r(v,w)$ be  children of $v$ instead of children of $s(v,w)$ in the definition of the macro tree. 

 Let $S \subseteq V(T)$ be any subset of nodes of $T$.
A \emph{micro-macro node array} (abbreviated node array) $X$ representing $S$ is an array of size $n_{M}$. The $i$th entry, denoted $X[i]$, represents the subset of nodes in $C(i)$, that is, $X[i] =  V(C(i)) \cap S$. The set $X[i]$ is encoded using the same bit representation as in Section~\ref{sec:preprocessing}. By our choice of parameter in the clustering the space used for this representation is $O(n_{T}/\log n_{T})$.

We can now explain the reason for using macro tree numbers to identify
the nodes instead of preorder numbers.  Consider a node array
representing a deep set. If a left node and the corresponding spine
node are both non-empty, then all nodes in the left node are to the left of the node in the spine node. Formally,
\begin{prop}\label{prop:macrotreeorder}
Consider a node array $X$ representing a deep set $\mc{X}$. For any pair of nodes $v, w \in \mc{X}$, such that $v\in X[i]$ and $w\in X[j]$, $i\neq j$, we have 
$$ v \lhd w \Leftrightarrow i<j \;.$$ 
\end{prop}
\begin{proof}
By Proposition~\ref{lem:orderlemma}(v) the claim is true for $i \lhd j$.  The remaining cases are $i=l(v',w')$ and $j=s(v',w')$ (Proposition~\ref{lem:orderlemma} (iii)) and $i=s(v',w')$ and $j=r(v',w')$ (Proposition~\ref{lem:orderlemma}(iv)). In both cases $i<j$ and it follows immediately that $v \lhd w \Rightarrow i <j$. For the other direction, it follows from the structure of the macro tree that in both cases either $v \lhd w$ or $w \prec v$. But $\mc{X}$ is deep and thus $v \lhd w$. 
\end{proof}
Thus, by using macro tree numbers we encounter the nodes in $X$ according to their preorder number in the original tree $T$. This simplifies the implementation of all the procedures except \Deep, since they all get deep sets as input.
%

We now present the detailed implementation of the set procedures on node arrays. As in Section~\ref{simple} we assume that the input to all of the procedures, except \Deep, represent a deep set.  Let $X$ be a node array.

\paragraph{Implementation of {\sc Parent}}
Procedure \Parent\ takes a node array $X$ representing a deep set as input.

\begin{procedure}[H]
\SetProcNameSty{textnormal}
Initialize an empty node array $R$ of size $n_{M}$ ($R[i]:=\emptyset$ for $i=1,\ldots n_M$) and set $i:=1$.

\While{$i \leq n_M$}{
	\lWhile{$X[i]= \emptyset$}{$i:=i+1$.}

	There are three cases depending on the type of $i$:

	\uCase{{\bf 1.} $i\in \{l(v,w), r(v,w)\}$}{Compute 
		$N := \parentc_{C(i, s(v,w), v)}(X[i])\;.$

		\ForEach{$j \in \{i, s(v,w), v\}$}{
                  $R[j] := R[j] \cup (N \cap V(C(j)))$.}
	}

 	\uCase{{\bf 2.} $i = l(v)$} {
		Compute 
                $N := \parentc_{C(i, v)}(X[i])\;.$
  
                        \ForEach{$j \in \{i, v\}$} {
                          $R[j] :=R[j] \cup (N \cap V(C(j)))$.}
  	}

	\uCase{{\bf 3.} $i \not\in \{l(v,w), r(v,w), l(v)\}$} {
		Compute 
                $N := \parentc_{C(i)}(X[i])\;.$
  
 	 	\uIf{$N \neq \emptyset$} {set $R[i]:=R[i] \cup N$.} 
		\ElseIf{$j := \parent_{M}(i) \neq \bot$}
		{set
		    $R[j] := R[j] \cup \{\first(j)\}$.}
		}
	Set $i:=i+1$.
}
Return $R$.
\caption{\Parent($X$)}
\end{procedure}

Procedure  $\Parent$ has three cases. Case 1 handles
the fact that  left or right nodes may have a node on a spine or the top
boundary node as parent. Since no left or right nodes can have their
parent outside their cluster there is no need to compute parents
in the macro tree. Case 2 handles the fact that a leaf node may have the boundary node as parent. Since no leaf node can have its parent outside its cluster
there is no need to compute parents in the macro tree. Case 3
handles boundary and spine nodes. In this case there is either a parent within the micro forest or we can use
the macro tree to compute the parent of the root of the micro
tree.
Since the input to \Parent\ is deep we only need to do one of the two
things. If the computation of parent in the micro tree returns a
nonempty set, this set is added to the output (line 18). Otherwise
(the returned set is empty), we compute parent of $i$ in the macro
tree (line 19).
If the computation of parent in the macro tree returns a node $j$,
this will either be a spine node or a boundary node. To take care of
the case where $j$ is a spine node, we add the lowest node
($\first(j)$) in $j$ to the output (line 20). If $j$ is a boundary node this is just $j$ itself. 

\paragraph{Implementation of {\sc Nca}}
We now give the implementation of procedure \Nca.
The input to procedure \Nca\ is two node arrays $X$ and $Y$ representing two subsets $\mc{X}, \mc{Y} \subseteq V(T)$,  $\norm{\mc{X}}=\norm{\mc{Y}}=k$. 
 The output is a node array $R$ representing the set $\Deep(\{\nca(\mc{X}_i,\mc{Y}_i) \mid 1\leq i\leq k\})$, where $\mc{X}_i$ and $\mc{Y}_i$ is the $i$th element of  $\mc{X}$ and $\mc{Y}$, w.r.t.\ their preorder number in the tree, respectively. We also assume that we have $\mc{X}_i \lhd \mc{Y}_i$ for all $i$ (since $\Nca$ is always called on a set of minimum ordered pairs). Note, that 
$\mathcal{X}_l $ and $\mathcal{Y}_l$ can belong to different clusters/nodes in the macro tree, i.e., we might have $\mathcal{X}_l \in X[i]$ and $\mathcal{Y}_l \in Y[j]$ where $i\neq j$. 

\begin{procedure}[H]
  \SetProcNameSty{textnormal}
  Initialize an empty node array $R$ of size $n_{M}$, set $i:=1$ and
  $j:=1$.
  
  \While{$i\leq n_M$ and $j \leq n_M$}{
    \lWhile{$X[i]= \emptyset$}{$i:=i+1$.}
        
    \lWhile{$Y[j] = \emptyset$}{$j:=j+1$.}

    Set $n:=\min(\sizes{X[i]},\sizes{Y[j]})$,  $X_i:=\leftn(n,X[i])$, and $Y_j:=\leftn(n,Y[j])$.
    
    Compare $i$ and $j$.  There are two cases:

    \uCase{{\bf 1.} $i=j$.}{ Set 
      \begin{equation*}
        S:=
        \begin{cases}
          C(i,v), &  \textrm{if } i=l(v),\\
          C(i, s(v,w), v), & \textrm{if } i\in\{l(v,w),r(v,w)\}. 
        \end{cases}
      \end{equation*}

       Compute $N:=\ncac_S(X_i,Y_j)$.

      \ForEach{macro node $h=c(s)$ where $s \in V(S)$} {set 
      $R[h]:=R[h] \cup (N\cap V(C(h)))$.}
    }

    \uCase{{\bf 2.}  $i \neq j$. }{Compute $h:=\ncac_M(i,j)$. There
      are two subcases:

    \uCase{{\bf (a)} $h$ is a boundary node} {Set $R[h]:=1$.}
   \uCase{{\bf (b)} h is a spine node $s(v,w)$}{
     There are three
     subcases:
      
      \uCase{{\bf i.} $i \in \{l(v,w),s(v,w)\}$ and $j\in \{s(v,w),r(v,w)\}$}{ 
        Compute $N:=\ncac_{C(i,j,s(v,w),v)}(X_i,Y_j)$.}
      
      \uCase{{\bf ii.} $i=l(v,w)$ and $w \preceq j$}{
        Compute $N:=\ncac_{C(i,s(v,w),v,w)}(\rn(1,X_i),w)$.}
      \uCase{{\bf iii.} $j=r(v,w)$ and $w \preceq i$}{
        Compute $N:=\ncac_{C(j,s(v,w),w,v)}(w,\leftn(1,Y_j))$.}
      
      Set $R[h]:=R[h]\cup (N\cap V(C(h)))$ and 
              $R[v]:=R[v]\cup (N\cap V(C(v)))$.
    }
  }
   Set $X[i]:=X[i]\setminus X_i$ and $Y[j]:=Y[j]\setminus Y_j$.
}
Return $\Deep(R)$.
\caption{\Nca($X$,$Y$)}
\end{procedure}

In the main loop of procedure \Nca\ (line 2--27) we first find the
next non-empty entries in the node arrays $X[i]$ and $Y[j]$ (line 3
and 4). 
We then compare the sizes of $X[i]$ and $Y[j]$ and construct two sets
of equal sizes $X_i$ and $Y_j$ consisting of the
$\min(\sizes{X[i]},\sizes{Y[j]})$ leftmost nodes from $X[i]$ and
$Y[j]$ (line 5).  
In Section~\ref{sec:correctsetmacro} we prove the following invariant on $X_i$ and $Y_j$
\begin{equation*}
  \leftn(1,X_i)=\mathcal{X}_l \textrm{ and } \leftn(1,Y_j)=\mathcal{Y}_l \textrm { for some } l \;. 
\end{equation*}
The procedure has two main cases. 
\begin{itemize}
\item If $i=j$ (Case 1) then $i$ is either a leaf, left, or right node
  due to the invariant and the assumption on the input that $\mc{X}_l
  \lhd \mc{Y}_l$ (for a formal proof see
  Section~\ref{sec:correctsetmacro}). If $i$ is a leaf node the
  nearest common ancestors of all pairs in $X_i$ and $Y_j$ are in the
  leaf node or the boundary node. If $i$ is a left or right node the
  nearest common ancestors of all the pairs are in $i$, on the spine,
  or in the top boundary node. In line 9 we compute $\ncac$ in the appropriate cluster depending on the type of $i$.
  
\item If $i\neq j$ (Case 2) we first compute the nearest common
  ancestor $h$ of $i$ and $j$ in the macro tree (line 14). Due to the
  structure of the macro tree $h$ is either a spine node or a boundary
  node (left, right, and leaf nodes have no descendants). If $h$ is a
  boundary node all pairs in $X_i$ and $Y_j$ have the same nearest
  common ancestor, namely $h$ (Case 2(a)). If $h$ is a spine node
  there are three cases depending on the types of $i$ and $j$. 
  \begin{itemize}
  \item In Case 2(b)i we have  $i=l(v,w)$ and $j\in \{s(v,w), r(v,w)\}$ (see
    Figure~\ref{fig:propositions}(d) and (f)), or $i=s(v,w)$ and
    $j=r(v,w)$ (see Figure~\ref{fig:propositions}(e)). In this case we
    compute $\ncac$ in the cluster containing $i, j, s(v,w), v$. 
  \item  In case 2(b)ii $i$ is a left node $l(v,w)$ and $j$ is a (not
    necessarily proper) descendant of $w$ (see
    Figure~\ref{fig:propositions}(g)). In this case we compute $\ncac$
    on the rightmost node in $X_i$ and $w$ in the cluster containing
    $i, v, w, s(v,w)$. We can restrict the computation to $\rn(1,X_i)$
    because we always run \Deep\ on the output from \Nca\ before using
    it in any other computation and all nearest common ancestors of
    the pairs in $X_i$ and $Y_j$ will be on the spine, and the deepest
    one will be the nearest common ancestor of the rightmost nodes in
    $X_i$ and $Y_j$ (see Section~\ref{sec:correctsetmacro} for a
    formal proof). 
    \item Case 2(b)iii is similar to Case 2(b)ii.
    \end{itemize}
\end{itemize}
In the end of the iteration we have computed the nearest common
ancestors of all the pairs in $X_i$ and $Y_j$ and the nodes from these
pairs are removed from $X[i]$ and $Y[j]$. 

\paragraph{Implementation of {\sc Deep}}  The implementation of \Deep\ resembles the previous implementation, but takes advantage of the fact that the input list is in macro tree order. 

\begin{procedure}[h]
  \SetProcNameSty{textnormal}
  Initialize an empty node array $R$ of size $n_{M}$. \\
  Find the smallest $j$ such that $X[j]\neq \emptyset$. If no such $j$
  exists stop. Set $i:=j+1$.
  
  \While{$i \leq n_M$}{
    \lWhile{$X[i]=\emptyset$}{$i:=i+1$.}

    Compare $j$ and $i$. There are three cases:
    
    \uCase{{\bf 1.} $j \lhd i$.}{ Set \begin{equation*}
        S:=
        \begin{cases}
          C(j,v), &  \textrm{if } j=l(v),\\
          C(j, s(v,w), v), & \textrm{if } j\in\{l(v,w),r(v,w)\}, \\
          C(j), & \textrm{otherwise}.
        \end{cases}
      \end{equation*}

      Set $R[j] := \deepc_S (X[j])$.}
    \uCase{{\bf 2.} $j \prec i$. }{

      \If{$j=s(v,w)$ and $i=r(v,w)$}{compute
        $N := \deepc_{C(r(v,w), s(v,w), v)}(X[i] \cup X[j]).$
 
        Set $R[j]:=X[j]\cap N$.}

    }
    \uCase{{\bf 3.} $i \prec j$ (can happen if $i=s(v,w)$ and $j=l(v,w)$).}{Compute
      $N := \deepc_{C(l(v,w), s(v,w), v)}(X[i] \cup X[j]).$ 
      
      Set $R[j]:=X[j]\cap N$, $X[i]:=X[i] \cap N$.}
    Set $j:=i$ and $i:=i+1$.
}
Set $R[j] := \deepc_S (X[j])$, where $S$ is set as in Case 1.

Return $R$.
\caption{\Deep($X$)}
\end{procedure}

The procedure \Deep\  has three cases. In case 1 node $i$ is to the
right of our "potential output node" $j$. Since any node $l$ that is a
descendant of $j$ must be to the left of $i$ ($l<i$) it cannot
appear later in the list $X$ than $i$. We can thus safely add
$\deepc_S(X[j])$ to $R$ at this point. To ensure that the cluster we
compute \Deep\ on is a tree we include the top boundary node if $j$ is
a leaf node and the top and spine node if $j$ is a left or right
node. We add the result to $R$ and set $i$ to be our new potential
output node. 

In case 2 node $j$ is an ancestor of
$i$ and therefore no node from $C(j)$ can be in the output list unless
$j$ is a spine node and $i$ is the corresponding right node. If this
is the case we compute \Deep\ of $X[j]$ and $X[i]$ in the cluster
containing $i$ and $j$ and add the result for $j$ to the output and set $i$ to be our new potential output node.

In case 3 node $i$ is an ancestor of $j$. This can only happen if $j$
is a left node and $i$ the corresponding spine node. We compute \Deep\
of $X[j]$ and $X[i]$ in the cluster containing $i$ and $j$ and 
add the result for $j$ to the output. We restrict $X[i]$ to the
nodes both in $X[i]$ and the result $N$ of the \Deep\ computation, and
let $i$ be our potential output node. The results for $X[i]$ cannot be
added directly to the input since
there might be nodes later in the input that are descendants of $i$.
Since a left node has no children we can safely add the result for
$j$ to the output.

After iterating through the whole node array $X$ we add the last potential node $j$ to the output after computing \Deep\ of it as in Case 1.

\paragraph{Implementation of {\sc \Mop}}
We now give the implementation of procedure \Mop.
Procedure $\Mop$ takes a pair of node arrays $(X,Y)$ and another node array $Z$ as input.  The pair $(X,Y)$ represents a set of minimum ordered pairs, where the first coordinates are in $X$ and the second coordinates are in $Y$.
To simplify the implementation of procedure $\Mop$ it calls two
auxiliary procedures $\Mopsim$ and $\Match$ defined below. Procedure
$\Mopsim$ computes $\mop$ of $Y$ and $Z$, and procedure $\Match$
computes the first coordinates from $X$ corresponding to the first
coordinates from the minimum ordered pairs of $Y$ and $Z$ computed by
$\Mopsim$. 

\begin{procedure}[h]
  \SetProcNameSty{textnormal}
  Compute $M:=\Mopsim(Y,Z)$. 
  
  Compute $R:=\Match(X,Y,\restrict{M}{1})$. 

  Return $(R,\restrict{M}{2})$.
  \caption{\Mop(($X$,$Y$),$Z$)}
\end{procedure}

\ignore{ 
              \begin{relate}
              \item[$\Mop((X,Y),Z)$] Compute $M:=\Mopsim(Y,Z)$. 
                
                Compute $R:=\Match(X,Y,\restrict{M}{1})$. 
                
                Return $(R,\restrict{M}{2})$.
              \end{relate}
}
Procedure $\Mopsim$ takes two node arrays as input and computes $\mop$ of these.

\begin{procedure}[H]
  \SetProcNameSty{textnormal}
  Initialize two empty node arrays $R$ and $S$ of size $n_M$. \\
  Set $i:=1$,
  $j:=1$,  $(r_1,r_2):=(0,\emptyset)$, $(s_1,s_2):=(0,\emptyset)$.
  
  \Repeat{$i>n_M$ or $j>n_M$}{
    \lWhile {$X[i] = \emptyset$}{
      set $i:=i+1$.
    } 
   
    There are four cases:
    
    \uCase{{\bf I.} $i=l(v,w)$ for some $v, w$.} {{\bf Until} $Y[j]\neq
      \emptyset$ and either $i \lhd j$, $i=j$, or $j=s(v,w)$ {\bf do} set $j:=j+1$.}
    \uCase{{\bf II.} $i=s(v,w)$ for some $v, w$.} {{\bf Until} $Y[j]\neq
      \emptyset$ and either $i \lhd j$ or $j=r(v,w)$ {\bf do} set $j:=j+1$.}
    \uCase{{\bf III.} $i \in \{r(v,w), l(v)\}$ for some $v, w$.} {{\bf
        Until} $Y[j]\neq \emptyset$ and either $i \lhd j$ or  $i=j$
      {\bf do}  set $j:=j+1$. }
    \uCase{{\bf IV.} $i$ is a boundary node.} {{\bf Until} $Y[j]\neq
      \emptyset$ and $i \lhd j$ {\bf do}  set $j:=j+1$.} 
    
    Compare $i$ and $j$. There are two cases:
    
    \uCase{{\bf 1.} $i\lhd j$.} {
      \If{$s_1 < j$} {set
        $R[r_1]:=R[r_1] \cup r_2$, $S[s_1]:=S[s_1] \cup s_2$, and
        $(s_1,s_2):= (j, \leftn_{C(j)}(1,Y[j]))$.} 
      
      Set
      $(r_1,r_2):=(i, \rn_{C(i)}(1,X[i]))$ and $i=i+1$.
    }
    
    \Other( // {\bf case 2.} ){Compute $(r,s):=\mopc_{C(i,j,v)}(X[i],Y[j])$,
      where $v$ is the top boundary node in the cluster $i$ and $j$
      belong to. \\
      \If{$r\neq \emptyset$}{ 
        \If{$s_1 < j$ or {\bf if} $s_1=j$
          and $\leftof_{C(i,j)}(X[i],s_2)=\emptyset$}  {set
          $R[r_1]:=R[r_1] \cup r_2$, $S[s_1]:=S[s_1] \cup s_2$.}
        Set $(r_1,r_2):=(i,r)$ and $(s_1,s_2):=(j,s)$.}
      There  are two subcases:\\
      \uCase{{\bf (a)} $i=j$, or $i=l(v,w)$ and $j=s(v,w)$.}{
        Set  $X[i]:=\rn_{C(i)}(1,\rightof_{C(i)}(X[i],r))$ and 
        $j:=j+1$.}
      \uCase{{\bf (b)} $i=s(v,w)$ and $j=r(v,w)$.} {\lIf {$r=\emptyset$} {set 
          $j:=j+1$} \lElse{set $i:=j$.}}
    }
  }
  Set $R[r_1]:=R[r_1] \cup r_2$ and $S[s_1]:=S[s_1] \cup s_2$. 
  
  Return $(R,S)$.
  
  \caption{\Mopsim($X$,$Y$)}
\end{procedure}

Procedure $\Mopsim$ is somewhat similar to the previous implementation of the procedure $\Mop$ from Section~\ref{implementationsimple}. As in the previous implementation we have a "potential pair" $((r_1,r_2), (s_1,s_2))$, where $r_1$ and $s_1$ are macro nodes, $r_2 \subseteq X[r_1]$, $s_2 \subseteq Y[s_1]$,  where $r_2=\{r^1 \lhd \cdots \lhd r^k \}$ and $s_2=\{s^1\lhd \cdots \lhd s^k \}$ such that 
$r^l\lhd s^l$ for $l=1,\ldots k$. 
Furthermore, for any $l$ there exists no node $y \in Y[j]$, for $j<s_1$, such that $r^l \lhd y \lhd s^l$ and no node $x \in X[i]$, for $i<r_1$, such that $r^l \lhd x \lhd s^l$. 

We have the following invariant at the beginning of each iteration:
\begin{equation}
\nexists x \in X[i], \textrm{ such that } x \unlhd x', \textrm{ for any } x' \in  r_2.
\end{equation}

We first find the next non-empty macro node $i$. We then have 4 cases
depending on which kind of node $i$ is. 
\begin{itemize}
\item In Case I $i$ is a left
  node. Due to Proposition~\ref{lem:orderlemma} we can have $\mop$ in
  $i$ (case (i), see Fig.~\ref{fig:propositions}(b)), in the spine
  (case (iii), see Fig.~\ref{fig:propositions}(d)), or in a node to
  the right of $i$ (case(v)). 
\item In Case II $i$ is a spine node. Due to
  Proposition~\ref{lem:orderlemma} we can have $\mop$  in the right
  node (case (iv) , see Fig.~\ref{fig:propositions}(e)) or in a node to the right of $i$ (case(v)). 
\item In Case III $i$ is a right node or a leaf node. Due to
  Proposition~\ref{lem:orderlemma} we can have $\mop$ in $i$ (case (i)
  and (ii), see Fig.~\ref{fig:propositions}(b)-(c)) or in a node to the right of $i$ (case(v)).
\item  In the last case (Case IV) $i$ must be a boundary node and $\mop$ must be in a node to the right of $i$.
\end{itemize}
We then compare $i$ and $j$. The case where $i \lhd j$ is similar to
the previous implementation of the procedure. We compare $j$ with our
potential pair (line 16). If $s_1 < j$ then $s_1 \lhd j$ since the
input is deep, and we can insert $r_2$ and $s_2$ into our output node
arrays $R$ and $S$, respectively. We also set $s_1$ to $j$ and $s_2$
to the leftmost node in $Y[j]$ (if $s_1=j$ we already have
$(s_1,s_2)=(j, \leftn_{C(j)}(1,Y[j]))$). Then---both if $s_1 \lhd j$
or $s_1=j$---we set $r_1$ to $i$ and $r_2$ to the rightmost node in
$X[i]$ (line 19). That we only need the rightmost node in $X[i]$ and the leftmost node in $Y[j]$ follows from the definition of $\mop$ and the structure of the macro tree.

Case 2 ($i \ntriangleleft j$) is more complicated. In this case we
first  compute $\mop$ in the cluster $i$ and $j$ belong to (line
21). If this results in any minimum ordered pairs ($r \neq \emptyset$)
we must update our potential pair (line 22--27). Otherwise we leave the
potential pair as it is and only update $i$ and $j$.  If $r \neq
\emptyset$ we compare $s_1$ and $j$ (line 23). 
As in Case 1 of the procedure we add our potential pair to the output and update the potential pair with $r$ and $s$ if $s_1 < j$, since this implies $s_1 \lhd j$. If $s_1=j$ and no nodes in $X[i]$ are to the left of the leftmost node in $s_2$  we also add the potential pair to the output and update it. We show in the next section that in this case $|s_2|=1$. Therefore we can safely add the potential pair to the output. In all other cases the pair $(r,s)\neq (\emptyset,\emptyset)$ shows a contradiction to our potential pair and we update the potential pair without adding anything to the output. 

Finally, in Case 2, we update $X[i]$, $i$, and $j$ (line 28--32). There are two cases depending on $i$ and $j$. In Case (a) either $i=j$  or $i$ is a left node and $j$ is the corresponding spine node. In both cases we can have nodes in $X[i]$ that are not to the left of any node in $Y[j]$. These nodes could be in a minimum ordered pair with nodes from another macro node. We show in the next section that this can only be true for the rightmost node in $X[i]$. $X[i]$ is updated accordingly. 
After this update all nodes in $Y[j]$ are to the left of all nodes in $X[i]$ in the next iteration and therefore $j$ is incremented.
In Case (b) $i$ is a spine node and $j$ is the corresponding right node. Since the input lists are deep, there is only one node in $X[i]$. If $r= \emptyset$ then no node in $Y[j]$ is to the right of the single node in $X[i]$. Since the input arrays are deep, no node later in the array $X$ can be to the left of any node in $Y[j]$ and we therefore increment $j$. If $r \neq \emptyset$ then $(r_1,r_2)=(i,X[i])$ and we update $i$. Instead of incrementing $i$ by one we set $i:=j$, this is correct since all macro nodes with macro node number between $i$ and $j$ are descendants of $i$, and thus contains no nodes from $X$, since $X$ is deep. 

When reaching the end of one of the arrays we add our potential pair
to the output and return (line 35--36).

As in Section~\ref{implementationsimple} we can implement \MopLeft\ similarly to \Mop. 
\\\\
Recall that proceudre \Mop\ calls \Match\ to find the first coordinates from $X$
corresponding to the first coordinates from the minimum ordered pairs
computed by \Mopsim.
Procedure \Match\ takes three node arrays $X$, $Y$, and $Y'$ representing deep sets $\mc{X}$, $\mc{Y}$, and $\mc{Y}'$, where 
$\norm{\mc{X}}=\norm{\mc{Y}}$, and $\mc{Y}' \subseteq \mc{Y}$. The output is a node array representing the set $\{\mc{X}_j \mid \mc{Y}_j \in \mc{Y'}\}$.

\begin{procedure}[H]
  \SetProcNameSty{textnormal}
  Initialize an empty node array $R$  of size $n_M$.

  Set $X_L:=\emptyset$, $Y_L:=\emptyset$, $Y_L':=\emptyset$, $x:=0$,
  $y:=0$, $i:=1$ and $j:=1$.

  \Repeat{$i>n_M$ or $j>n_M$}{

    \lWhile{$X[i]=\emptyset$} {set $i:=i+1$.}
  
    \lWhile{$Y[j]=\emptyset$} {set $j:=j+1$.}

    Set $x:=\sizes{X[i]}$ and $y:=\sizes{Y[j]}$.
    
    Compare $Y[j]$ and $Y'[j]$. There are two cases:
  
    \uCase{{\bf 1.} $Y[j]=Y'[j]$} {Compare $x$ and $y$. There are three
      subcases:
    
      \uCase{{\bf (a)} $x=y$.} {Set $R[i]:=R[i] \cup X[i]$, $i:=i+1$, and
        $j:=j+1$.}

      \uCase{{\bf (b)} $x < y$.} {Set $R[i]:=R[i] \cup X[i]$, $
        Y[j]:=\rn(y-x,Y[j])$, $Y'[j]:=Y[j]$,  and $i:=i+1$.}

      \uCase{{\bf (c)} $x > y$.} {Set $X_L:=\leftn(y,X[i])$, $R[i]:=R[i] \cup X_L$, $X[i]:=X[i]\setminus X_L$, and $j:=j+1$.}
      
    }
    \uCase{{\bf 2.} $Y[j]\neq Y'[j]$} {Compare $x$ and $y$. There are
      three subcases:

      \uCase{{\bf (a)} $x=y$.} {Set $R[i]:=R[i] \cup
        \match(X[i],Y[j],Y'[j])$, $i:=i+1$, and $j:=j+1$.}
      
      \uCase{{\bf (b)} $x < y$.} {Set $Y_L:=\leftn(x,Y[j])$, $Y'_L:=Y'[j] \cap Y_L$,
        $R[i]:=R[i] \cup \match(X[i],Y_L,Y'_L)$, 

        $Y[j]:=Y[j]\setminus Y_L$, $Y'[j]:=Y'[j] \setminus Y'_L$,  and $i:=i+1$.}
      
      \uCase{{\bf (c)} $x > y$.} {Set $X_L:=\leftn(y,X[i])$,
        $R[i]:=R[i] \cup \match(X_L,Y[j],Y'[j])$, 

        $X[i]:=X[i]\setminus X_L$, and $j:=j+1$.}
    }
  }
  Return $R$.
  \caption{\Match($X$,$Y$,$Y'$)}
\end{procedure}

Procedure $\Match$ proceeds as follows. First we find the first
non-empty entries in the two node arrays $X[i]$ and $Y[j]$ (line
4--5). We then compare $Y[j]$ and $Y'[j]$ (line 7). 

If they are equal we keep all nodes in $X$ with the same rank as the
nodes in $Y[j]$ (case 1). We do this by splitting into three cases. If
there are the same number of nodes $X[i]$ and $Y[j]$ we add all nodes
in $X[i]$ to the output and increment $i$ and $j$ (case 1(a)). If
there are more nodes in $Y[j]$ than in $X[i]$ we add all nodes in
$X[i]$ to the output and update $Y[j]$ and $Y'[j]$ to contain only the
$y-x$ leftmost nodes in $Y[j]$ (case 1(b)). We then increment $i$ and
iterate. If there are more nodes in $X[i]$ than in $Y[j]$ we add the
first $y$ nodes in $X[i]$ to the output, increment $j$,  and update
$X[i]$ to contain only the nodes we did not add to the output (case 1(c)). 

If $Y[j] \neq Y'[j]$ we call the cluster procedure \match\ (case
2). Again we split into three cases depending on the number of nodes
in $X[i]$ and $Y[j]$. If they have the same number of nodes we can
just call \match\ on $X[i]$, $Y[j]$, and $Y'[j]$ and increment $i$ and
$j$ (case 2(a)). If $\sizes{Y[j]} > \sizes{X[i]}$ we call match with
$X[i]$ the leftmost $\sizes{X[i]}$ nodes of $Y[j]$ and with the part
of $Y'[j]$ that are a subset of these leftmost $\sizes{X[i]}$ nodes of
$Y[j]$ (case 2(b)). We then update $Y[j]$ and $Y'[j]$ to contain only
the nodes we did not use in the call to \match\ and increment $i$. If
$\sizes{Y[j]} < \sizes{X[i]}$ we call \match\ with the leftmost
$\sizes{Y[j]}$ nodes of $X[i]$, $Y[j]$, and $Y'[j]$ (case 2(c)). We then update $X[i]$ to contain only the nodes we did not use in the call to \match\ and increment $j$.

\paragraph{Implementation of {\sc Fl}} 
Procedure \Fl\ takes as input a node array $X$ representing a
deep set and a label~$\alpha$.

\begin{procedure}[H]
  \SetProcNameSty{textnormal}q
  Initialize an empty node array $R$ of size $n_M$ and two node lists $L$ and $S$.
  
  \Repeat{$i>n_M$}{
    \lWhile{$X[i]=\emptyset$} {set $i:=i+1$. }
  
    There are three cases depending on the type of $i$:
    
    \uCase{{\bf 1.} $i \in \{l(v,w), r(v,w)\}$} {Compute 
      $N := \flc_{C(i, s(v,w), v)}(X[i], \alpha)$.
      
      \uIf{$N \neq \emptyset$}{
       \lForEach{$j \in \{i, s(v,w), v\}$}{set $R[j] = R[j] \cup (N \cap V(C(j)))$.}} 
      \lElse{set $L := L \circ \parent_{{M}}(v)$.}
    }
    \uCase{{\bf 2.} $i = l(v)$} {Compute $N := \flc_{C(i, v)}(X[i],\alpha)$.
  
      \uIf{$N \neq \emptyset$}{\lForEach{ $j \in \{i, v\}$}{set $R[j] :=R[j] \cup (N \cap V(C(j)))$.}}
      \lElse{set $L := L \circ \parent_{{M}}(v)$.}
    }
    \uCase{{\bf 3.} $i \not\in \{l(v,w), r(v,w),l(v)\}$}{Compute $N := \flc_{C(i)}(X[i], \alpha)$.
  
      \uIf{$N \neq \emptyset$}{set $R[i] := R[i] \cup N$.}
      \lElse{set $L := L \circ \parent_{{M}}(i)$.}
    }
  }
Compute the list $S := \flc_{M}(L, \alpha)$. 

\lForEach{node $i \in S$} {set $R[i] := R[i] \cup \flc_{C(i)}(\first(i), \alpha))$.} 

Return $\Deep(R)$.
\caption{\Fl($X$, $\alpha$)}
\end{procedure}

The $\Fl$ procedure is similar to $\Parent$. The cases 1, 2 and 3
compute $\Fl$ on a micro forest. If the result is within the micro
tree we add it to $R$ and otherwise we store in a node list $L$ the node in the
macro tree which contains the parent of the root of the micro forest. Since we always call \Deep\
on the output from $\Fl(X,\alpha)$ there is no need to
compute \Fl\ in the macro tree if $N$ is nonempty. We then compute $\Fl$ in the macro tree on the
list $L$, store the results in a list $S$, and use this to compute the final result.

Consider the cases of procedure \Fl.  In case 1 $i$ is a left or right node. Due to Proposition~\ref{lem:ancestorlemma} case (i) and (ii) $\fl$ of a node in $i$ can be in $i$ or on the spine or in the top boundary node. If this is not the case it can be found  by a computation of $\Fl$ of the parent of the top boundary node of $i$'s cluster in the macro tree (Proposition~\ref{lem:ancestorlemma} case (iii)).
In case 2 $i$ is a leaf node. Then $\fl$ of a node in $i$ must either
be in $i$, in the top boundary node, or can be found  by a computation
of $\Fl$ of the parent of the top boundary node of $i$'s cluster
in the macro tree. If $i$ is a spine node or a boundary node (case 3), then
$\fl$ of a node in $i$ is either in $i$ or can be found by a computation of $\Fl$ of the parent of $i$ in the macro tree.


\subsection{Correctness of the Set Procedures}\label{sec:correctsetmacro}
The following lemmas show that the set procedures are correctly implemented. 

\begin{lemma}
Procedure \Parent\ is correctly implemented.
\end{lemma}
\begin{proof}
We will prove that in iteration $i$ the procedure correctly computes
the parents of all nodes in the macro node $i$.
There are four cases depending on the type of $i$. 
\begin{itemize}
\item Consider the case $i\in\{l(v,w),r(v,w)\}$, i.e., $i$ is a left or right node. For all nodes $x$ in $C(i)$, $\parent(x)$ is either in $C(i)$, on the spine $s(v,w)$, or is the top boundary node $v$. The parents of all input nodes in $C(i)$ is thus in $N$ computed in Case 1 in the procedure. The last line in Case 1 ("For each $j\in \{i, s(v,w),v\},\ldots$") adds the set of parents to the appropriate macro node in the output array. 

\item If $i$ is a leaf node $l(v)$ then for any node $x \in C(i)$, $\parent(x)$ is either in $C(i)$ or is the boundary node $v$. The parents of all input nodes in $C(i)$ is thus in $N$ computed in Case 2 in the procedure. The last line in Case 2 ("For each $j\in \{i,v\}, \ldots$") adds the set of parents to the appropriate macro node in the output array. 

\item If $i$ is a spine node $s(v,w)$ then the input contains at most one
node in $C(i)$, since the input to the procedure is deep. For any
$x\in C(i)$, $\parent(x)$ is either a node on the spine or the top
boundary node $v$. This is handled by Case 3 in the procedure. Let $x$
be the node in $X[i]$. If $\parent(x)=v$, then $N=\emptyset$, and we
compute $j$, which is the parent $v$ of $i$ in the macro tree, and add
$j$ to the output array (since $j=v$ is a boundary node
$\first(j)=v$). If $\parent(x)$ is another node $y$ on the spine, then $N=\{y\} \neq \emptyset$ and $y$ is added to the output array.

\item If $i$ is a boundary node $v$, then $\parent(v)$ is either another
boundary node $v'$, the bottom node on a spine, or $\bot$ if $v$ is
the root. This is handled by Case 3 in the procedure. In all three
cases $N=\emptyset$ and we compute the parent $j$ of $i$ in the macro
tree. If $i$ is the root, then $j=\bot$ and we do nothing. Otherwise,
we add $\first(j)$ to the output. If $parent(v)$ is a boundary node
then $\first(j)=j$. If $j$ is a spine node then $\first(j)$ is the bottom node on $j$. 
\end{itemize}
In each iteration of the procedure we might add nodes to the output, but we never delete anything written to the output in earlier iterations. 
Procedure \Parent\ thus correctly computes the parents of all nodes in~$X$.    
\end{proof}

Before proving the correctness of procedure \Nca\ we will prove the
following invariant on the variables $X_i$ and $Y_j$ in the procedure.
\begin{lemma}\label{inv:nca}
In procedure \Nca\ we have the following invariant of $X_i$ and $Y_j$:
\begin{equation*}
\leftn(1,X_i)=\mathcal{X}_l \textrm{ and } \leftn(1,Y_j)=\mathcal{Y}_l \textrm { for some } l \;. 
\end{equation*}
\end{lemma}
\begin{proof}
The proof is by induction on the number of iterations of the outer
loop. After the while loop on $X$ in the first iteration (line 3), $i$ is the smallest integer such that $X[i]\neq \emptyset$. Due to the macro tree order of the array $X$, $X[i]$ contains the first nodes from $X$ 
w.r.t.\ the preorder of the original tree (Proposition~\ref{prop:macrotreeorder}). 
Similarly, $Y[j]$ contains the leftmost node in $\mc{Y}$. The invariant now follows immediately from the assignment of $X_i$ and $Y_j$.

For the induction step consider iteration $m$ and let $i'$ and $j'$ be
the values of $i$ and $j$ after the while loops in the previous
iteration, i.e, after line 4. By the induction hypothesis
$\leftn(1,X_{i'})=\mathcal{X}_l \textrm{ and }
\leftn(1,Y_{j'})=\mathcal{Y}_l \textrm { for some } l$. Let
$n'=\min(\sizes{X[i']},\sizes{Y[j']})$. Then $X_{i'}$ contains
$\mc{X}_l, \ldots, \mc{X}_{l+n'}$ and $Y_{j'}$ contains $\mc{Y}_l,
\ldots, \mc{Y}_{l+n'}$. We will show that
$\leftn(1,X_i)=\mc{X}_{l+n'+1}$. In the end of the previous iteration
we removed $X_{i'}$ from $X[i']$ (line 26). There are two cases
depending on wether $X[i']$ is empty or not at the beginning of
iteration $m$. 
\begin{itemize}
\item If $X[i']\neq
  \emptyset$ then it clearly contains $\mc{X}_{l+n'+1}$ as its leftmost
  node. Since a spine node can only contain one node from $\mc{X}$, $i'$
  cannot be a spine node. Thus $i=i'$, when we get to line 5 in the
  current iteration 
  It follows that $\leftn(1,X_i)=\mc{X}_{l+n'+1}$.
\item $X[i']=\emptyset$. 
  It follows
  from the macro tree order of $X$ that $X[i]$ contains
  $\mc{X}_{l+n'+1}$ as its leftmost node.
\end{itemize}
It follows by a similar argument that
$\leftn(1,Y_j)=\mc{Y}_{l+n'+1}$. 
\end{proof}

\begin{lemma} 
Let $X$ and $Y$ be two node arrays representing the deep sets $\mc{X}$ and $\mc{Y}$, $\norm{\mc{X}}=\norm{\mc{Y}}=k$, and let $\mc{X}_i$ and $\mc{Y}_i$ denote the $i$th element of $\mc{X}$ and $\mc{Y}$, w.r.t.\ their preorder number in the tree, respectively. For all $i=1,\ldots , k$, assume $\mc{X}_i \lhd \mc{Y}_i$.
Procedure $\Nca(X,Y)$ correctly computes  $\Deep(\{\nca(\mc{X}_i,\mc{Y}_i) | 1\leq i\leq k\})$.
\end{lemma}
\begin{proof} 
We are now ready to show that the procedure correctly takes care of all possible cases from Proposition~\ref{lem:ncalemma}. The proof is split into two parts. First we will argue that some of the cases from the proposition cannot occur during an iteration of the outer loop of \Nca. Afterwards we prove that the procedure takes care of all the cases that can occur. 

Case (iii) cannot happen since if $i=j$ is a spine node then $\mc{X}_l$ is either a descendant or an ancestor of $\mc{Y}_l$ contradicting the assumption on the input that $\mathcal{X}_l \lhd \mathcal{Y}_l$. Case (vi) can only happen if $i\neq j$: 
If $i=j$ and we are in case (vi) then $i=j$ is a boundary node, and this would imply that $C(i)$ only consists of one node, i.e., $\mathcal{X}_l= X[i]=Y[j]=\mathcal{Y}_l$ contradicting the assumption on the input that $\mathcal{X}_l \lhd \mathcal{Y}_l$. 
Due to this assumption on the input we also have that in case (iv) of the proposition $i$ is either a left node or a spine node and 
$j$ is a spine node or a right node. For case (v) either $i$ is a left node and $j$ is a descendant of the bottom boundary node of $i$'s cluster or $j$ is a right node and $i$ is a descendant of the bottom boundary node of $j$'s cluster. All the other cases from case (v) would contradict the assumption that $\mathcal{X}_l \lhd \mathcal{Y}_l$.

The procedure first constructs two sets $X_i$ and $Y_j$ containing the
elements $\mc{X}_l,\ldots,\mc{X}_{l+n}$ and $\mc{Y}_l,\ldots,\mc{Y}_{l+n}$ for some $l$, respectively, where $n=\min(\sizes{X[i]},\sizes{Y[j]})$.
The procedure \Nca\ has two main cases depending on whether $i=j$ or not. Case 1 ($i=j$) takes care of cases (i)--(ii) from Proposition~\ref{lem:ncalemma}. Case 2 ($i\neq j$) takes care of the remaining cases from Proposition~\ref{lem:ncalemma} (iv)--(vi) that can occur.

First consider Case 1: We compute nearest common ancestors $N$ of the  $n$ nodes in $X_i$ and $Y_j$ in a cluster $S$ depending on what kind of node $i$ is. We need to show that Case 1 handles Case (i) and (ii) from the Proposition correctly. 
\begin{itemize}
\item Case (i). $i=j$ is a leaf node. By the Proposition the nearest common ancestors of the pairs in $(\mc{X}_l,\mc{Y}_l), \ldots,(\mc{X}_{l+n},\mc{Y}_{l+n})$ from $X_i$ and $Y_j$ is either in $c(i)$ or in the boundary node, i.e., in $C(i,v)$.  
\item Case (ii). $i=j$ is a left or right node.  By the Proposition the nearest common ancestors of the pairs in $\{(\mc{X}_l,\mc{Y}_l), \ldots,(\mc{X}_{l+n},\mc{Y}_{l+n})\}$ from $X_i$ and $Y_j$ is either in $c(i)$, on the spine, or in the top boundary node, i.e., in $C(i,s(v,w),v)$. 
\end{itemize} Thus $S$ is correctly set in both cases. After the
computation of $N$ in line 9 the output is then added to the entries in the
output array $R$ for each of the macro nodes belonging to nodes in
$V(S)$ (line 10--12). Case 1 thus handles Case (i)-(ii) (and only these two cases) from Proposition~\ref{lem:ncalemma}.

Next consider Case 2 ($i \neq j$). We first compute the nearest common ancestor $h$ of $i$ and $j$ in the macro tree. The macro node $h$ is either a boundary node or a spine node due to the structure of the macro tree (see also Proposition~\ref{lem:ncalemma}).  We will show that Case 2 takes care of the remaining cases.
\begin{itemize}
\item Case (iv). From the above discussion it follows that we have one of the three following cases.
$i=l(v,w)$ and $j=s(v,w)$, $i=l(v,w)$ and $j=r(v,w)$, or $i=s(v,w)$ and $j=r(v,w)$. All three cases are handled in Case 2(b)i of the procedure. It follows from the proposition that \ncac\ is computed in the correct cluster.
\item Case (v). It follows from the discussion above that either $i=l(v,w)$ and $w\preceq j$, or $j=r(v,w)$ and $w\preceq i$. These two cases are handled by Case~2(b)ii and 2(b)iii of the procedure. It follows from the Proposition that \ncac\ is computed in the correct cluster. We need to argue that we can restrict the computation of \ncac\ to the pair $(\rn(1,X_i),w)$ instead of computing \ncac\ for all nodes in  $\{\mc{X}_l, \ldots,\mc{X}_{l+n}\}$. Consider the case where $i=l(v,w)$ and $w\preceq j$ (Case 2(b)ii of the procedure). Since $w \preceq\mc{Y}_r$ for all $r=l,\ldots l+n$, and $\mc{X}_l \lhd \mc{X}_{l+1} \lhd \ldots \lhd \mc{X}_{l+n}$, then $\nca(\mc{X}_{r},\mc{Y}_{r})\preceq \nca(\mc{X}_{l+n},\mc{Y}_{l+n})$ for all $r=l,\ldots l+n$. Thus we do not need to compute $\nca(\mc{X}_{r},\mc{Y}_{r})$ for $r \neq n+l$, since the output of the procedure is $\Deep(\{\nca(\mc{X}_i,\mc{Y}_i) | 1\leq i\leq k\})$. A similar argument shows that we can restrict the computation to $(w,\leftn(1,Y_j))$ in Case 2(b)iii.
\item Case (vi). It follows from the discussion above and the proposition that $i\neq j$ and $i$ and $j$ are in different clusters, and we are not in any of the cases from (iv) and (v). Thus $h$ must be a boundary node and all the pairs $\{(\mc{X}_l,\mc{Y}_l), \ldots,(\mc{X}_{l+n},\mc{Y}_{l+n})\}$ have the same nearest common ancestor, namely $h$. This is handled by Case 2(a). 
\end{itemize}
We have now argued that the procedure correctly takes care of all possible cases from Proposition~\ref{lem:ncalemma}. It remains to show that all pairs from $\{\nca(\mc{X}_i,\mc{Y}_i) | 1\leq i\leq k\}$ are considered during the computation. It follows from the invariant that we only consider pairs from the input. In the last lines we remove the nodes from the input that we have computed the \nca s of in this iteration. It follows from the proof of the invariant that no entry in the input arrays is left nonempty. Thus all pairs are taken care of. 
\end{proof}

To prove that procedure \Deep\ is correctly implemented we will use the following fact about preorder and postorder numbers in the macro tree.
\begin{prop}\label{prop:preordermacro}
Let $i$ and $j$ be nodes in the macro tree identified by their macro tree number such that $i<j$. For all $x \in C(i), y \in C(j)$ we have
\begin{enumerate}
\item $\pre(x) < \pre(y)$ unless $i=l(v,w)$ and $j=s(v,w)$.
\item $\post(y) > \post(x)$ unless $i=s(v,w)$ and $j=r(v,w)$.
\end{enumerate}
\end{prop}

\begin{prop}\label{prop:preordermacrotree}
Let $x_1,\ldots,x_n$ be nodes from the macro tree associated with their macro tree number such that $x_1 < x_2 < \cdots < x_n$.
If $x_i \lhd x_j$ for some $i$ and $j$ then $x_i \lhd x_k$ for all 
$x_k>x_j$.
\end{prop}
\begin{proof}
From $x_i \lhd x_j$ we have $\pre(x_i)<\pre(x_j)$ and $\post(x_i) < \post(x_j)$. 
Since $x_k>x_j$ we have $\pre(x_j)<\pre(x_k)$ unless $x_k=s(v,w)$ and $x_j=l(v,w)$. In that case, $\pre(x_k)+1=\pre(x_j)>\pre(x_i)$. Since $x_i \lhd x_j$ we have $x_i\neq x_j$ and thus $\pre(x_k)>\pre(x_i)$.

It remains to show that $\post(x_i) < \post(x_k)$. Assume for the sake of contradiction that $\post(x_k)<\post(x_i) <\post(x_j)$. This implies $x_i \prec x_k$ and  $x_j \prec x_k$ contradicting $x_i \lhd x_j$.
\end{proof}
%
%
We will first prove the following invariants on $i$ and $j$ in procedure \Deep.
\begin{lemma}\label{inv:deepi}
In line 5 of procedure \Deep\
we have the following invariant on $i$ and $j$: 
For all $l$ such that $j < l < i$ we have $X[l]=\emptyset$.
%
\end{lemma}
\begin{proof}
Let $i'$ be the value of $i$ in line 5 of the previous iteration of
the outer loop (line 3--18). 
Then $i$ is the smallest index greater than $i'$ such that the
corresponding entry in $X$ is nonempty. This is true since $i$ was set
to $i'+1$ in the end of the previous iteration (line 17),
and in line 4 of this iteration $i$ was incremented until we found a nonempty entry.  Since $j=i'$ (this
was also set in line 17 of the previous iteration), $i$ is the first nonempty entry greater than $j$ and the claim follows.
%
%
\end{proof}

\begin{lemma}\label{inv:deep}
At the beginning of each iteration of the main loop of procedure
\Deep\ (line 3) we have the following invariant on $j$:
%
For all nodes $x\in X[j]$ and $y \in X[l]$, where $1 \leq l < j$, we have $x \not \prec y$.
\end{lemma}
\begin{proof}
Recall that $x \prec y \Leftrightarrow \pre(x)<\pre(y) \textrm{ and }
\post(y) < \post(x)$. By Proposition~\ref{prop:preordermacro} the only
case where we can have $\pre(x) < \pre(y)$ is if $l=l(v,w)$ and
$j=s(v,w)$ for some $v,w$. Assume this is the case. If $X[l]=\emptyset$ the claim follows trivially.  Otherwise, let $i'$ and $j'$ be the values of $i$ and $j$ in the previous iteration, respectively (since $l<j$ and $X[l]\neq \emptyset$ there must be such an iteration). We have $j=l+1$, $i'=j=s(v,w)$ and $j'=l=l(v,w)$. Thus in the previous iteration the procedure entered case 3, where $X[i']$ was set to $X[i']\cap \deepc_{C(l(v,w),s(v,w),v)}(X[i']\cup X[j'])$, and thus $X[j]$ contains no nodes that are ancestors of nodes in $X[j']=X[l]$.  
\end{proof}

\begin{lemma}
Procedure \Deep\ is correctly implemented.
\end{lemma}
\begin{proof}
We will prove that $x\in \Deep(X)$ iff $x\in X$ and there exists no $y\in X$ such that $x\prec y$.

Assume $x \in \Deep(X)$. Consider the iteration when $x$ is assigned to the output. There are three cases depending on which case we are in when $x$ is added to the input. If $j \lhd i$ (Case 1 of the procedure) then $x \in \Deep_S(X[j])$
and it follows from the invariant on $j$ (Lemma~\ref{inv:deep}) that $x$ has no descendants in any nodes $y \in X[l]$, $l<j$. 
For $j<l<i$ the claim follows directly from Lemma~\ref{inv:deepi}. It remains to show that $x$ has no descendants in $X[l]$ for $l\geq i$. By Proposition~\ref{prop:preordermacrotree} we have $j\lhd l$ for all $l>i$ and the claim follows from Proposition~\ref{lem:ancestorlemma}. 
 
If $j \prec i$ (Case 2 of the procedure) then $j$ is a spine node $s(v,w)$ and $i$ is the corresponding right node $r(v,w)$, and we compute $N:= \Deep_{C(r(v,w),s(v,w),v)}(X[i]\cup X[j])$. Since $x \in \Deep(X)$ we have $x \in R[j]=X[j]\cap N$. It follows from the invariant (Lemma~\ref{inv:deep}) and the computation of $N$ that $x$ has no descendants in $X[l]$ for any $l\leq j$. For $l>j$ it follows from the structure of the macro tree that for any $l>i$ we have $j \lhd l$. For $j<l<i$ the claim follows directly from Lemma~\ref{inv:deepi}. The claim follows from Proposition~\ref{lem:ancestorlemma}. For $j<l<i$ the claim follows directly from Lemma~\ref{inv:deepi}.

If $i \prec j$ (Case 3 of the procedure) then $i$ is a spine node $s(v,w)$ and $j$ is the corresponding left node $l(v,w)$, and we compute $N:= \Deep_{C(l(v,w),s(v,w),v)}(X[i]\cup X[j])$. Since $x \in \Deep(X)$ we have $x \in R[j]=X[j]\cap N$. It follows from the computation of $N$ that $x$ has no descendants in $X[i]\cup X[j]$. Since $l(v,w)$ has no descendants in the macro tree it follows from Proposition~\ref{lem:ancestorlemma} that $x$ has no descendants in $X[l]$ for any $l \neq j$. 

If $x$ is assigned to the output in line 19 
then it follows from the invariant on $j$ (Lemma~\ref{inv:deep}) and the computation of $\Deep_S(X[j])$ that $x$ has no descendants in $X$.

For the other direction let $x \in X$ be a node such that $X\cap
V(T(x))=\{x\}$. Let $l$ be the index such that $x\in X[l]$. All
nonempty entries in $X$ are $i$ in line 5 at some iteration. Consider the iteration when $i=l$. Unless $i=l(v,w)$ and $j=s(v,w)$ (Case 3 of the procedure) $X[i]$ is not changed in this iteration. If we are in Case 3, then $N$ is computed and $X[i]$ is set to $X[i]\cap N$. Since $x$ has no descendants in $X$ we have $x \in N$ and thus $x \in X[i]$ after the assignment. At the end of this iteration $j$ is set to $i$.  Consider the next iteration when $j=l$. If $j \lhd i$ or $i>n_M$ then $x \in \Deep_S(X[j])=R[j]$.  If $j \prec i$ we have $j=s(v,w)$ and $i = r(v,w)$ since $x$ has no descendants in $X$. For the same reason we have $x\in N$ and thus $x\in X[j]\cap N=R[j]$.  If $i \prec j$ we have $i=s(v,w)$ and $j=l(v,w)$.  Again $x\in N$ and thus $x\in X[j]\cap N=R[j]$.  
\end{proof}
We now consider procedures \Mopsim\ and \Match.
\begin{lemma}\label{lem:mopInv1}
Let $((r_1,r_2), (s_1,s_2))$ be as defined in procedure \Mopsim. Then $r_1$ and $s_1$ are macro nodes, $r_2 \subseteq X[r_1]$, $s_2 \subseteq Y[s_1]$,  where $r_2=\{r^1 \lhd \cdots \lhd r^k \}$ and $s_2=\{s^1\lhd \cdots \lhd s^k \}$. For any $l=1,\ldots, k$ we have 
\begin{enumerate}
\item $r^l\lhd s^l$,
\item for all $j\leq s_1$ there exists no node $y \in Y[j]$ such that $r^l \lhd y \lhd s^l$, 
\item for all $i \leq r_1$ there exists no node $x \in X[i]$ such that $r^l \lhd x \lhd s^l$. 
\end{enumerate}
\end{lemma}
\begin{proof}
It follows immediately from the code that $r_1$ and $s_1$ are macro nodes and that  $r_2 \subseteq X[r_1]$, $s_2 \subseteq Y[s_1]$,  where $r_2=\{r^1 \lhd \cdots \lhd r^{k_1} \}$ and $s_2=\{s^1\lhd \cdots \lhd s^{k_2} \}$. Due to the macro tree order of the tree and the fact that $X$ represents a deep set, no node in $X[i]$ can be to the right of any node in $X[r_1]$ for $i<r_1$. To prove condition 3 it is thus enough to prove it for $i=r_1$.
We proceed by induction on the number $k$ of iterations of the outer
loop (line 3--34). We consider the time right after the $k$th iteration of the loop, i.e., right before the $(k+1)$th iteration. The base case ($k=0$) is trivially satisfied. 

For the induction step let 
$i^*$ and $j^*$ be the values of $i$ and $j$ at line 14 in iteration
$k$. Let $r_i'$ and $s_i'$ for $i=1,2$ be the values of $r_i$ and $s_i$, respectively, after the $(k-1)$th round. There are 3 cases: 
\begin{enumerate}
\item $r_2'=r_2$ and $s_2'=s_2$: the claim follows directly from the induction hypothesis. 
\item $r_2' \neq r_2$ and $s_2'=s_2$: condition 2 from the lemma follows directly from the induction hypothesis. Since $s_2$ and thus also $s_1$ were not changed, $r_2$ was set in case 1 of the procedure and $j^*=s_1$. Therefore, $i^*\lhd j^*$, $r_1=i^*$, and $\norm{r_2}=1$. Let $r_2=\{r^1\}$ and $s_2=\{s^1\}$. We have $r_1=i^* \lhd j^* = s_1$ and thus $r^1\lhd s^1$ satisfying condition 1 from the lemma.  To prove condition 3 is satisfied we only have to consider the case $i=r_1$. Since $r_2$ was set in case 1 of the procedure, $r^1$ is the rightmost node in $X[r_1]$ and it follows immediately that there  exists no node $x \in X[r_1]$ such that $r^1 \lhd x \lhd s^1$.
\item $r_2' \neq r_2$ and $s_2' \neq s_2$:  
We first prove condition 1 and 3. If the potential pair was set in
case 1 (line 15--19)
of the procedure then $r_1=i^*\lhd j^* = s_1$ and $|r_2|=1$ implying $r^1 \lhd
s^1$ (condition 1). The node $r^1$ is the
rightmost node in $X[r_1]$ (line 19) and it follows that there exists no node
$x\in X[r_1]$ such that $r^1 \lhd x \lhd s^1$ proving condition 3.
If  the potential pair was set in case 2 then both condition 1 and 3 follows from the
correctness of the implementation of \mop\ and the computation
$(r,s)=\mop_{C(i,j,v)}(X[i^*],Y[j^*])=\mop_{C(i,j,v)}(X[i^*],Y[s_1])$
in line 21.

Let $y \in Y[j]$, for $j
  \leq s_1$, be a node such that $y \not \in s_2$. To prove condition
  2 is satisfied we will show that
  $r^l$ is not to the left of $y$. There are two cases
\begin{itemize}
\item $j=s_1$. Since $s_2' \neq
  s_2$ there are two cases depending on which case of the procedure
  the potential pair was set in. 
If the potential pair was set in case 2 of the procedure the
claim follows from the
correctness of the implementation of \mop\ and the computation
$(r,s)=\mop_{C(i,j,v)}(X[i^*],Y[j^*])=\mop_{C(i,j,v)}(X[i^*],Y[s_1])$
in line 21.

If the potential pair was set in case 1, then $r_1 =i^* \lhd
j^*=s_1$. Since $s_2\neq s_2'$, $s_2$ was changed in the $k$th
iteration and is therefore the
leftmost node in $Y[s_1]$ (line 17
). The claim follows. 

\item $j < s_1$. We will use that we just proved the claim
  for $j=s_1$. Assume for the sake of contradiction that there exists
  a $y \in Y[j]$ such that $r^l \lhd y$. Since $Y$ is representing a
  deep set and due to the macro tree order of $Y$ this implies $r^l \lhd
  y \lhd y'$ for all $y' \in Y[s_1]$ contradicting that the claim is
  true for $j=s_1$.
\end{itemize} 
\end{enumerate}
\end{proof}

\begin{lemma}\label{lem:mopInv2}
We have the following invariant at the beginning of each iteration of
the main loop (line 3) of \Mopsim:
\begin{equation*}
\nexists x \in X[i], \textrm{ such that } x \unlhd x', \textrm{ for any } x' \in  r_2.
\end{equation*}
\end{lemma}
\begin{proof}
By induction on the number of iterations of the outer loop. In the base case $r_2=\emptyset$ and the condition is trivially satisfied. Note that $X$ is representing a deep set and thus either $x\lhd x'$ or $x' \lhd x$ for all $x \in X[i]$.  For the induction step let $i'$, $j'$, and $r_2'$ be the values of $i$, $j$, and $r_2$ respectively in the iteration before this. By the induction hypothesis $x'\lhd x$ for all $x \in X[i']$ and $x' \in r_2'$. Due to the macro tree order of $X$ and the fact that $X$ represents a deep set, all nodes in $X[i']$ are to the left of all nodes in $X[i]$. Thus, if $r_2=r_2'$ it follows from the induction hypotheses that $x'\lhd x$ for all $x \in X[i]$ and $x' \in r_2'=r_2$. For $r_2'\neq r_2$ there are two cases: If $i' \lhd j'$ then $r_2=\rn_{C(i')}(1,X[i'])$ and $i>i'$ and thus the condition is satisfied. Otherwise $r_2$ was set in case 2 of the procedure. Since $r_2\neq r_2'$ we have $r_2=r \subseteq X[i']$ and $r\neq \emptyset$. There are two subcases: If $i=j$ or $i=l(v,w)$ and $j=s(v,w)$ (Case 2(a) of the procedure) then $X[i]$ either contains a single node, which is the rightmost of the nodes in $X[i']$ that are to the right of all nodes in $r_2$ or if there are no such nodes $X[i]=\emptyset$. In both cases the condition is satisfied. If $i=s(v,w)$ and $j=r(v,w)$ then $i>i'$ and the condition is satisfied.
\end{proof}

\begin{lemma}\label{lem:mopsim}
Procedure \Mopsim\ is correctly implemented.
\end{lemma}
\begin{proof}
Let $\mathcal{X}$ and $\mathcal{Y}$ be the sets represented by $X$ and $Y$, respectively. Let $R= \restrict{\Mopsim(X,Y)}{1}$ and $S=\restrict{\Mopsim(X,Y)}{2}$. For simplicity we will slightly abuse the notation and write $(x,y)\in \Mopsim(X,Y)$ iff there exists an $i$ such that  $x \in R[i]$   and $y \in S[i]$.
We want to show that $$(x,y)\in \Mopsim(X,Y) \Leftrightarrow (x,y)\in \mop(\mathcal{X},\mathcal{Y})\,.$$
Assume $(x,y)\in \Mopsim(X,Y)$. Consider the round where $x$ and $y$ were added to $R$ and $S$, respectively.  We have $x = r^l \in r_2$ and $y = s^l \in s_2$. We want to show that there is no node $x' \in X[i]$ for any $i$ such that $x \lhd x' \lhd y$ and no node $y' \in Y[j]$ for any $j$ such that $x \lhd y' \lhd y$. By Lemma~\ref{lem:mopInv1} this is true for $i\leq r_1$ and $j \leq s_1$. By the macro tree order of $Y$ we have that $y \lhd y'$ for any $y' \in Y[j]$ when $j>s_1$. Let $i'$ be the value of $i$ in the round where $x$ and $y$ is added to the output. We will show that no node in $X[i']$ is to the left of any node in $s_2$. Due to the macro tree order of $X$ this implies that no node in $X[i]$ is to the left of any node in $s_2$ for any $i \geq i'$. 
If $i'=r_1$ then it follows directly from Lemma~\ref{lem:mopInv1}.
If $i'>r_1$ it follows from the implementation of the procedure that $i'$ is the first non-empty entry in $X$ greater than $r_1$. Thus the claim follows for any $j$. We now return to show that no node in $X[i']$ is to the left of any node in $s_2$. There are two cases depending on whether $j=s_1$ or $j>s_1$. 
If $j > s_1$ then $j$ was changed either in one of the four cases I--IV, or in the previous iteration in case 2. If $j$ was equal to $s_1$ at the beginning of this iteration then $j$ was incremented in one of the four cases I--IV. Thus none of the cases applied to $s_1$. By Proposition~\ref{lem:orderlemma} no node in $X[i']$ can be to the left of a node in $X[s_1]$. Since $s_2 \subseteq X[s_1]$ the claim follows. If $j=s_1$ it follows from case 2 that $\leftn(X[i'],s_2)=\emptyset$ (otherwise the potential pairs would not have been added to the output in this iteration) and the claim follows immediately.

Now assume $(x,y)\in \mop(\mathcal{X},\mathcal{Y})$. We will deal with each of the cases from Proposition~\ref{lem:orderlemma} separately.
\begin{enumerate}
\item Case (i): $c(x)=c(y)=r(v,w)$.
\item Case (i): $c(x)=c(y)=l(v,w)$.
\item Case (ii): $c(x)=c(y)=l(v)$.
\item Case (iii): $c(x)=l(v,w)$ and $c(y)=s(v,w)$.
\item Case (iv): $c(x)=s(v,w)$ and $c(y)=r(v,w)$.
\item Case (v): $c(x)=l(v,w)$ and $c(y)=r(v,w)$.
\item Case (v): $c(x)\lhd c(y)$ and $c(x)$ and $c(y)$ belong to different clusters.
\end{enumerate}
Note that if $c(x)\lhd c(y)$ then $x$ is the rightmost node in $X[c(x)]$ and $y$ is the leftmost node in $Y[c(y)]$.
We first show that in all cases we will have $x= r^l \in r_2$ and $y=s^l \in s_2$ for some $l$ at some iteration. Consider the first iteration where either $x\in X[i]$ or $y\in Y[j]$. Let $i'$ and $j'$ be the values of $i$ and $j$, respectively, in this iteration. There are three cases:
\begin{itemize}
\item[(a)] $x \in X[i']$ and $y \in Y[j']$. For case 1--5 the procedure goes into case 2. From the correctness of \mopc\ we get $x \in r$ and $y \in s$. Thus $r\neq \emptyset$ and we set $(r_1,r_2)=(i',r)$ and $(s_1,s_2)=(j',s)$ and the claim follows. For case 6--7 the procedure goes into case 1. Since this iteration is the first where $y\in Y[j]$ we have $j'>s_1$ and we set $(r_1,r_2)=(i',\rn_{C(i')}(1,X[i']))$ and $(s_1,s_2)=(j',\leftn_{C(j')}(1,Y[j']))$. Since $x$ is the rightmost node in $X[i']$ and $y$ is the leftmost node in $Y[c(j')]$ the claim follows. 

\item[(b)] $x \in X[i']$ and $y \not \in Y[j']$. Since $(x,y)\in \mop(\mathcal{X}, \mathcal{Y})$ this implies $j'<c(y)$ and there exists no node $y' \in Y[j']$ such that $x \lhd y'$. Assume that there existed such a $y'$. Then $x \lhd y' \lhd y$ due to the macro tree order of $Y$ contradicting $(x,y)\in \mop(\mathcal{X}, \mathcal{Y})$. Thus $i' \not\!\!\lhd\;  j'$. From case I--IV of the procedure it follows that either $i'=j'$, $i'=l(v,w)$ and $j'=s(v,w)$, or $i'=s(v,w)$ and $j'=r(v,w)$. From this and $j'<c(y)$ it follows that we are in case 4 or 7 from above.

The procedure enters case 2 in this iteration. If we are in case 4 then $i'=l(v,w)=j'$ and $c(y)=s(v,w)$. If $r=\emptyset$ then $i=i'$, $X[i']$ is unchanged, and $j=j'+1=s(v,w)=c(y)$ at the end of this iteration. If $r\neq \emptyset$ then $x$ must be to the right of all nodes in $x'\in r$. Assume that there is a $x' \in r$ such that $x \lhd x'$. Since $x' \in r$ there exists a node $y' \in s$ such that $x' \lhd y' \lhd y$. That $y' \lhd y$ follows from $y' \in l(v,w)$ and $y\in s(v,w)$ and the assumption that $\mathcal{Y}$ is deep. Thus $x \lhd x' \lhd y' \lhd y$ contradicting that $(x,y)\in \mop(\mathcal{X}, \mathcal{Y})$. Therefore, $i=i'$, $x \in X[i']$ and $j=j'+1=s(v,w)=c(y)$ at the end of this iteration. From case I of the procedure and the analysis of case (a) it follows that $x= r^l \in r_2$ and $y=s^l \in s_2$ for some $l$.

Now assume we are in case 7. By the same argument as before $i=i'$, $x
\in X[i]$, and $j >j'$ at the end of this iteration. Unless
$i'=l(v,w)=j'$ this implies that $i \lhd j$ at line 14 ("Compare $i$
and $j$") in the next iteration.  If $i'=l(v,w)=j'$ then either $i
\lhd j$ after the first loop in the next iteration (line 7), and the claim follows as before, or $i=l(v,w)$ and $j=s(v,w)$. In the last case we get into case (b) again, but it follows from the analysis that in the iteration after the next we will have $i \lhd j=c(y)$. The claim follows from the analysis of case (a).

\item[(c)] $x \not \in X[i']$ and $y \in Y[j']$. It follows by inspection of the cases that unless we are in case 1 we have $i' \lhd j'$.  If we are in case 1 ($j'=c(x)=r(v,w)=c(y)$) we have either $i' \lhd j'$ or $i'=s(v,w)$.  First we consider the cases 2--7. Since $i' \lhd j'$ the procedure enters case 1 in this iteration. Thus $i$ is incremented and $j$ stays the same. This happens until $i=c(x)$. Now consider case 1. If $i' \lhd j'$ the procedure enters case 1 in this iteration. Thus $i$ is incremented and $j$ stays the same. In the next iteration either the same happens or $i'=s(v,w)$. If $i'=s(v,w)$ the procedure enters case 2. Since $i'$ is a spine node and $\mathcal{X}$ is deep, $X[i]$ contains only one node $x'$. By the structure of the macro tree and the assumption that $\mathcal{X}$ is deep $x' \lhd x$. Since $x \lhd y \in Y[j']$ this implies $r \neq \emptyset$. It follows from case 2(b) of the procedure that $i$ is incremented while $j$ stays the same. At line 14 ("Compare $i$ and $j$") in the next iteration we will have $j=i=r(v,w)$ since all entries in $X$ between $j'$ and $r(v,w)$ are empty due to the assumption that $\mathcal{X}$ is deep. The claim follows from the analysis in case (a).
\end{itemize}
It remains to show that once $x=r^l \in r_2$ and $y=s^l \in s_2$ they will stay this way until added to the output.
Consider the iteration where $x$ and $y$ are assigned to $r_2$ and $s_2$. At the end of this iteration either $i$ or $j$ or both are incremented. Assume $j$ is incremented while the potential pairs are still unchanged. Since $j$ is incremented we have $s_1 < j$ until $s_1$ is changed. It follows from case 1 and 2 of the procedure that in this case $(r_1,r_2)$ is only changed if at the same time $(s_1,s_2)$ are changed and right before that  $(r_1,r_2)$ and $(s_1,s_2)$ are added to the output.

Consider first case 1--3. If $i$ is incremented then $j$ is incremented in one the cases I--V in the next  iteration since $i'=j'$. By the above argument $x$ and $y$ are added to the output. For case 4 $j$ is incremented (case 2(a) of the procedure) and the claim follows as before. 
For case 5--7 first note that $r_2$ and $s_2$ contain only one node each, i.e., $x= r_2$ and $y= s_2$. For case 5 $i$ is incremented (case 2(b) of the procedure). Since $\mathcal{X}$ is deep we have $i\geq r(v,w)=j'$ at line 14 ("Compare $i$ and $j$") in the next iteration. If $i > r(v,w)$ then $j>j'$ and the claim follows. If $i=r(v,w)$ the procedure enters case 2. If $r = \emptyset$ then $j$ is incremented and the claim follows. If $r \neq \emptyset$ then $s_1 =j$ and $(x,y) \in \mop(\mathcal{X},\mathcal{Y})$ implies $\leftn_{C(i,j)}(X[i], s_2) =\leftof_{C(i,j)}(X[i], y)= \emptyset$. Thus  $(r_1,r_2)$ and $(s_1,s_2)$ are added to the output.
If we are in case 6 and 7, $i$ is incremented. Consider case 6. Since $(x,y) \in \mop(\mathcal{X},\mathcal{Y})$ all entries in $X$ between $l(v,w)$ and $r(v,w)$ are empty. Thus at line 14 ("Compare $i$ and $j$") in the next iteration $i\geq r(v,w)$. The proof is equivalent to the one for case 5. Consider case 7. If $j$ is a boundary node then all entries in $X$ between $c(x)$ and $j$ are empty. Thus $j$ is incremented in the second loop of the next iteration. For all other cases for $j$ the proof is similar to the proof of case 5.
\end{proof}

\begin{lemma}\label{inv:match}
In procedure \Match\ we have the following invariant of $X[i]$ and $Y[j]$ in line 6:
\begin{equation*}
\leftn(1,X[i])=\mathcal{X}_l \textrm{ and } \leftn(1,Y[j])=\mathcal{Y}_l \textrm { for some } l \;. 
\end{equation*}
\end{lemma}
\begin{proof}
Induction on the number of iterations of the outer loop. Base case: In the first iteration $X[i]$ and $Y[j]$ are the first nonempty entries in $X$ and $Y$ and thus $\leftn(1,X[i])=\mathcal{X}_1 \textrm{ and } \leftn(1,Y[j])=\mathcal{Y}_1$. For the induction step let $i'$ and $j'$ be the values of $i$ and $j$ in the previous iteration. By the induction hypothesis $\leftn(1,X[i'])=\mathcal{X}_{l'} \textrm{ and } \leftn(1,Y[j'])=\mathcal{Y}_{l'}$. If $x=|X[i']|= |Y[j']|$ both $i$ and $j$ were incremented and $\leftn(1,X[i])=\mathcal{X}_{l'+x}$ and $\leftn(1,Y[j])=\mathcal{Y}_{l'+x}$. If $x=|X[i']|< |Y[j']|$ then $i$ was incremented implying $\leftn(1,X[i])=\mathcal{X}_{l'+x}$. In that case $j=j'$ and $Y[j]=\leftn(x,Y[j'])$ implying $\leftn(1,Y[j])=\mathcal{Y}_{l'+x}$. Similarly, if $|X[i']|> |Y[j']|=y$ we have $\leftn(1,X[i])=\mathcal{X}_{l'+y}$. In that case $j=j'$ and $\leftn(1,Y[j])=\mathcal{Y}_{l'+y}$.
\end{proof}

\begin{lemma}\label{lem:matchmacro}
Procedure \Match\ is correctly implemented.
\end{lemma}
\begin{proof}
We need to show that for all $1\leq k\leq |\mathcal{X}|$: $\mathcal{X}_{k} \in \Match(X,Y,Y') \Leftrightarrow \mathcal{X}_{k} \in \{\mathcal{X}_j | \mathcal{Y}_j \in \mathcal{Y}\}$. Consider the iteration where $\mathcal{X}_{k} \in X[i]$ and $\mathcal{Y}_{k} \in Y[j]$. By Lemma~\ref{inv:match}  such an iteration exists. If $Y[j]=Y[j']$ then $\mathcal{Y}_{k} \in \mathcal{Y'}$ implying $\mathcal{X}_{k} \in \{\mathcal{X}_j | \mathcal{Y}_j \in \mathcal{Y}\}$. It follows from the implementation of case 1(a) and 1(b) that if $x \leq y$ all nodes in $X[i]$ are added to the output and thus  $\mathcal{X}_{k} \in \Match(X,Y,Y')$. If $x >y$ then $\mathcal{X}_{k} \in \leftn(y,X[i])$ since $\mathcal{Y}_{k} \in Y[j]$ and thus $\mathcal{X}_{k} \in \Match(X,Y,Y')$.

If $Y[j]\neq Y'[j]$ the procedure calls $\match$ with  some subset of $X[i]$, $Y[j]$, and $Y'[j]$ depending on the size of $x$ and $y$. By Lemma~\ref{inv:match} and the correctness of \match\ it follows that $\mathcal{X}_{k} \in \Match(X,Y,Y') \Leftrightarrow \mathcal{X}_{k} \in \{\mathcal{X}_j | \mathcal{Y}_j \in \mathcal{Y}\}$.
\end{proof}

\begin{lemma}
Procedure \Mop\ is correctly implemented.
\end{lemma}
\begin{proof}
Follows from the correctness of \Mopsim\ (Lemma~\ref{lem:mopsim}) and \Match\ (Lemma~\ref{lem:matchmacro}).
\end{proof}
Finally, we consider correctness of the \Fl\ procedure.

\begin{lemma}
Procedure \Fl\ is correctly implemented.
\end{lemma}
\begin{proof}
%
Let $\mathcal{X}$ denote the set represented by $X$ and let $F=\{\fl(x,\alpha)| x\in \mathcal{X}  \}$. To show $\Fl(X,\alpha) \subseteq F$ we will first show that for any node $x$ added to $R$ during the computation $x \in F$. Consider a node $x \in R[i]$ for some $i$. Either $x$ was added directly to $R$ after a computation of $N$ in one of the three cases of the procedure or it was added after the computation of $S$. In the first case $x \in F$ follows from the correctness of $\Fl_C$. If $x$ was added after the computation of $S$ it follows from the correctness of $\Fl_M$ that $x \in C(i)$ for some $i \in S$. Due to the correctness of $\Fl_C$ we have $x \in F$.

To show $\Deep(F) \subseteq \Fl(X,\alpha)$ we use Proposition~\ref{lem:ancestorlemma}.  Let $x$ be a node in $\Deep(F)$ and let $x'$ be a node in $X$ such that $\fl(x',\alpha)=x$. We have $x' \in X[i]$ for some $i$. If $i$ is a left or right node then according to Proposition~\ref{lem:ancestorlemma} $x$ can be in $i$ (case (i)), on the spine (case (ii)), in the top boundary node (case (ii)), or in an ancestor of $i$ in the macro tree (case (iii)). If $x$ is in the same cluster as $x'$ then it follows from the correctness of $\Fl_C$ that $x \in N$. Thus $x$ is added to $R$ and due to the correctness of $\Deep$ we have $x \in \Fl(X,\alpha)$. If $c(x)$ is in a different cluster than $c(x')$ then $c(x)$ is an ancestor of $c(x')$ in the macro tree due to Proposition~\ref{lem:ancestorlemma}. Since $x \in \Deep(F)$ we have $N=\emptyset$ and thus $\parent(v) \prec_M c(x')$ is added to $L$. It follows from the correctness of $\Fl_M$ that $c(x) \in S$. Due to the structure of the macro tree $c(x)$ is either a boundary node or a spine node and thus $x=\fl_{C(c(x))}(\first(c(x)),\alpha)=\Fl_{C(c(x))}(\first(c(x)),\alpha)$. The last equality follows from the correctness of $\Fl_C$. That $x \in \Fl(X,\alpha)$ now follows from the above analysis showing that only nodes from $F$ are added to $R$ and the correctness of \Deep.

If $i$ is a leaf node then $x$ can be in $i$ (case (i)), in the top boundary node (case (iii)), or in an ancestor of $i$ in the macro tree (case (iii)). The correctness follows by an analysis similar to the one for the previous case. If $i$ is a spine node or a boundary node, then $x$ is either in $i$ (case (i)) or  in an ancestor of $i$  in the macro tree (case (iii)). The correctness follows by an analysis similar to the one for the first case.
\end{proof}


\subsection{Complexity of the Tree Inclusion Algorithm}
To analyze the complexity of the node array implementation we first bound the running time of the above implementation of the set procedures. All procedures scan the input from left-to-right while gradually producing the output. In addition to this procedure $\Fl$ needs a call to a node list implementation of $\Fl$ on the macro tree. Given the data structure described in Section~\ref{sec:preprocessing} it is easy to check that each step in the scan can be performed in $O(1)$ time giving a total of $O(n_{T}/\log n_T)$ time. Since the number of nodes in the macro tree is $O(n_{T}/\log n_T)$, the call to the node list implementation of $\Fl$ is easily done within the same time. Hence, we have the following lemma.
\begin{lemma}\label{lem:auxmacro}
For any tree $T$ there is a data structure using $O(n_T)$ space
and $O(n_T\log n_{T})$ preprocessing time which supports all of the
set procedures in $O(n_T/\log n_T)$ time.
\end{lemma}
Next consider computing the deep occurrences of $P$ in $T$ using
the procedure $\Emb$ of Section~\ref{sec:recursion} and
Lemma~\ref{lem:auxmacro}. The following lemma bounds the space usage.
\begin{lemma}\label{lem:lin_space_wc}
The total size of the saved embeddings at any time during the computation of $\Emb(\roots (P))$ is 
$O(n_T)$.
\end{lemma}
\begin{proof}
Let $v$ be the node for which we are currently computing \Emb. Let $p$ be the path from the root to $v$ and let $w_0,\ldots, w_l$ be the light nodes on this path. We have $l=\lightdepth(v)$. 
As in the proof of Lemma~\ref{lem:lin_space}  it
suffices to bound $\norm{\Emb(\heavy(\parent(w_i)))}$ for all $i$. Assume that $l_P \leq l_T$
(otherwise we can check this in linear time and conclude that $P$ cannot
be included in $T$). Each of the node arrays use $O(n_T/\log n_T)$ space
and therefore by Corollary~\ref{cor:ldepth}  we have that   $\sum_{i=1}^{l}\norm{\Emb(\heavy(\parent(w_i)))}= O(n/\log n_T
\cdot \log l_P) = O(n_T)$.
\end{proof}
For the time complexity note that
during the computation of $\Emb(\roots(P))$ each node $v \in V(P)$
contributes a constant number of calls to the set procedures. Hence,
the total time used by the algorithm is $O(n_Pn_T/\log n_T + n_{T}\log n_{T})$. Thus we have shown
the following.
\begin{theorem}\label{thm:complex}
For trees $P$ and $T$ the tree inclusion problem can be solved in
$O(n_Pn_T/\log n_T + n_{T}\log n_{T})$ time and $O(n_T)$ space.
\end{theorem}
Combining the results in Theorems~\ref{thm:simple},
\ref{thm:complex} and Corollary~\ref{cor:simple} we
have the main result of Theorem~\ref{thm:main}.

\section{Conclusion}
We have presented three algorithms for the tree inclusion problem, which match or improve the best known time complexities while using only linear space. We believe that some of the new ideas are likely to be of both practical and theoretical value in future work. From a practical perspective, space is a common bottleneck for processing large data sets and hence reducing the space can significantly improve performance in practice. From a theoretical perspective, we have introduced several non-trivial algorithms to manipulate sets of nodes in trees that may have applications to other problems. For instance, the $\Nca$ procedure from Section~\ref{micromacro} computes multiple nearest common ancestor queries in time \emph{sublinear} in the size of input sets.

\section*{Acknowledgments}
We would like to thank the anonymous reviewers of earlier drafts of this paper for many valuable comments that greatly improved the quality of the paper. We would also especially like to thank the reviewer who discovered the error in the space complexity of the original draft.

\bibliographystyle{abbrv}
\bibliography{pub}

\end{document}